\documentclass[aps,prl,twocolumn,superscriptaddress]{revtex4-1} 
\usepackage{graphicx}
\usepackage{xcolor}
\usepackage{subfigure}
\usepackage{amsmath}
\usepackage{braket}
\usepackage{mathtools}
\usepackage{bm} 
\usepackage[colorlinks= true,urlcolor=blue,linkcolor= blue,citecolor=blue,bookmarks=false,pdfstartview=]{hyperref}

\begin{document}

\title{Anomalous magnetic exchange in a dimerized quantum-magnet \\composed of unlike spin species}

\author{S. P. M. Curley}
\affiliation{Department of Physics, University of Warwick, Gibbet Hill Road, Coventry, CV4 7AL, UK}
\author{B. M. Huddart}
\affiliation{Centre for Materials Physics, Durham University, Durham, DH1 3LE, UK}
\author{D. Kamenskyi}
\affiliation{Experimental Physics V, Center for Electronic Correlations and Magnetism,
Institute of Physics, University of Augsburg, 86135 Augsburg, Germany}
\affiliation{Molecular Photoscience Research Center, Kobe University, 657-8501 Kobe, Japan}
\author{M. J. Coak}
\affiliation{Department of Physics, University of Warwick, Gibbet Hill Road, Coventry, CV4 7AL, UK}
\author{R. C. Williams}
\affiliation{Department of Physics, University of Warwick, Gibbet Hill Road, Coventry, CV4 7AL, UK}
\author{S. Ghannadzadeh}
\affiliation{Department of Physics, Clarendon Laboratory, Oxford University, Parks Road,
Oxford, OX1 3PU, UK}
\author{A. Schneider}
\affiliation{Experimental Physics VI, Center for Electronic Correlations and Magnetism,
Institute of Physics, University of Augsburg, 86135 Augsburg, Germany}
\author{S. Okubo}
\affiliation{Molecular Photoscience Research Center, Kobe University, 657-8501 Kobe, Japan}
\author{T. Sakurai}
\affiliation{Molecular Photoscience Research Center, Kobe University, 657-8501 Kobe, Japan}
\author {H. Ohta}
\affiliation{Molecular Photoscience Research Center, Kobe University, 657-8501 Kobe, Japan}
\author{J. P. Tidey}
\affiliation{Department of Chemistry, University of Warwick, Gibbet Hill, Coventry CV4 7AL, UK}
\author{D. Graf}
\affiliation{National High Magnetic Field Laboratory, Florida State University, Tallahassee, Florida 32310, USA}
\author{S. J. Clark}
\affiliation{Centre for Materials Physics, Durham University, Durham DH1 3LE, UK}
\author{S. J. Blundell}
\affiliation{Department of Physics, Clarendon Laboratory, Oxford University, Parks Road,
Oxford, OX1 3PU, UK}
\author{F. L. Pratt}
\affiliation{ISIS Facility, Rutherford Appleton Laboratory, Chilton, Oxfordshire, OX11 0QX, UK}
\author{M. T. F. Telling}
\affiliation{ISIS Facility, Rutherford Appleton Laboratory, Chilton, Oxfordshire, OX11 0QX, UK}
\author{T. Lancaster}
\email{tom.lancaster@durham.ac.uk}
\affiliation{Centre for Materials Physics, Durham University, Durham, DH1 3LE, UK}
\author{J. L. Manson}
 \email{jmanson@ewu.edu}
\affiliation{Department of Chemistry and Biochemistry, Eastern Washington University, Cheney, Washington 99004, USA}
\author{P. A. Goddard}
 \email{p.goddard@warwick.ac.uk}
\affiliation{Department of Physics, University of Warwick, Gibbet Hill Road, Coventry, CV4 7AL, UK}



\begin{abstract}

We present here a study of the magnetic properties of the antiferromagnetic dimer material CuVOF$_4$(H$_2$O)$_6\cdot$H$_2$O, in which the dimer unit is composed of two different $S = 1/2$ species, Cu(II) and V(IV). An applied magnetic field of $\mu_0H_{\rm c1} = 13.1(1)$\,T is found to close the singlet-triplet energy gap, the magnitude of which is governed by the antiferromagnetic intradimer, $J_0 \approx 21$\,K, and interdimer, $J' \approx 1$\,K, exchange energies, determined from magnetometry and electron-spin resonance measurements. The results of density functional theory (DFT) calculations are consistent with the experimental results and predicts antiferromagnetic coupling along all nearest-neighbor bonds, with the magnetic ground state comprising spins of different species aligning antiparallel to one another, while spins of the same species are aligned parallel. The magnetism in this system cannot be accurately described by the overlap between localized V orbitals and magnetic Cu orbitals lying in the Jahn-Teller (JT) plane, with a tight-binding model based on such a set of orbitals incorrectly predicting that interdimer exchange should be dominant. DFT calculations indicate significant spin density on the bridging oxide, suggesting instead an unusual mechanism in which intradimer exchange is mediated through the O atom on the Cu(II) JT axis.


\end{abstract}

\maketitle

\section{Introduction}

Cooperative phenomena in materials known to exhibit quantum critical points (QCPs) have been the subject of consistent interest in condensed matter physics \cite{Gegenwart2008,Stockert2011,Zapf2014a}. In particular, systems of antiferromagnetically (AFM) coupled $S$ = 1/2 dimers have been known to exhibit two magnetic-field-induced phase-transitions, the first of which, at least, involves the system passing through a QCP which, under certain conditions, belongs to the Bose-Einstein Condensate (BEC) universality class \cite{Zapf2014a, Lancaster2014b}.

In zero-field (ZF) and at low-temperatures, weakly interacting $S = 1/2$ AFM dimers exist in a state of quantum-disorder, the ground-state being composed of a sea of spin-singlets ($S=0$) situated against a backdrop of quantum fluctuations. Above this singlet ground-state resides a degenerate excited triplet-state ($S =1$), with the size of the singlet-triplet energy gap dictated by the strength of the intradimer AFM Heisenberg exchange interaction, $J_0 > 0$. The presence of any interdimer exchange, $J'$, serves to disperse the excited triplet, giving a band of excitations and reducing the size of the singlet-triplet energy gap relative to the case for an isolated dimer.

Upon application of an external field, the system moves through the first QCP, at $H_{\rm{c1}}$, as the Zeeman interaction splits the degenerate $S = 1$ triplet and lowers the energy of the $S_z = +1$ state below that of the $S = 0$ singlet ground-state, such that at $H_{\rm{c1}}$ the system enters a long-range $XY$-AFM ordered state. Under certain conditions, the triplet-excitations in the ordered state can be described as bosonic quasi-particles \cite{Zapf2014a}. Further application of field eventually fully polarizes the spins along the field direction, as the system enters a ferromagnetic (FM) saturated state above $H_{\rm{c2}}$.

In order for the excited triplet-state to effectively map onto a BEC of magnons picture, the transverse component of the spins must spontaneously break the rotational $O$(2) symmetry [analogous to the $U$(1) symmetry present in an atomic BEC] of the system at $H_{\rm{c1}}$ \cite{Giamarchi2008}. Thus, any term which breaks the rotational symmetry of the spin Hamiltonian prohibits the system from being described within the BEC universality class \cite{Zheludev2013}. Dimers which exhibit an excited triplet-state where the crystal structure breaks the $O$(2) symmetry have been reported previously \cite{Nawa2011,Sebastian2006}.

However, we present here the magnetic properties of the dimer system CuVOF$_4$(H$_2$O)$_6\cdot$H$_2$O \cite{Donakowski2012}, where the rotational symmetry is broken not only by the structure, but also by the spin-species which make up the dimer-unit. Within an applied magnetic field, the system can be modelled as a lattice of weakly coupled $S$ = 1/2 AFM dimers interacting via Heisenberg exchange, with the magnetic properties summarised by:
\begin{multline}\label{eq:Hamiltonian}
\mathcal{H} = J_0\sum\limits_{i} {\bf{\hat{S}}}_{1,i}\cdot{\bf{\hat{S}}}_{2,i}
+ \sum\limits_{<mnij>}J'_{mnij}
{\bf{\hat{S}}}_{m,i}\cdot{\bf{\hat{S}}}_{n,j} \\
- g\mu_{\rm{B}}\mu_{0}H\sum \limits_{i} {\hat{S}}^{z}_{\textit{m,i}}
\end{multline}
where $i$ and $j$ denote dimers and $m,n$ = 1,2 label magnetic sites \cite{Lancaster2014b,Giamarchi2008}. We note that as the dimer-unit lacks a center of inversion symmetry, there is the possibility of an additional Dzyaloshinskii-Moriya interaction (DMI) term in the Hamiltonian of the form ${\boldsymbol{D}} \cdot ({\bf{S}}_1\times {\bf{S}}_2)$. The DMI term is expected to be small. An order-of-magnitude estimate can be obtained from the departure of the g-factor from the free electron value $|D| \sim (\Delta g/g) J_0$ \cite{Moriya_DMI_1960}. For the dimer compound Ba$_3$Cr$_2$O$_8$ $(\Delta g/g) J_0 \sim 1$~K with $J_0 = 27.6(2)$~K and we expect similar values for our material \cite{Kofu2009,Aczel2009a}.

In this paper, we present ZF muon-spin relaxation data that indicate an absence of magnetic order down to temperatures of 100~mK, typical behavior for a system of weakly interacting dimers \cite{Lancaster2014b}. In addition, radio-frequency (RF) susceptometry measurements confirm the existence of two field-induced phase-transitions akin to behavior observed in other BEC class dimers \cite{Jaime2004,Aczel2009}, and allow the magnetic phase diagram to be elucidated.

Due to the exceptional energy resolution and relevant frequency range, electron-spin resonance (ESR) is one of the most appropriate experimental techniques to probe the singlet-triplet excitations. Such transitions have been observed by high-frequency ESR in many AFM spin-dimers, such as: SrCu$_2$(BO$_3$)$_2$  \cite{SrCuBO_1,SrCuBO_2} and CuTe$_2$O$_5$ \cite{CuTeO} based on Cu(II) (3$d^9$, $S$ = 1/2) ions; Ba$_3$Cr$_2$O$_8$~\cite{BaCrO,BaCrO_SrCrO} and Sr$_3$Cr$_2$O$_8$~\cite{BaCrO_SrCrO,SrCrO} based on Cr$^{5+}$ (3d$^1$, $S$ = 1/2). Here, ESR measurements directly observe the closure of the singlet-triplet energy gap in CuVOF$_4$(H$_2$O)$_6\cdot$H$_2$O and highlight several excitations in the system, including a so-far-unidentified resonance which appears to be unique to this system.


As detailed in Ref. \cite{Donakowski2012}, CuVOF$_4$(H$_2$O)$_6\cdot$H$_2$O is composed of the two unlike $S = 1/2$ ions, Cu(II) (3$d^9$) and V(IV) (3$d^1$), linked via a lambda-shaped Cu---O---V bond. Work in Ref. \cite{Donakowski2012} showed that the formation of this Cu---O---V bond relies on the Jahn-Teller (JT) distortion of the Cu(II) octahedra, and replacing Cu(II) with other $M$(II) transition-metal ions ($M$ = Ni, Zn, Co) results in the $M$(II) and V(IV) ions forming isolated octahedra. In this work, we demonstrate that the JT-active Cu is not only responsible for the polar structure, but also the low-dimensional magnetism in the system. Density functional theory calculations show that the unlike spin species likely play an important role in the intradimer exchange mechanism in this compound, which appears to be distinct from the exchange coupling picture, typical for Cu(II) magnets, of overlapping $d$ orbitals within the JT plane.


\section{Results}

\subsection{Structure}

\begin{figure}[t]
\centering
\includegraphics[width=0.8\columnwidth]{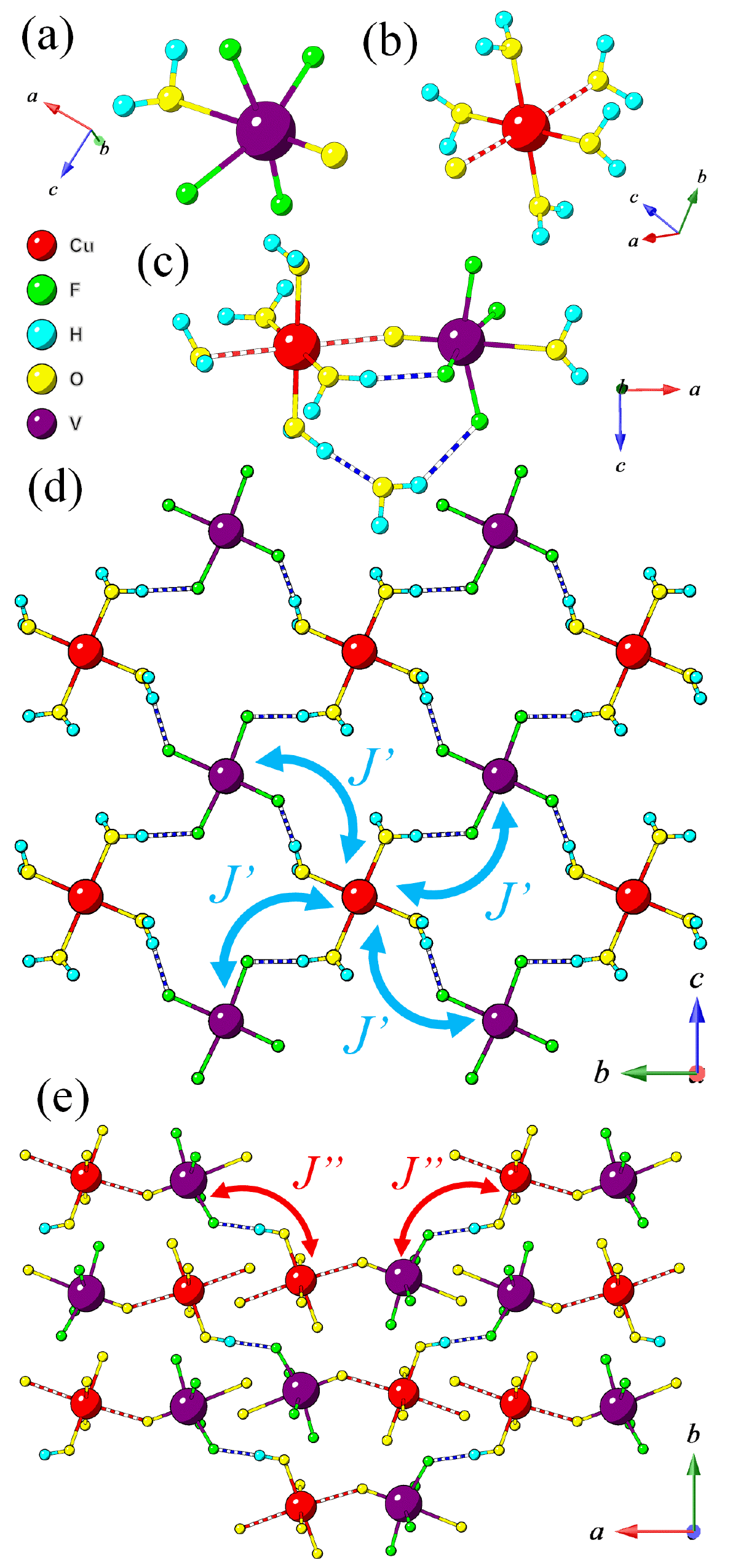}
\caption[width=1 \linewidth]{\small Local octahedral environment of (a) V(IV) and (b) Cu(II). Red-striped bonds indicate the Jahn-Teller axis of the Cu(II) octahedra. (c) Dimer unit with intradimer H-bonds (blue-striped bonds) through equatorial ligands and uncoordinated water molecule. (d) Interdimer H-bond network within the $bc$-plane expected to mediate primary interdimer exchange ($J'$). Uncoordinated waters are omitted for clarity. (e) Packing of the dimers within the $ab$-plane showing equatorial H-bonds expected to mediate secondary interdimer exchange ($J''$) assumed to be very small; see text. H-bonds between axial and uncoordinated waters omitted for clarity. Structure is determined from single-crystal X-ray diffraction data collected at 150~K; see SI~\cite{SI}.}  \label{Fig:Struc}
\vspace{-0cm}
\end{figure}
\noindent

Single-crystal X-ray diffraction data indicate CuVOF$_4$(H$_2$O)$_6\cdot$H$_2$O crystallises into an orthorhombic structure with polar space-group $Pna2_{1}$, in agreement with Ref.~\cite{Donakowski2012}. The magnetic structure of CuVOF$_4$(H$_2$O)$_6\cdot$H$_2$O is based on a lattice of weakly interacting antiferromagneticaly coupled $S = 1/2$ dimers. The dimer-unit itself is composed of two differing $S$ = 1/2 ions, V(IV) and Cu(II), both of which reside in octahedral environments of the form VF$_4$O$_2$, Fig.~\ref{Fig:Struc}(a), and CuO$_6$, Fig.~\ref{Fig:Struc}(b). Figure~\ref{Fig:Struc}(c) shows that the Cu(II) and V(IV) ions are linked via a bent Cu---O---V bond [with a bond-angle of $142.87(5)^{\circ}$ and a through-bond distance of 3.942(2)\,\AA] and by a single Cu---O---H$\,\cdots$F---V bond. 

The Jahn-Teller axis of the Cu lies along the bridging Cu---O bond. As such, the unpaired electron (3$d^9$) of the metal centre is expected to reside in the $d_{x^2-y^2}$ orbital, oriented in the plane perpendicular to this bond and directed along the shorter Cu---O bonds which lie within the JT plane.
For V, density functional theory (DFT) calculations (outlined below) indicate significant spin-density between the V---F bonds, also perpendicular to the bridging V---O bond. 
It should be noted that JT-active Cu(II) systems often exhibit extreme low-dimensionality, as the reduced orbital overlap along JT direction typically leads to strong superexchange interactions only along pathways perpendicular to the JT axis, such as in the quasi-two-dimensional [Cu(HF$_2$)(pyz)$_2$]SbF$_6$ \cite{Manson2009a} and the quasi-one-dimensional Cu(NO$_3$)$_2$(pyz)$_3$ \cite{Huddart2019} (pyz = pyrazine = C$_4$H$_4$N$_2$) molecule-based magnets. 
As the magnetic orbitals of both the Cu and V lie within the plane of the equatorial ligands of each octahedral environment, one might expect that the minimal orbital overlap along the Cu---O---V bond direction would lead to the intradimer exchange coupling being mediated along the intradimer Cu---O---H$\,\cdots$F---V H-bond pathway, seen in Fig.~\ref{Fig:Struc}(c). H-bonds have previously been shown to be highly effective mediators of superexchange interactions in low-dimensional magnets, such as [CuF$_2$(H$_2$O)$_2$(pyz)] \cite{CuF2H2Opyz} and CuSO$_4$(C$_2$H$_8$N$_2$)$\cdot$2H$_2$O \cite{Kravchina2011}. However, it is shown later, that this is not the case for this material.


Weak H-bonds between the dimers form a complex 3D interdimer network, outlined in detail in \cite{Donakowski2012}. Only the interdimer H-bonds between equatorial ligands are expected to mediate significant magnetic exchange, as the magnetic orbitals of both transition-metal ion species lie within the plane of the equatorial ligands, with no spin-density located on the axial water ligands of either species (see DFT below). As such, the primary interdimer exchange is expected to act within the $bc$-plane via the H-bond network shown in Fig.~\ref{Fig:Struc}(d), resulting in each dimer having four nearest dimer neighbors $n = 4$. 
There may also be some very weak exchange within the $ab$-plane ($J''$). Most interdimer H-bond pathways within the $ab$-plane involve the JT (pseudo-JT) axis of the Cu (V). The only interdimer pathways which do not are the equatorial H-bonds highlighted in Fig.~\ref{Fig:Struc}(e) (which bridge adjacent Cu and V ions), making them the most probable $J''$ exchange pathways.
It should be noted that adjacent Cu octahedra throughout the lattice are arranged in a staggered fashion (likewise for adjacent V octahedra), as seen in Fig.~\ref{Fig:Struc}(e), indicating a staggered $g$-tensor within the system.

\subsection{Magnetometry}

\subsubsection{SQUID magnetometry}

\begin{figure}[t]
\centering
\includegraphics[width=1 \linewidth]{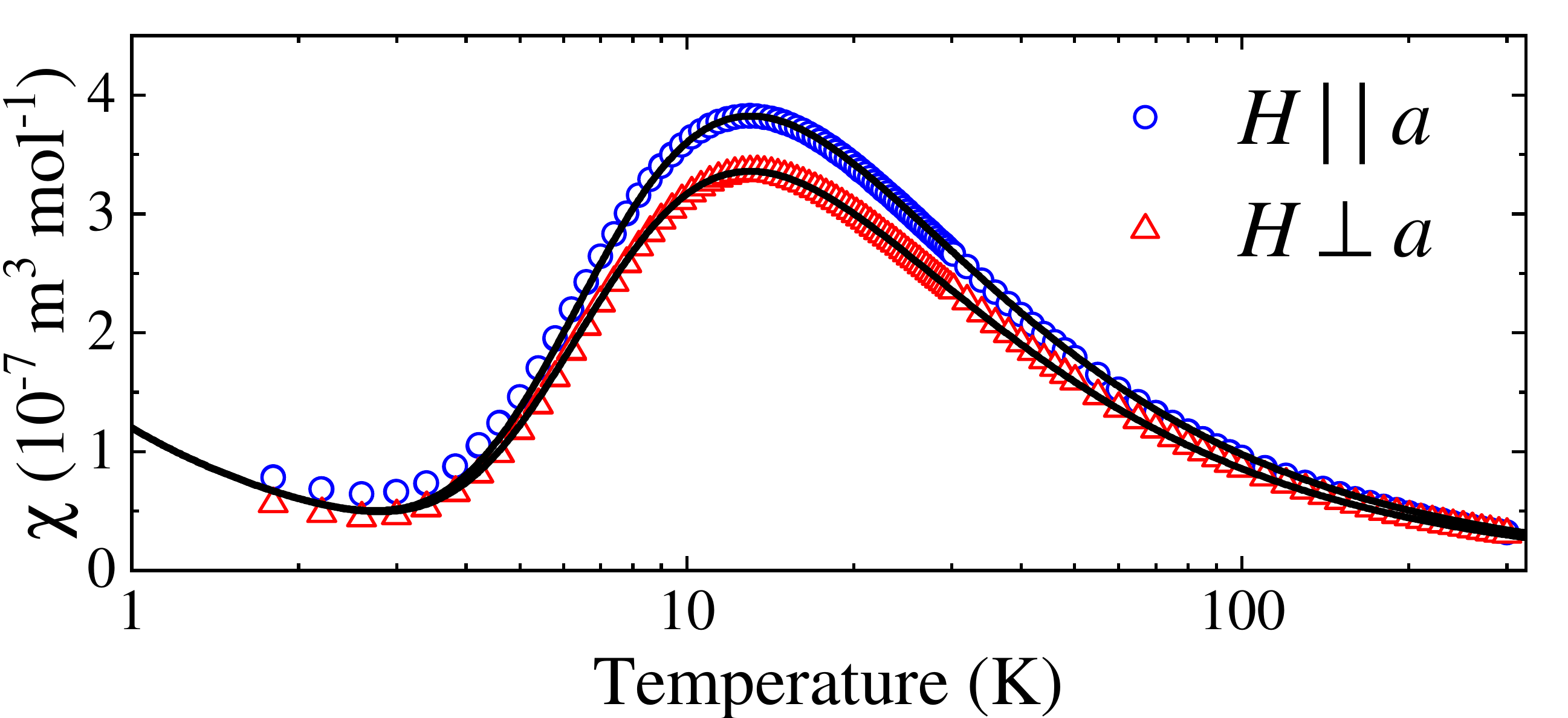}
\caption[width=1 \linewidth]{\small Quasi-static DC-field magnetic susceptibility $\chi(T)$ for an orientated single-crystal of CuVOF$_4$(H$_2$O)$_6\cdot$H$_2$O measured in an applied field of $\mu_{0}H$ = 0.1 T. Solid lines are a global fit of Eq.~\ref{eq:BB} to both data sets as described in the text.}
\vspace{-0cm} \label{fig:CuV_susc}
\end{figure}

Figure~\ref{fig:CuV_susc} shows the static magnetic susceptibility [$\chi$($T$)] for a single crystal of CuVOF$_4$(H$_2$O)$_6\cdot$H$_2$O with field orientated parallel and perpendicular to the crystallographic $a$ axis (which lies close to parallel with the Cu---O---V bond). 
Upon decreasing temperature, $\chi$($T$) data in both orientations rise to a broad hump centered around 15~K, decrease down to $T \approx 3$~K, and then exhibit a slight upturn at $T < 3$~K; behavior typical of AFM coupled spin-half dimers.
Over the measured temperature range, $1.8 \leq T \leq~300~\rm{K}$, $\chi$($T$) can well described using a Bleaney-Bowers model with mean-field interactions $\chi_{\rm{b}}(T)$ \cite{Aczel2009, Aczel2009a} plus a $\chi_{\rm{pm}}(T)$ term to model the low-temperature paramagnetic tail, of the form,
\noindent
\begin{equation}\label{eq:BB}
    \chi = (1-\rho)  \chi_{\rm{b}}(T) + \rho\chi_{\rm{pm}}(T)
\end{equation} 
\noindent
where $\rho$ captures the fraction of the sample attributable to uncoupled $S$ = 1/2 spins due to impurities and broken dimers, or, possibly arising from the staggering of the local $g$-tensor as seen in staggered $S = 1/2$ chains \cite{Feyerherm2000,Liu2019} (the full form of Eq.~\ref{eq:BB} can be found in the SI~\cite{SI}).
The $\chi(T)$ datasets were fit simultaneously sharing $J_{0}$, $J'$ and $\rho$ as global parameters, but with $g$-factors free to vary for each dataset. The resultant fit is shown in Fig.~\ref{fig:CuV_susc} (solid lines) and returns parameters of $J_{0} = 21.3(1)$~K, $J' = 1.3(1)$~K (taking $n = 4$ from the structure) and $\rho = 2.5(1)\%$. The extracted $g$-factors of $g_{a} = 2.1(1)$ and $g_{bc} = 2.0(1)$ are in excellent agreement with the values determined from ESR measurements discussed below.
\noindent
\subsubsection{Radio frequency susceptometry}

\begin{figure}[t]
\centering
\includegraphics[width = \linewidth]{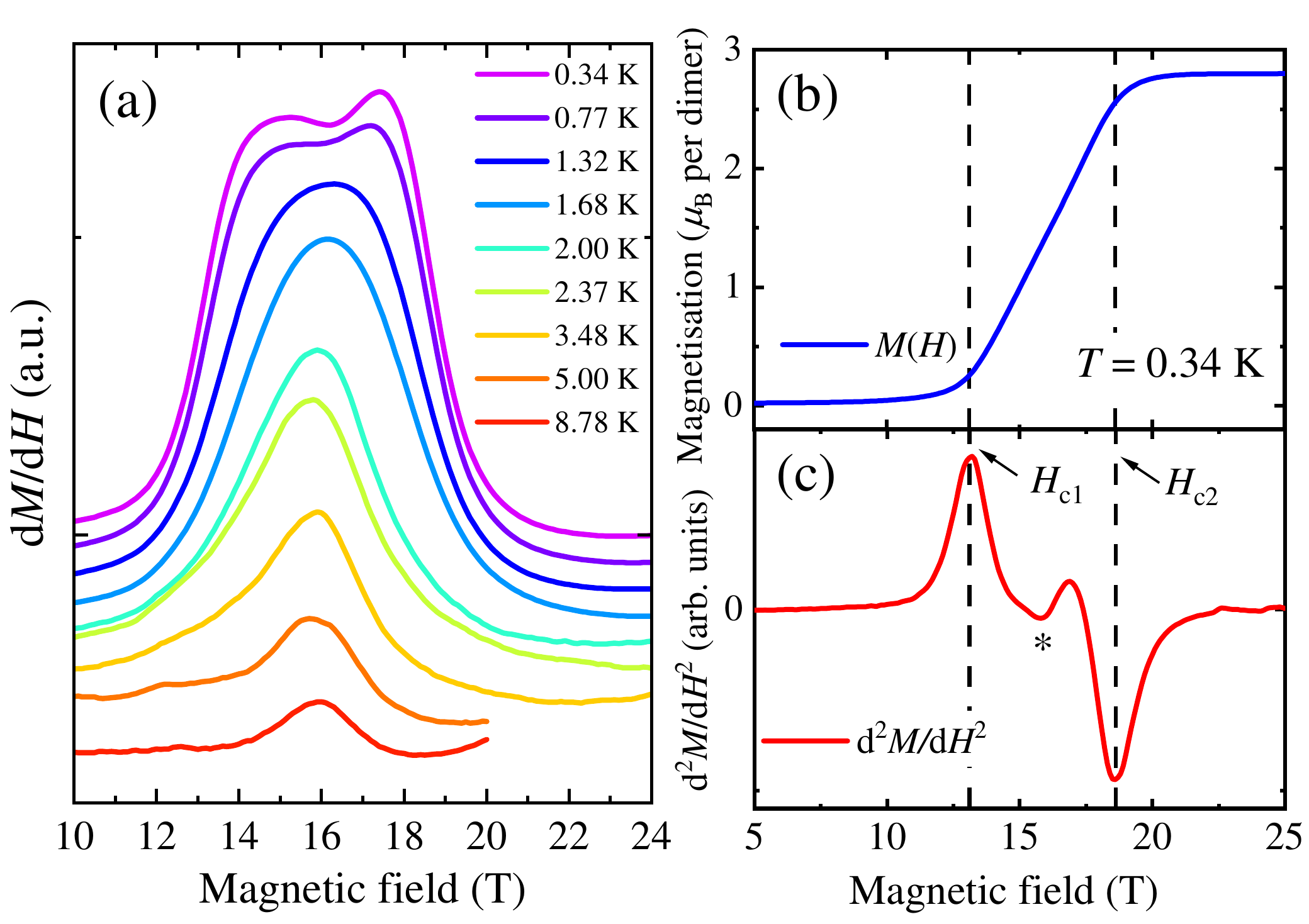}
\caption[width=1 \linewidth]{\small (a) Dynamic susceptibility d$M$/d$H$ measured at several temperatures using radio-frequency (RF) susceptometry with field parallel to the $a$-axis. Data are offset at each temperature for clarity. Magnetization $M$($H$) (b) and its second derivative (d$^2 M$/d$H^2$) (c) measured at $T = 0.34$~K extracted from the RF susceptometry. The positions of the first and second critical fields, $H_{\rm{c1}}$ and $H_{\rm{c2}}$, derived from d$^2 M$/d$H^2$, are marked with dashed lines in (b) and (c). The minimum feature in d$^2 M$/d$H^2$, discussed in text, is marked with an asterisk.}  \label{fig:dynamic_susc}
\vspace{-0cm}
\end{figure}

Figure~\ref{fig:dynamic_susc}(a) shows the field dependence of the differential magnetization (d$M$/d$H$) measured at various temperatures for a single-crystal of
CuVOF$_4$(H$_2$O)$_6\cdot$H$_2$O, with field parallel to $a$, measured using a radio-frequency oscillator technique \cite{Athas1993, Coffey2000}. 
Measurements were performed using quasi-static fields to mitigate magnetocaloric effects known to be present in dimer systems in rapidly changing magnetic fields \cite{Brambleby2017b, Ghannadzadeh2011}.

\noindent

At low temperatures, d$M$/d$H$ exhibits two peak-like features centered around 14.5~T and 18.5~T, which coalesce and become unresolvable as separate peaks for 1.68~$< T \leq~2$~K.
Typically in $S$ = 1/2 dimer systems, sharp cusps are observed in d$M$/d$H$ at the critical fields $H_{\rm c1}$ and $H_{\rm c2}$ \cite{Lancaster2014b,Brambleby2017b}. The reason for the broad nature of the features in d$M$/d$H$ here is unknown, but could arise due to H-bond disorder within the complex interdimer exchange network, giving rise to a distribution in the superexchange between neighboring transition metal sites and a smearing of the transition features in d$M$/d$H$ \cite{Manson2018}.

The magnetization $M$($H$), shown in Fig.~\ref{fig:dynamic_susc}(b), is extracted by integrating the measured d$M$/d$H$ response and calibrated using similar temperature DC-field SQUID data. The magnetization saturates at $M_{\rm{sat}} = 2.7(1)~ \mu_{\rm{B}}$ per dimer unit, in excellent agreement with the calculated effective moment of 2.7 $\mu_{\rm{B}}$ per dimer determined in \cite{Donakowski2012}. The low-temperature $M$($H$) response is typical for a system of weakly interacting $S$ = 1/2 AFM dimers \cite{Lancaster2014b,Jaime2004}, with a sharp upturn at $H = H_{\rm c1}$ corresponding to the closing of the singlet-triplet energy gap and a levelling off at $H = H_{\rm c2}$ indicating the spins are fully polarised along the field direction at $H_{\rm c2}$.

Fig.~\ref{fig:dynamic_susc}(c) shows the peaks in the second derivative of the $M$($H$) (d$^2 M$/d$H^2$) that we use to track the positions of $H_{\rm c1}$ and $H_{\rm c2}$, as done previously for other dimer systems \cite{Nakajima2006}. The minimum in d$^2 M$/d$H^2$ between the two critical fields, marked with an asterisk, disappears for 1.68~$< T \leq~2$~K. This is the same temperature range where the two peak features in d$M$/d$H$ coalesce [Fig.~\ref{fig:dynamic_susc}(a)] and provides a consistent estimate for the temperature limit beyond which the two critical fields can no longer be resolved.



\noindent
\noindent

\subsection{Electron-spin resonance}
\noindent
\subsubsection{$X$-band ESR}
\begin{figure}[t]
	\includegraphics[width=0.45\textwidth]{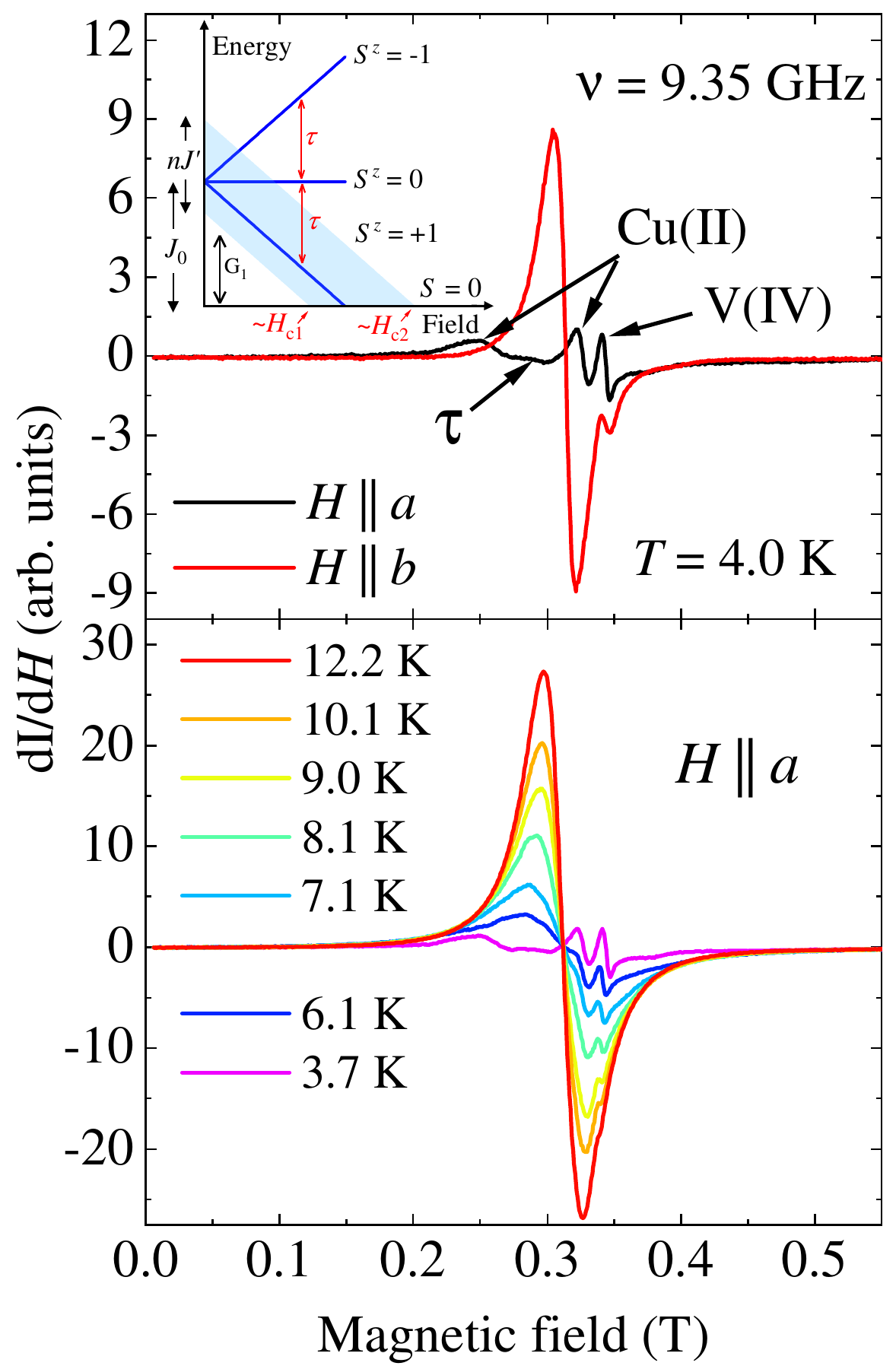}
	\caption{Top panel shows $X$-band ESR spectra of CuVOF$_4$(H$_2$O)$_6\cdot$H$_2$O for $H \parallel a$ (black curve) and H $\parallel b$ (red curve) measured at 4.0~K. Inset shows schematic energy-level diagram for interacting $S = 1/2$ dimers. Bottom panel shows the temperature evolution of ESR spectra for $H \parallel a$.}
	\label{fig:X_band_spectra}   
\end{figure}
\noindent

Figure~\ref{fig:X_band_spectra} displays the low-temperature electron-spin resonance spectra (the derivative of the absorption) measured at a frequency of $\nu = 9.35$~GHz. The top panel compares spectra for field along the $a$- and $b$-axis, respectively. 
For field along $a$ (black curve) the spectrum consists of four peaks indicated with arrows. Upon increasing temperature the magnitude of the three most intense peaks diminishes, indicating these resonances can be attributed to free paramagnetic ions with $g$-factors: 1.95, originating from V(IV); 2.37, originating from Cu(II); and 2.05, which is likely due to Cu(II) impurities. However, we note the additional possibility of an uncompensated moment on the dimer due to the non-identical spins. This would likely manifest as a $g$-factor between that of the Cu and V values. The weak peak marked $\tau$ corresponds to transitions ($\Delta S_z = \pm 1$) that occur within the excited triplet state. 
For field along $b$ only two peaks can be resolved: that arising from V(IV) with $g=1.965$ and a very intense absorption with $g=2.15$ which derives from transitions within the triplet and Cu(II) impurities. These results are in agreement with the magnetometry, which also suggests the presence of a small component resembling free $S$ = 1/2 ions within the sample.


The temperature evolution of the ESR spectra with $H \parallel a$ is shown in the bottom panel of Fig.~\ref{fig:X_band_spectra}. As expected, upon increasing temperature the intensity of the transitions within the excited triplet ($\tau$-peak) increases rapidly as their population increases, whilst the intensities of the other transition peaks decreases. Above 10~K, the triplet transitions dominate the excitation spectrum exhibiting a maximum intensity at 17~K, in agreement with the hump around 15~K seen in $\chi(T)$ (Fig.~\ref{fig:CuV_susc} above).
At 10~K, the $g$-factor of the $\tau$-mode with $H \parallel a$ is 2.14 and grows to 2.155 upon warming to room temperature $T = 295$~K. The $g$-factors along the principal axes are therefore $g_a=2.145$, $g_b=2.046$ and $g_c=2.055$ at 20~K and  $g_a = 2.155$, $g_b = 2.039$ and $g_c = 2.047$ at room temperature; see Supplemental Information (SI)~\cite{SI}.
\noindent
\subsubsection{High-frequency ESR}
\begin{figure}[t]
	\includegraphics[width=0.45\textwidth]{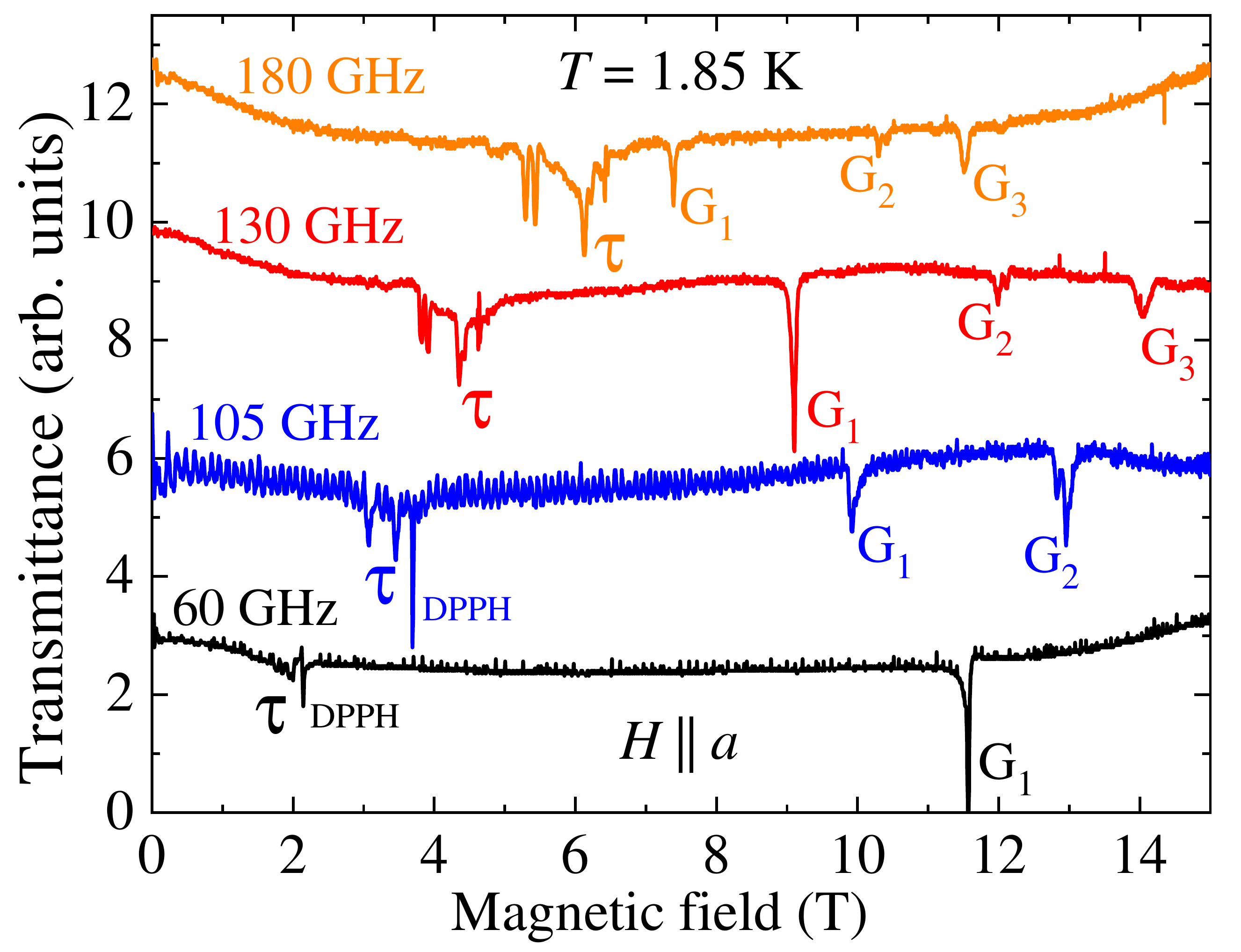}
	\caption{ESR spectra of CuVOF$_4$(H$_2$O)$_6\cdot$H$_2$O at different radiation frequencies measured at 1.85~K with $H \parallel a$. A sharp peak marked ``DPPH'' corresponds to the absorption by the field marker compound. The peaks labelled $\tau$ and G$_{1,2,3}$ are as described in the text.}
	\label{fig:spectra_a}   
\end{figure}

\noindent
Figure~\ref{fig:spectra_a} shows the ESR spectra measured at frequencies of $\nu=60$, 105, 130 and 180~GHz for field applied approximately along the dimer-axis ($H \parallel a$) with multiple modes observed in the spectra ($\tau$, G$_1$, G$_2$, and G$_3$) that can be categorised based on their field dependence.

The field position of the $\tau$-mode, and the several satellite peaks arising from free paramagnetic spins, increases linearly with the radiation frequency, $h \nu =g \mu_{\rm B} \mu_{0} H$; 
behavior typical for paramagnetic ions and transitions within the excited triplet of a dimer system. These modes are high-field extensions of the excitations observed in the $X$-band spectra discussed above.

The G$_{n}$ modes have a finite energy in zero-field, $\Delta_{\rm G_{\textit{n}}}$, such that $\Delta_{\rm G_{\textit{n}}}$ decreases upon increasing field. This is typical behavior of singlet-triplet transitions whose frequency follows $h \nu = h \Delta_{\rm G_{\textit{n}}}-g \mu_{\rm B} \mu_{0} H$, where $\Delta_{\rm{G}_\textit{n}}$ corresponds to the energy-gap between the $S = 0$ singlet ground-state and $S_z = +1$ triplet state. Often, singlet-triplet transitions are forbidden in simple dimers, however, the presence of a non-zero DMI or an alternating $g$-tensor can serve to mix the wave functions of the spin-singlet ground state and spin-triplet exited states, permitting singlet-triplet transitions \cite{BaCrO}. 

Normally, ESR experiments probe transitions at the center of Brillouin zone ($\Gamma$ point), at $k = 0$. However, under certain circumstances transitions at non-zero $k$ can be observed due to Brillouin zone folding \cite{BaCrO,BaCrO_SrCrO,CsCuBr}. Therefore, we assign the most intense mode, G$_1$, to singlet-triplet excitations at the $\Gamma$ point whilst the other modes, G$_{2}$ and G$_{3}$, are attributed to similar transitions which occur at non-zero $k$.

\begin{figure}[t]
\centering
\includegraphics[width=1 \linewidth]{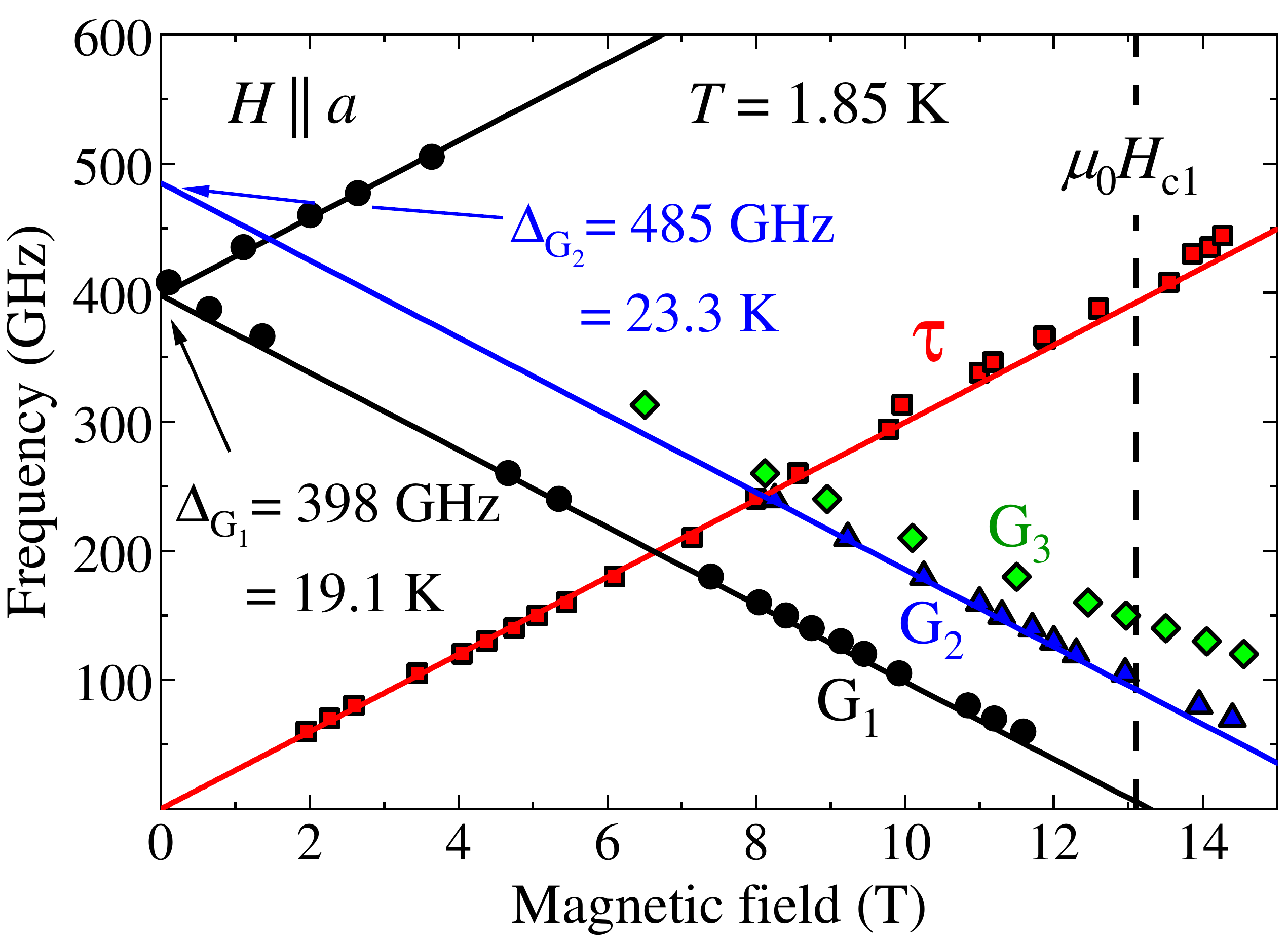}
\caption[width=1 \linewidth]{\small Frequency-field dependence of the ESR transitions in CuVOF$_4$(H$_2$O)$_6\cdot$H$_2$O measured at 1.85~K with $H \parallel a$. Solid lines are the best fit of the data with $g=2.14$ as obtained from the X-Band measurements (Fig.~\ref{fig:X_band_spectra}). Dashed line marks the first critical field, $\mu_0H_{\rm{c1}}$ = 13.1 K, obtained from RF-susceptometry measurements at $T = 0.34~$K with $H \parallel a$.}  
\vspace{-0cm} \label{fig:FFD_a}
\end{figure}

The frequency dependence of the peak positions at 1.85~K is shown in Fig.~\ref{fig:FFD_a}. Solid lines show the best fit of the data obtained using $g=2.14$
taken from the X-Band measurements, the fit returns values of $\Delta_{\rm\rm{G_1}}=398(3)~\rm{GHz}$ [19.1(1)~K] and $\Delta_{\rm{G_2}} = 485(3)$~GHz [23.3(1)~K].

\noindent
\noindent

Typically, the ESR spectra of similar dimer compounds \cite{BaCrO_SrCrO,BaCrO} exhibits two modes, $\Delta_{\rm{G_1}}$ and $\Delta_{\rm{G_2}}$, which provide information about the strength of the intradimer and interdimer coupling. The singlet-triplet zero-field splitting is dictated by the intradimer coupling $J_0$ whilst the interdimer interaction $J'$ determines the dispersion of the triplet state. Values of $J_0$ and $J'$ can be estimated within the random phase approximation \cite{RFA,Kofu} (where $h \Delta_{\rm{G_1,G_2}} = J_0 \mp 2 J'$ for $n = 4$ nearest neighbors) which returns parameters of $J_0=21.0(2)$~K and $J'=1.0(1)$~K, in excellent agreement with magnetometry data.
Extrapolating the G$_1$ mode to zero-frequency estimates values of $\mu_0H_{\rm{c1}}^a = 13.3(2)$~T and $\mu_0H_{\rm{c1}}^{b} = 14.0(2)$~T for field along the $a$ and $b$ axis, respectively, which is close to the critical field  of 13.1(1)~T, for field along the $a$-axis, extracted from RF susceptometry [dashed line Fig.~\ref{fig:FFD_a}]. 

The phase transition occurs at the point when the Zeeman interaction closes the singlet-triplet energy gap. Upon increasing field, the singlet-triplet energy gap need not strictly close at the $\Gamma$ point but may close at another point within the Brillouin zone. Whilst ESR measurements are sensitive to the $\Gamma$ point, the bulk magnetometry probes the whole Brillouin zone. Thus, the critical fields extracted from ESR data must fall between the magnetic-field range outlined by the values of $H_{\rm c1,2}$ determined from magnetometry measurements, as reported in the dimer compound Ba$_3$Cr$_2$O$_8$ \cite{BaCrO} and in line with the results in this work.

The observation of a third mode G$_3$ in the ESR spectrum is unexpected, as only two modes are observed in the ESR spectrum of similar dimer materials \cite{BaCrO_SrCrO,BaCrO,SrCrO}. Whilst the origin of this mode is presently unclear, further ESR experiments are planned to investigate the possible existence of a DMI, and to elucidate the effect this may have on the field-dependence of the dimer energy-levels, as outlined in \cite{Matsumoto2008}.
\noindent
\noindent
\subsection{Muon-spin relaxation}

\begin{figure}[t]
	\includegraphics[width=\columnwidth]{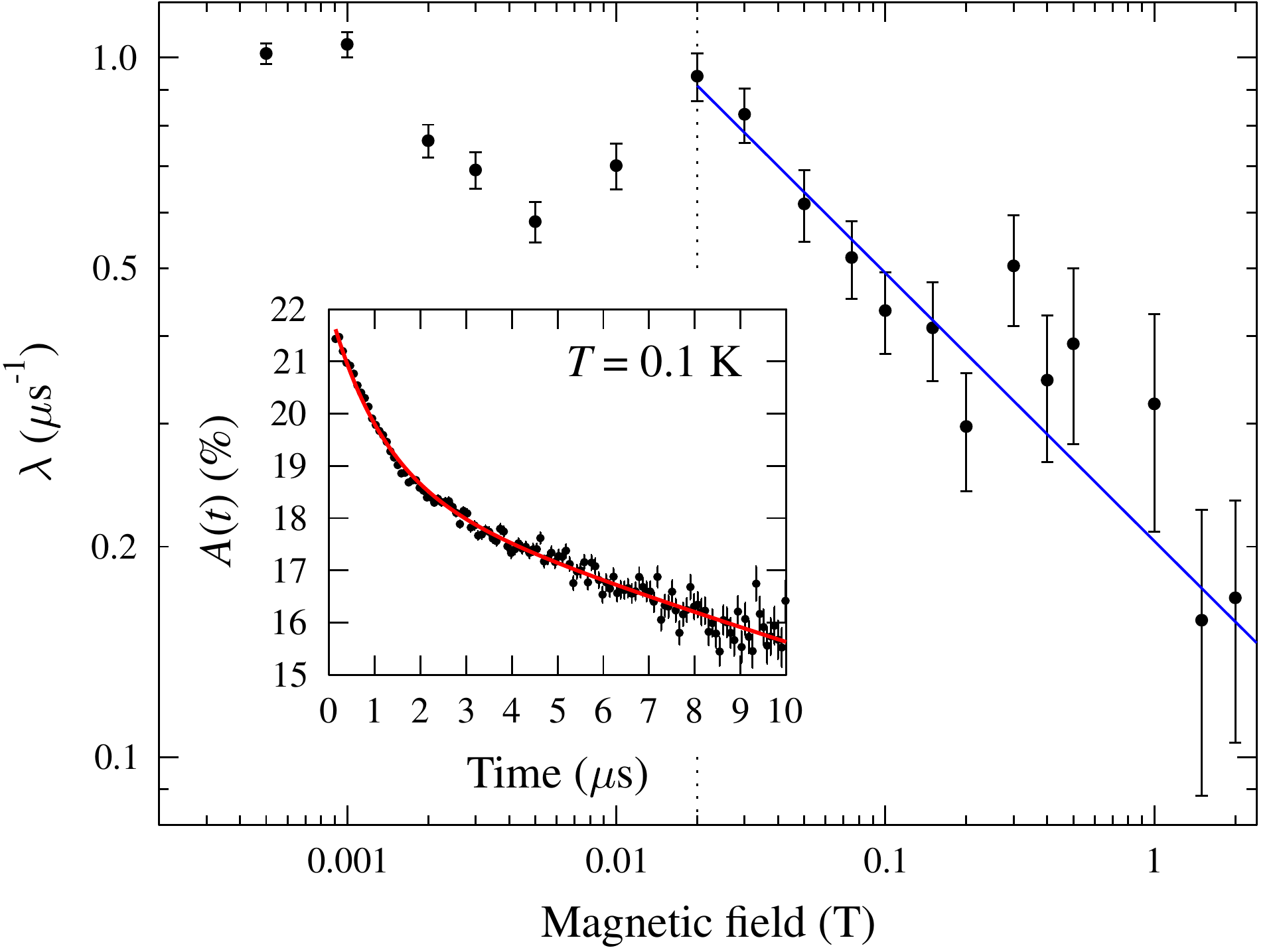}
	\caption{Field-dependence of the longitudinal-field relaxation rate alongside a power-law fit, described in the text, appropriate for diffusive spin transport. Inset: ZF $\mu^+$SR spectrum measured at $T=0.1$~K.}
	\label{fig:muSR_combined}
\end{figure}

Zero-field positive-muon-spin-relaxation ($\mu^+$SR) spectra (inset Fig.~\ref{fig:muSR_combined}) show very little temperature-dependence, and do not show any oscillations in the asymmetry (that would be characteristic of magnetic order) down to 0.1~K; see SI~\cite{SI}. The spectra are instead characterised by exponential relaxation due to fluctuating electronic moments and a slowly-relaxing contribution due to muons implanting at sites not sensitive to these electronic moments. The observation of exponential relaxation due to fluctuating electron moments is distinct from the behavior of the $S=1/2$ dimer system [Cu(pyz)$_{0.5}$(gly)]ClO$_4$ (gly = C$_2$H$_5$NO$_2$), for which only Gaussian relaxation, due to disordered nuclear magnetic moments, is observed \cite{Lancaster2014b}. This implies that either the amplitude of the fluctuating field at the muon site is larger in the Cu-V system or (assuming a fast-fluctuation limit typical of this temperature regime) that the characteristic fluctuation rate of the electronic moments is lower.

We also carried out longitudinal-field (LF) $\mu^+$SR measurements on CuVOF$_4$(H$_2$O)$_6\cdot$H$_2$O at $T=1.2$~K for $0.5 \le B \le 2000$~mT ($B = \mu_0H$) to investigate the spin dynamics. The field-dependence of the LF relaxation rate can be used to determine the nature of transport of the spin excitations (i.e., ballistic or diffusive) as the spin autocorrelation functions have different spectral densities in the two cases. For one-dimensional (1D) diffusive transport, the spectral density $f(\omega)$ has the form $f(\omega) \sim {\omega}^{1/2}$ \cite{butler}, which leads to a $\lambda \propto B^{-1/2}$ power-law relation. In contrast, for ballistic transport, $f(\omega)$ follows a logarithmic
relation $f(\omega) \propto~$ln$(c/{\omega})$, or $\lambda \propto~$ln$(c/B)$, where $c$ is a constant. In our case, for applied fields greater than 20~mT, above which the relaxation due to quasistatic nuclear moments is sufficiently quenched, the field-dependence of the relaxation rate is well-described by a power-law fit of the form $\lambda=aB^{-n}$ (see Fig.~\ref{fig:muSR_combined}). We note that neither a logarithmic field-dependence describing ballistic spin transport (or 2D diffusive transport \cite{butler}) nor a Redfield model \cite{Hayano1979}, appropriate for a dense array of randomly, dynamically fluctuating spins, can accurately describe the data.
We obtain an exponent $n=0.38(4)$, which is in reasonable agreement with the theoretical value $n=0.5$ for 1D diffusive transport and similar to the values $n = 0.35$ and $n=0.42$ measured for the 1D Heisenberg antiferromagnetic chain compounds DEOCC-TCNQF$_4$ \cite{Pratt2006} and Cu(pyz)(NO$_3$)$_2$ \cite{Xiao2015}, respectively, suggesting that the low frequency excitations in our material at this temperature are diffusive and primarily constrained to one dimension.

\noindent
\noindent

\subsection{Density functional theory}

\begin{figure}[t]
	\includegraphics[width=\columnwidth]{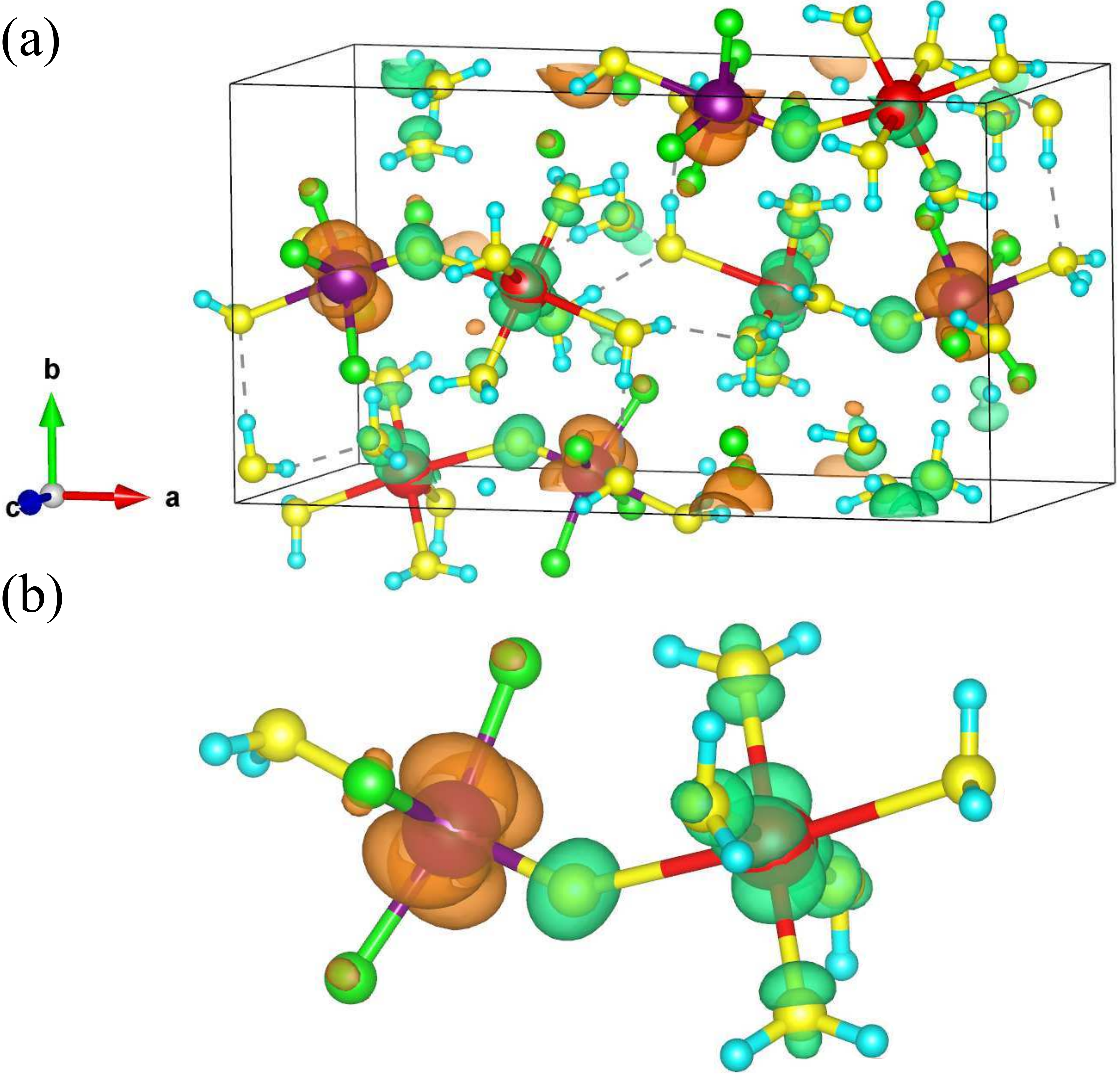}
		\caption{Spin density distribution for the ground-state magnetic structure obtained from DFT. Orange and green isosurfaces represent regions with significant up and down spin density, respectively. (b) View of a single dimer in an antiferromagnetic configuration. The color coding of the atoms here is the same as it is in Fig.~\ref{Fig:Struc}.}
	\label{fig:spin_density}
\end{figure}

To identify the significant exchange pathways within the system, we carried out spin-polarized density functional theory (DFT+$U$) calculations using the plane-wave basis-set electronic structure code \textsc{castep} \cite{CASTEP}. Details of these calculations can be found in the SI~\cite{SI}. The spin-density distribution for the ground-state magnetic structure identified by our calculations is shown in Fig.~\ref{fig:spin_density}. As shown in Fig.~\ref{fig:spin_density}(a), the spins of the Cu and V ions within a dimer are aligned antiferromagnetically, Cu and V spins belonging to neighboring dimers are also aligned antiferromagnetically. As a result, the magnetic ground state can be thought of as comprising interpenetrating Cu and V ferromagnetic sublattices. In Fig.~\ref{fig:spin_density}(b) we show the spin density distribution across a single dimer when the system is in the ground state. There is significant spin density on the Cu and V ions and also on the O atom joining the Cu and V within a dimer, with this spin density having the opposite sign to the V ion within the dimer. As anticipated from the crystal structure, DFT finds that the magnetic orbitals of the Cu ion lie along the Cu---O bonds within the Jahn-Teller plane, inducing spin density in these O atoms. On the other hand, the O atoms on either end of the dimer carry very little spin density. For the V ion, the magnetic orbitals lie in-between the V---F bonds and the spin density on the F atoms is relatively small. The central O atom lies along the JT-axis of the Cu ion, suggesting that its spin instead results from an AFM interaction with the V atom, with which it shares a short bond ($\approx 1.6$ \AA).

By considering the DFT energies corresponding to several collinear spin configurations, we calculated the exchange constants associated with each of the exchange pathways. The calculated exchange constants depend very strongly on the value used for the Hubbard $U$, as has previously been found for systems based on Cu \cite{IOThomas} and V \cite{tsirlin}. For $U=5$~eV on both the Cu and V $d$-orbitals, we obtained a value $J_0=24.7(6)$~K for the intradimer exchange, which is broadly consistent with the experimental value. This value of $U$ results in an interdimer coupling constant $J'=8.6(15)$~K for the exchange within the $bc$ plane, which is significantly larger than experiment, though we note that smaller values of $J'$ are obtained when using larger values of $U$ (see SI~\cite{SI}). The interdimer exchange within the $ab$-plane between Cu and V ions on adjacent dimers [likely via the H-bonds shown in Fig.~\ref{Fig:Struc}(e)] was found to be $< 0.6$~K and hence its sign cannot be determined unambiguously within the uncertainties associated with these calculations. (Any coupling within the $ab$-plane between two Cu, or two V, ions on adjacent dimers would be expected to be even weaker than this, as these H-bond pathways involve one or multiple JT or pseudo-JT axes.)
%
%
These results show that each dimer couples antiferromagnetically to its four nearest-neighbors in the $bc$ plane, with only very weak coupling between dimers within the $ab$-plane. This is in agreement with the exchange network posited from inspecting the structure.

\begin{figure}[t]
	\includegraphics[width=\columnwidth]{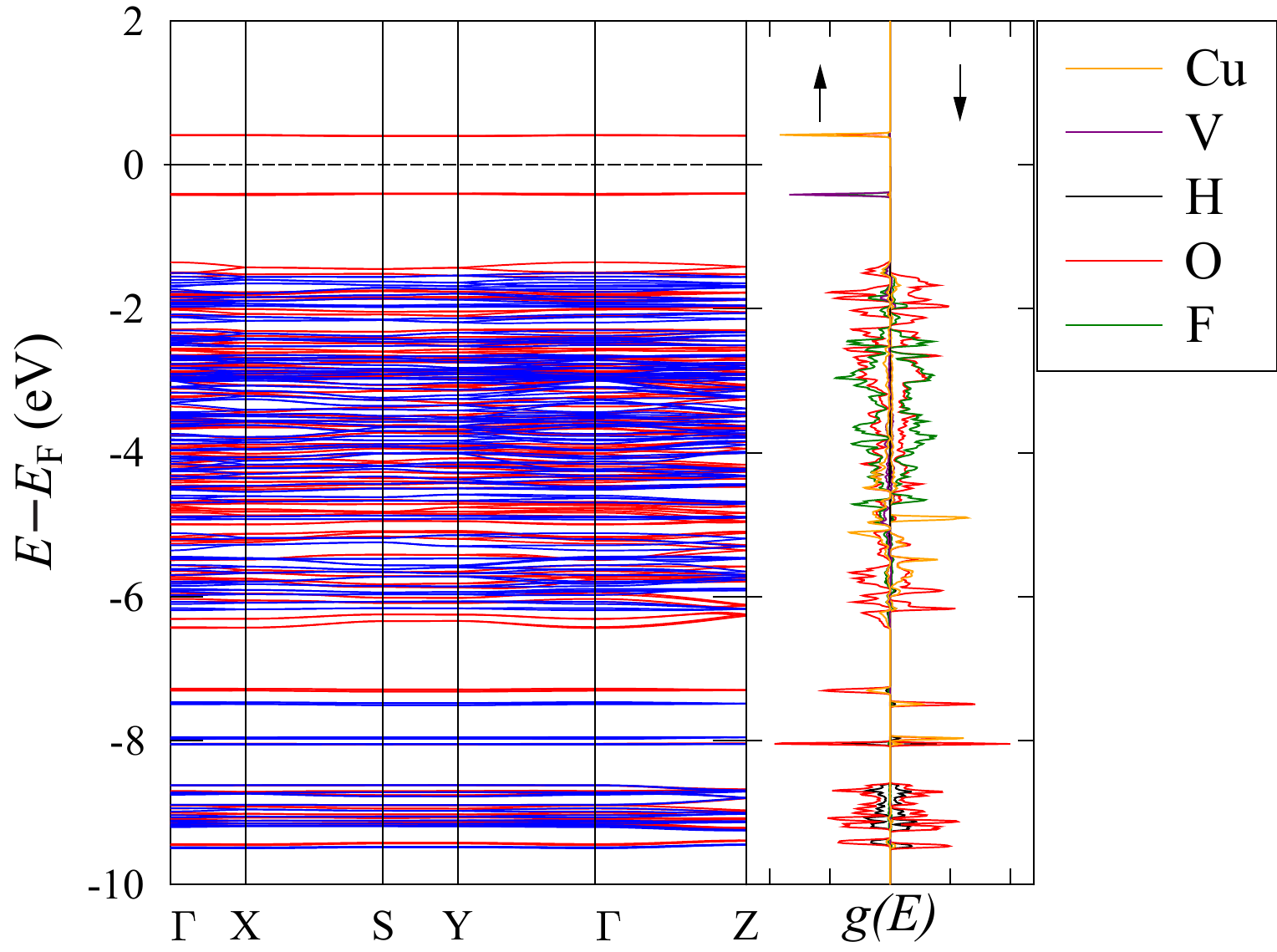}
	\caption{Spin-polarized band structures for each of the magnetic ground state and the density of states for each spin channel. Bands corresponding to spin-up and spin-down are indicated by red and blue lines respectively. The density of states are shown projected onto each atomic species.}
	\label{fig:spectral_gs}
\end{figure}

More insight into how the magnetism in this system derives from the electronic structure can be obtained from the spin-polarized band structure and density of states (DOS), shown for the ground state magnetic configuration in Fig.~\ref{fig:spectral_gs}. We note that this band structure and DOS is not typical of those obtained for an AFM system (see, for example, those for other AFM configurations in Fig.~S9 in the SI), which typically exhibit pairs of degenerate spin-up and spin-down bands, and equal DOS in the two spin channels. This is due to the fact that the spin centers in the present system belong to two distinct species, and in the ground state all of the spins belonging to the same species point in the same direction. Despite this, and the fact that the ordered moments on Cu and V are not equal, the integrated DOS of the occupied states in each spin channel is equal, such that the system has no net magnetization, consistent with ESR data and the absence of any hysteresis in $M$($H$) (Fig.~S1 in \cite{SI}). Within 1~eV of the Fermi energy, the band structure is characterized by two sets of dispersionless bands, indicative of localized states, 0.4~eV above and below the Fermi energy, corresponding to states localized on V and Cu, respectively. The V-centered states also have contributions from the F atoms the V is bonded to, while the Cu-centered states have contributions from the O atoms lying in the JT plane. These localized Cu and V states occupy the same spin channel despite the system being in an antiferromagnetic state. We can rationalize this by noting that these Cu orbitals are unoccupied and it is instead occupied Cu states that lie well below the Fermi energy that give rise to the moments on the Cu. These form part of an overlapping set of bands located around 2 to 6 eV below the energy and are not as well-localized as the unoccupied Cu states. As we discuss later, these lower-lying Cu states are likely to play a key role in determining the magnetism in this system.

\begin{figure}[t]
	\includegraphics[width=\columnwidth]{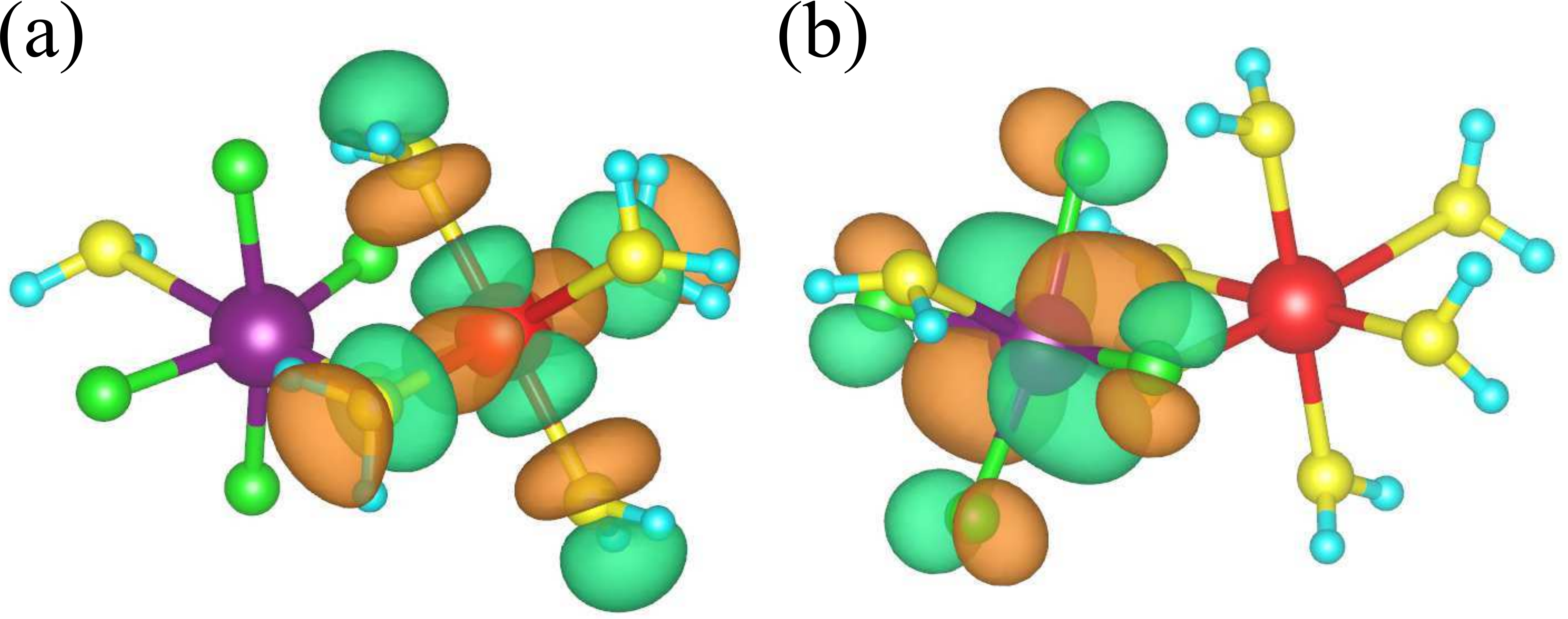}
	\caption{The two distinct classes of maximally-localized Wannier functions, which are localized either on (a) Cu or (b) V. Orange and green isosurfaces correspond to positive and negative values for the Wannier function, respectively.}
	\label{fig:wannier_small}
\end{figure}

To more directly assess the exchange between pairs of magnetic ions, we devised a tight-binding model using Wannier orbitals derived from the two sets of four bands just below and above the Fermi energy (the construction of this model is described in detail in the SI~\cite{SI}). Our model therefore includes eight Wannier orbitals, with four centered on Cu ions and four centered on V ions, as shown in Fig.~\ref{fig:wannier_small}. A tight-binding model constructed from these Wannier orbitals, including only those overlaps corresponding to hopping between nearest-neighbor ions, is able to successfully reproduce the main features of these bands. Within this model, the intradimer hopping $t=2.87$~meV (33.3~K) is found to be nearly an order of magnitude smaller than the effective interdimer hopping $t'=17.1$~meV (198.4~K) between dimers in $bc$ plane (variations in the interdimer hopping are discussed in detail in the SI).  The relative sizes of these hopping deviate strongly from the calculated and experimental exchange constants, but can be explained from the shapes of the Wannier orbitals. Both Cu- and V-centered orbitals lie perpendicular to the dimers and therefore the overlap between orbitals on adjacent dimers within the $bc$ plane is much stronger than the overlap between orbitals along the dimer. We note that a similar tight-binding model based on Wannier orbitals was able to describe the magnetic interactions in copper-pyrazine antiferromagnets \cite{powell}. However, in those systems superexchange is mediated by pyz ligands lying in the JT plane, with the shapes of the Wannier orbitals reflecting this. This is in stark contrast to our system, where the principal exchange is along the JT axis, and where it appears that the magnetism cannot be accurately described by considering only hopping between localized orbitals near the Fermi energy.

Taken together, the results of this suite of calculations indicate that the magnetism in this system cannot be simply described in terms of superexchange between localized Cu and V orbitals. While the occupied V orbitals are highly localized and close to the Fermi energy, the Cu orbitals that give Cu its magnetic moment lie among a set of overlapping bands much further below the Fermi energy. The DFT total-energy approach used to calculate exchange constants takes into account all of the electrons in the system and yields exchange constants that are qualitatively consistent with those from experiment, unlike the hopping parameters derived from our tight-binding model, which instead suggests intradimer exchange should be dominant. It is therefore likely other atoms within the dimer play a key role in determining the magnitude of the intradimer exchange. This is likely to include the central O atom joining Cu and V atoms within a dimer, as this atom is found to have a significant spin density and could therefore be responsible for mediating the exchange within a dimer.

\section{Discussion}

\begin{figure}[t]
\centering
\includegraphics[width=1 \linewidth]{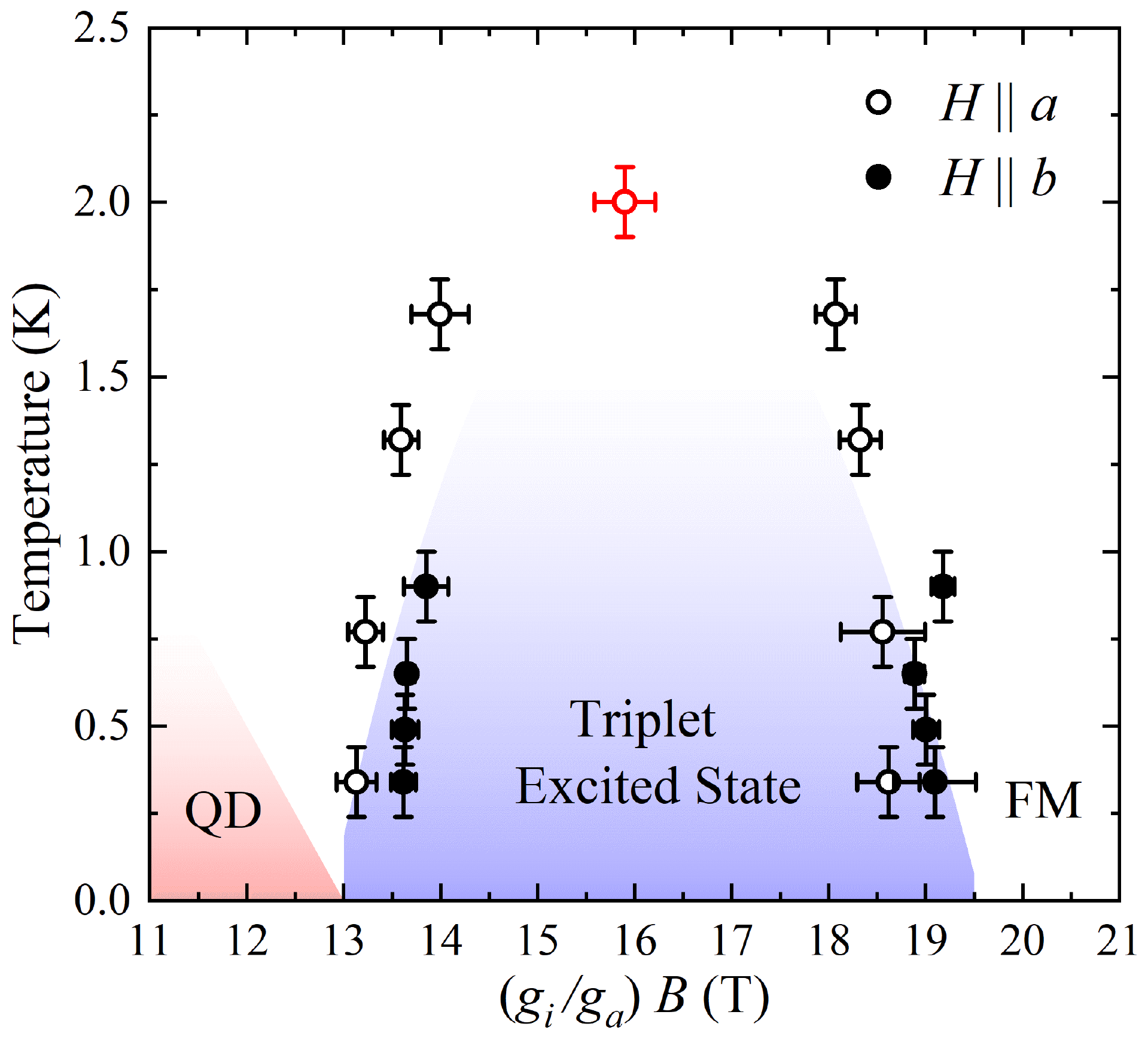}
\caption[width=1 \linewidth]{\small Temperature-field phase-diagram for CuVOF$_4$(H$_2$O)$_6\cdot$H$_2$O where $g_i$ denotes $g$-factors with field along $i = a, b$ as described in the text. Blue shaded region serves as a guide to the eye to highlight phase boundaries that enclose the excited triplet state. Red shaded region indicates the quantum disordered (QD) state at fields below $H_{\rm{c1}}$ whilst FM indicates the ferromagnetically saturated state above $H_{\rm c2}$. Upper limit for the dome marked with red circle at $T = 2.0(1)$~K.}  
\vspace{-0cm} \label{fig:BT}
\end{figure}

The positions of $H_{\rm c1,2}$ obtained from the magnetometry experiments show some dependence on the orientation of the sample relative to the applied magnetic field, which we ascribe predominantly to $g$-factor anisotropy.
Our orientation-dependent ESR measurements enable us to determine $g$-factors for measurements made with an applied field along all three principal crystallographic axes \cite{SI}. Scaling our values of $H_{\rm c1,2}$ measured with $H \perp a$ using $g_b = 2.046$ and plotting them as $(g_b/g_a) \mu_{\rm B} B_{\rm c1,2}$ (where $B_{\rm c1,2} = \mu_0 H_{\rm c1,2}$), we find that the values of $H_{\rm c1,2}$ for both orientations very nearly coincide with one another as shown in Fig.~\ref{fig:BT}. The red 2~K data-point (measured with $H \parallel a$) indicates the upper temperature limit for the existence of the triplet excited state, as estimated using RF susceptometry.

Despite correcting for $g$-factor anisotropy, a slight anisotropy is still observed between similar temperature $H_{\rm c1,2}$ values determined with $H \parallel a$ and $H \parallel b$. The cause of this may be the smearing out of the positions of $H_{\rm c1,2}$ due to dislocations in the extensive H-bond network, as described above. The lack of an inversion center between the Cu and V ions may also permit the existence of a small DM term, which could lead to an additional orientation dependence of $H_{\rm c1,2}$ \cite{Kofu2009,Nawa2011}.

The temperature-field phase diagram in Fig.~\ref{fig:BT} is a result of both the intradimer exchange $J_{0}$ and the interdimer exchange $J'$. As $\chi(T)$, ESR and DFT calculations suggest $J'$ is AFM, at low-temperature the critical fields relate to $J_{0}$ and $J'$ via \cite{Tachiki1970},
\noindent
\begin{align}\label{eq:Tachiki}
     g \mu_{\rm{B}} B_{\rm{c1}} &= J_0-nJ'/2, &&
     g \mu_{\rm{B}} B_{\rm{c2}} &= J_0+nJ' 
\end{align}
\noindent
where $n = 4$ is the number of nearest dimer neighbors, as determined from DFT calculations. 

Using the critical field values at $T = 0.34$~K and $g$-factors from ESR, Eqs.~\ref{eq:Tachiki} return parameters of $J_{0} = 22.1(6)$~K and $J' = 1.4(2)$~K ($n = 4$), in excellent agreement with $\chi(T)$ and ESR. The system can therefore be well approximated using an isotropic dimer-model with weak interdimer interactions \cite{Tachiki1970}. 



As described above, crystal architectures composed of JT-active Cu(II) octahedra have previously been shown to promote low-dimensional magnetic behavior \cite{Ghannadzadeh2013,Wang2013,Manson2009b,Brambleby2015}. The $d_{x^2-y^2}$ orbital of the Cu lies within a plane perpendicular to the JT axis, resulting in minimal orbital overlap between adjacent Cu ions bonded along a JT axis, and hence very weak magnetic exchange in this direction. This is reflected in our DFT calculations [Fig.~\ref{fig:spin_density}(b)], which show an increased spin density, arising from the $d_{x^2-y^2}$ orbital, along the equatorial Cu---OH$_2$ bonds, perpendicular to the axial O---Cu---OH$_2$ JT bond direction.

Similarly, the presence of a pseudo-JT distortion in six-coordinate V(IV) ($3d^1$) complexes can also facilitate the emergence of low-dimensional magnetism \cite{Waki2005Pb2V3O9,Mentre1999Pb2V3O9}. For non-polar octahedra, an axial elongation results in the $3d^1$ electron occupying the degenerate $d_{xz}$ and $d_{yz}$ orbitals, whilst an axial compression leads to the $3d^1$ electron inhabiting the $d_{xy}$ orbital. (Typically, axial compression is favoured in $3d^1$ octahedra, as occupying the $d_{xy}$ orbital offers the greater energy saving compared to occupying one of the degenerate $d_{xz}$ or $d_{yz}$ orbitals \cite{DalalJT}). In the present case, the shortened V=O double-bond [1.6083(8)\,\AA] and elongated V---OH$_2$ bond [2.2903(7)\,\AA] of the polar VF$_4$O$_2$ octahedra makes it difficult to ascertain which orbital the $3d^1$ electron occupies by inspecting the structure alone. However, DFT calculations in Fig.~\ref{fig:spin_density}(b) show an increased spin-density between the V---F ligands in the $bc$-plane, indicative that the $3d^1$ electron occupies the $d_{xy}$ orbital.

\noindent
\noindent

As anticipated, DFT calculations find negligible spin-density located on the axial water ligand of either the Cu or V octahedra. Therefore, whilst the axial waters do form H-bonds between adjacent dimers within the $ab$-plane, the interdimer magnetic exchange will occur primarily within the $bc$-plane via the H-bond network outlined in Fig.~\ref{Fig:Struc}(d). Furthermore, the only H-bond pathway within the $ab$-plane which is not via an axial water [blue-striped bonds, Fig.~\ref{Fig:Struc}(e)] is found by DFT to mediate a vanishingly weak exchange, such that the system can be considered to be a magnetically 2D network of $S = 1/2$ dimers.

The interdimer exchange within this 2D network ($b$c-plane) is estimated to be $J' \approx 1$~K, whilst the intradimer exchange is determined to be much greater at $J_0 \approx 21$~K. Therefore, the exchange pathway which mediates $J_0$ must be considerably more efficient than the $J'$ exchange pathways. The interdimer Cu---O---H$\,\cdots$F---V bonds which mediate $J'$ fall within the range $6.51-6.656$\,\AA, which is very similar in length to the intradimer  Cu---O---H$\,\cdots$F---V bond 6.553\,\AA  [Fig.~\ref{fig:CuV_susc}(c)]. It is therefore unlikely the strong $J_0$ is mediated along the intradimer H-bond pathway. Instead it seems apparent that the bridging oxide, which DFT shows as harboring significant spin-density and which is located along the JT-axis of the Cu, is involved in mediating the intradimer exchange. For the reasons outlined above, significant exchange through a JT bond is very rare. However, a similar scenario has been encountered once before in the 2D AFM chain compounds CuX$_2$(pyz); where X = Cl, Br, F \cite{Butcher2008,Lapidus2013,Lancaster2019}. In this case, however, magnetic exchange is mediated along Cu—--X$_2$--—Cu bibridges, for which only one of the X--—Cu--—X bridges lies along the JT axial direction, while the other lies in the plane containing the $d_{x^2-y^2}$ magnetic orbital \cite{Lancaster2019}. This means that, in the present case, the manner in which the intradimer exchange is mediated through the O atom on the Cu(II) JT axis remains unusual.

The two distinct $S = 1/2$ species in CuVOF$_4$(H$_2$O)$_6 \cdot$H$_2$O exhibit very different behavior, with the occupied magnetic orbitals for V being highly localized states located just below the Fermi energy, whereas the occupied Cu states lie significantly lower in energy and are more delocalized. The tight-binding model, based on Wannier orbitals localized around the Cu and V, successfully reproduces the electronic band structure of the system in the vicinity of the Fermi energy, but fails to accurately describe the magnetic properties of the system. In contrast, DFT calculations, which also consider the delocalized bands below the Fermi energy, are able to qualitatively describe the magnetism within the compound. We hypothesise that the strong V=O double-bond allows the vanadyl species to donate a sizable portion of its spin-density to the bridging oxide, leading to the Cu and V ions coupling antiferromagnetically, via the delocalized bands below the Fermi energy, to form an AFM spin-half dimer. This demonstrates that the JT distortion, coupled with the unlike nature of the spin-species, is not only responsible for the polar crystal structure~\cite{Donakowski2012} but also, by driving the formation of the Cu---O=V bond, establishes the effective intradimer exchange pathway within the dimer units. This is completely different to the usual way in which JT physics drives low-dimensional magnetism in other Cu(II) systems.

\section{Acknowledgements} This project has received funding from the European Research Council (ERC) under the European Union’s Horizon 2020 research and innovation program (Grant Agreement No. 681260). This work is supported by the EPSRC (under grants EP/N032128/1 and EP/N024028/1). We thank the STFC-ISIS facility for the provision of muon beamtime and K.J.A. Franke and M.C. Worsdale for experimental assistance. We acknowledge computing resources provided by STFC Scientific Computing Department’s SCARF cluster and Durham Hamilton HPC. Work at EWU was supported by the National Science Foundation under grant No. DMR-1703003. A portion of this work was performed at the National High Magnetic Field Laboratory, which is supported by the National Science Foundation Cooperative Agreement No. DMR-1644779 and the State of Florida. DK was supported by the PRIME program of the German Academic Exchange Service (DAAD) with funds from the German Federal Ministry of Education and Research (BMBF). We would like to acknowledge the members of the X-ray Diffraction Research Technology Platform at The University of Warwick for their assistance with the work described in this paper. Data presented in this paper resulting from the UK effort will be made available at XXXXXXX.

\providecommand{\noopsort}[1]{}\providecommand{\singleletter}[1]{#1}%
%



\begin{thebibliography}{0}%
\makeatletter
\providecommand \@ifxundefined [1]{%
 \@ifx{#1\undefined}
}%
\providecommand \@ifnum [1]{%
 \ifnum #1\expandafter \@firstoftwo
 \else \expandafter \@secondoftwo
 \fi
}%
\providecommand \@ifx [1]{%
 \ifx #1\expandafter \@firstoftwo
 \else \expandafter \@secondoftwo
 \fi
}%
\providecommand \natexlab [1]{#1}%
\providecommand \enquote  [1]{``#1''}%
\providecommand \bibnamefont  [1]{#1}%
\providecommand \bibfnamefont [1]{#1}%
\providecommand \citenamefont [1]{#1}%
\providecommand \href@noop [0]{\@secondoftwo}%
\providecommand \href [0]{\begingroup \@sanitize@url \@href}%
\providecommand \@href[1]{\@@startlink{#1}\@@href}%
\providecommand \@@href[1]{\endgroup#1\@@endlink}%
\providecommand \@sanitize@url [0]{\catcode `\\12\catcode `\$12\catcode
  `\&12\catcode `\#12\catcode `\^12\catcode `\_12\catcode `\%12\relax}%
\providecommand \@@startlink[1]{}%
\providecommand \@@endlink[0]{}%
\providecommand \url  [0]{\begingroup\@sanitize@url \@url }%
\providecommand \@url [1]{\endgroup\@href {#1}{\urlprefix }}%
\providecommand \urlprefix  [0]{URL }%
\providecommand \Eprint [0]{\href }%
\providecommand \doibase [0]{http://dx.doi.org/}%
\providecommand \selectlanguage [0]{\@gobble}%
\providecommand \bibinfo  [0]{\@secondoftwo}%
\providecommand \bibfield  [0]{\@secondoftwo}%
\providecommand \translation [1]{[#1]}%
\providecommand \BibitemOpen [0]{}%
\providecommand \bibitemStop [0]{}%
\providecommand \bibitemNoStop [0]{.\EOS\space}%
\providecommand \EOS [0]{\spacefactor3000\relax}%
\providecommand \BibitemShut  [1]{\csname bibitem#1\endcsname}%
\let\auto@bib@innerbib\@empty
\end{thebibliography}%


%


\begin{thebibliography}{56}%
\makeatletter
\providecommand \@ifxundefined [1]{%
 \@ifx{#1\undefined}
}%
\providecommand \@ifnum [1]{%
 \ifnum #1\expandafter \@firstoftwo
 \else \expandafter \@secondoftwo
 \fi
}%
\providecommand \@ifx [1]{%
 \ifx #1\expandafter \@firstoftwo
 \else \expandafter \@secondoftwo
 \fi
}%
\providecommand \natexlab [1]{#1}%
\providecommand \enquote  [1]{``#1''}%
\providecommand \bibnamefont  [1]{#1}%
\providecommand \bibfnamefont [1]{#1}%
\providecommand \citenamefont [1]{#1}%
\providecommand \href@noop [0]{\@secondoftwo}%
\providecommand \href [0]{\begingroup \@sanitize@url \@href}%
\providecommand \@href[1]{\@@startlink{#1}\@@href}%
\providecommand \@@href[1]{\endgroup#1\@@endlink}%
\providecommand \@sanitize@url [0]{\catcode `\\12\catcode `\$12\catcode
  `\&12\catcode `\#12\catcode `\^12\catcode `\_12\catcode `\%12\relax}%
\providecommand \@@startlink[1]{}%
\providecommand \@@endlink[0]{}%
\providecommand \url  [0]{\begingroup\@sanitize@url \@url }%
\providecommand \@url [1]{\endgroup\@href {#1}{\urlprefix }}%
\providecommand \urlprefix  [0]{URL }%
\providecommand \Eprint [0]{\href }%
\providecommand \doibase [0]{http://dx.doi.org/}%
\providecommand \selectlanguage [0]{\@gobble}%
\providecommand \bibinfo  [0]{\@secondoftwo}%
\providecommand \bibfield  [0]{\@secondoftwo}%
\providecommand \translation [1]{[#1]}%
\providecommand \BibitemOpen [0]{}%
\providecommand \bibitemStop [0]{}%
\providecommand \bibitemNoStop [0]{.\EOS\space}%
\providecommand \EOS [0]{\spacefactor3000\relax}%
\providecommand \BibitemShut  [1]{\csname bibitem#1\endcsname}%
\let\auto@bib@innerbib\@empty
\bibitem [{\citenamefont {Gegenwart}\ \emph {et~al.}(2008)\citenamefont
  {Gegenwart}, \citenamefont {Si},\ and\ \citenamefont
  {Steglich}}]{Gegenwart2008}%
  \BibitemOpen
  \bibfield  {author} {\bibinfo {author} {\bibfnamefont {P.}~\bibnamefont
  {Gegenwart}}, \bibinfo {author} {\bibfnamefont {Q.}~\bibnamefont {Si}}, \
  and\ \bibinfo {author} {\bibfnamefont {F.}~\bibnamefont {Steglich}},\ }\href
  {\doibase 10.1038/nphys892} {\bibfield  {journal} {\bibinfo  {journal} {Nat.
  Phys.}\ }\textbf {\bibinfo {volume} {4}},\ \bibinfo {pages} {186} (\bibinfo
  {year} {2008})}\BibitemShut {NoStop}%
\bibitem [{\citenamefont {Stockert}\ and\ \citenamefont
  {Steglich}(2011)}]{Stockert2011}%
  \BibitemOpen
  \bibfield  {author} {\bibinfo {author} {\bibfnamefont {O.}~\bibnamefont
  {Stockert}}\ and\ \bibinfo {author} {\bibfnamefont {F.}~\bibnamefont
  {Steglich}},\ }\href {\doibase 10.1146/annurev-conmatphys-062910-140546}
  {\bibfield  {journal} {\bibinfo  {journal} {Annual Review of Condensed Matter
  Physics}\ }\textbf {\bibinfo {volume} {2}},\ \bibinfo {pages} {79} (\bibinfo
  {year} {2011})}\BibitemShut {NoStop}%
\bibitem [{\citenamefont {Zapf}\ \emph {et~al.}(2014)\citenamefont {Zapf},
  \citenamefont {Jaime},\ and\ \citenamefont {Batista}}]{Zapf2014a}%
  \BibitemOpen
  \bibfield  {author} {\bibinfo {author} {\bibfnamefont {V.}~\bibnamefont
  {Zapf}}, \bibinfo {author} {\bibfnamefont {M.}~\bibnamefont {Jaime}}, \ and\
  \bibinfo {author} {\bibfnamefont {C.~D.}\ \bibnamefont {Batista}},\ }\href
  {\doibase 10.1103/RevModPhys.86.563} {\bibfield  {journal} {\bibinfo
  {journal} {Rev. Mod. Phys.}\ }\textbf {\bibinfo {volume} {86}},\ \bibinfo
  {pages} {563} (\bibinfo {year} {2014})}\BibitemShut {NoStop}%
\bibitem [{\citenamefont {Lancaster}\ \emph {et~al.}(2014)\citenamefont
  {Lancaster}, \citenamefont {Goddard}, \citenamefont {Blundell}, \citenamefont
  {Foronda}, \citenamefont {Ghannadzadeh}, \citenamefont {M\"oller},
  \citenamefont {Baker}, \citenamefont {Pratt}, \citenamefont {Baines},
  \citenamefont {Huang}, \citenamefont {Wosnitza}, \citenamefont {McDonald},
  \citenamefont {Modic}, \citenamefont {Singleton}, \citenamefont {Topping},
  \citenamefont {Beale}, \citenamefont {Xiao}, \citenamefont {Schlueter},
  \citenamefont {Barton}, \citenamefont {Cabrera}, \citenamefont {Carreiro},
  \citenamefont {Tran},\ and\ \citenamefont {Manson}}]{Lancaster2014b}%
  \BibitemOpen
  \bibfield  {author} {\bibinfo {author} {\bibfnamefont {T.}~\bibnamefont
  {Lancaster}}, \bibinfo {author} {\bibfnamefont {P.~A.}\ \bibnamefont
  {Goddard}}, \bibinfo {author} {\bibfnamefont {S.~J.}\ \bibnamefont
  {Blundell}}, \bibinfo {author} {\bibfnamefont {F.~R.}\ \bibnamefont
  {Foronda}}, \bibinfo {author} {\bibfnamefont {S.}~\bibnamefont
  {Ghannadzadeh}}, \bibinfo {author} {\bibfnamefont {J.~S.}\ \bibnamefont
  {M\"oller}}, \bibinfo {author} {\bibfnamefont {P.~J.}\ \bibnamefont {Baker}},
  \bibinfo {author} {\bibfnamefont {F.~L.}\ \bibnamefont {Pratt}}, \bibinfo
  {author} {\bibfnamefont {C.}~\bibnamefont {Baines}}, \bibinfo {author}
  {\bibfnamefont {L.}~\bibnamefont {Huang}}, \bibinfo {author} {\bibfnamefont
  {J.}~\bibnamefont {Wosnitza}}, \bibinfo {author} {\bibfnamefont {R.~D.}\
  \bibnamefont {McDonald}}, \bibinfo {author} {\bibfnamefont {K.~A.}\
  \bibnamefont {Modic}}, \bibinfo {author} {\bibfnamefont {J.}~\bibnamefont
  {Singleton}}, \bibinfo {author} {\bibfnamefont {C.~V.}\ \bibnamefont
  {Topping}}, \bibinfo {author} {\bibfnamefont {T.~A.~W.}\ \bibnamefont
  {Beale}}, \bibinfo {author} {\bibfnamefont {F.}~\bibnamefont {Xiao}},
  \bibinfo {author} {\bibfnamefont {J.~A.}\ \bibnamefont {Schlueter}}, \bibinfo
  {author} {\bibfnamefont {A.~M.}\ \bibnamefont {Barton}}, \bibinfo {author}
  {\bibfnamefont {R.~D.}\ \bibnamefont {Cabrera}}, \bibinfo {author}
  {\bibfnamefont {K.~E.}\ \bibnamefont {Carreiro}}, \bibinfo {author}
  {\bibfnamefont {H.~E.}\ \bibnamefont {Tran}}, \ and\ \bibinfo {author}
  {\bibfnamefont {J.~L.}\ \bibnamefont {Manson}},\ }\href {\doibase
  10.1103/PhysRevLett.112.207201} {\bibfield  {journal} {\bibinfo  {journal}
  {Phys. Rev. Lett.}\ }\textbf {\bibinfo {volume} {112}},\ \bibinfo {pages}
  {207201} (\bibinfo {year} {2014})}\BibitemShut {NoStop}%
\bibitem [{\citenamefont {Giamarchi}\ \emph {et~al.}(2008)\citenamefont
  {Giamarchi}, \citenamefont {R{\"{u}}egg},\ and\ \citenamefont
  {Tchernyshyov}}]{Giamarchi2008}%
  \BibitemOpen
  \bibfield  {author} {\bibinfo {author} {\bibfnamefont {T.}~\bibnamefont
  {Giamarchi}}, \bibinfo {author} {\bibfnamefont {C.}~\bibnamefont
  {R{\"{u}}egg}}, \ and\ \bibinfo {author} {\bibfnamefont {O.}~\bibnamefont
  {Tchernyshyov}},\ }\href
  {http://0-www.nature.com.pugwash.lib.warwick.ac.uk/articles/nphys893.pdf}
  {\bibfield  {journal} {\bibinfo  {journal} {Nat. Phys.}\ }\textbf {\bibinfo
  {volume} {4}},\ \bibinfo {pages} {198} (\bibinfo {year} {2008})}\BibitemShut
  {NoStop}%
\bibitem [{\citenamefont {Zheludev}\ and\ \citenamefont
  {Roscilde}(2013)}]{Zheludev2013}%
  \BibitemOpen
  \bibfield  {author} {\bibinfo {author} {\bibfnamefont {A.}~\bibnamefont
  {Zheludev}}\ and\ \bibinfo {author} {\bibfnamefont {T.}~\bibnamefont
  {Roscilde}},\ }\href
  {https://www.sciencedirect.com/science/article/pii/S1631070513001497}
  {\bibfield  {journal} {\bibinfo  {journal} {Comptes Rendus Physique}\
  }\textbf {\bibinfo {volume} {14}},\ \bibinfo {pages} {740} (\bibinfo {year}
  {2013})}\BibitemShut {NoStop}%
\bibitem [{\citenamefont {Nawa}\ \emph {et~al.}(2011)\citenamefont {Nawa},
  \citenamefont {Michioka}, \citenamefont {Yoshimura}, \citenamefont {Matsuo},\
  and\ \citenamefont {Kindo}}]{Nawa2011}%
  \BibitemOpen
  \bibfield  {author} {\bibinfo {author} {\bibfnamefont {K.}~\bibnamefont
  {Nawa}}, \bibinfo {author} {\bibfnamefont {C.}~\bibnamefont {Michioka}},
  \bibinfo {author} {\bibfnamefont {K.}~\bibnamefont {Yoshimura}}, \bibinfo
  {author} {\bibfnamefont {A.}~\bibnamefont {Matsuo}}, \ and\ \bibinfo {author}
  {\bibfnamefont {K.}~\bibnamefont {Kindo}},\ }\href {\doibase
  10.1143/jpsj.80.034710} {\bibfield  {journal} {\bibinfo  {journal} {J. Phys.
  Soc. Japan}\ }\textbf {\bibinfo {volume} {80}},\ \bibinfo {pages} {034710}
  (\bibinfo {year} {2011})}\BibitemShut {NoStop}%
\bibitem [{\citenamefont {Sebastian}\ \emph {et~al.}(2006)\citenamefont
  {Sebastian}, \citenamefont {Tanedo}, \citenamefont {Goddard}, \citenamefont
  {Lee}, \citenamefont {Wilson}, \citenamefont {Kim}, \citenamefont {Cox},
  \citenamefont {McDonald}, \citenamefont {Hill}, \citenamefont {Harrison},
  \citenamefont {Batista},\ and\ \citenamefont {Fisher}}]{Sebastian2006}%
  \BibitemOpen
  \bibfield  {author} {\bibinfo {author} {\bibfnamefont {S.~E.}\ \bibnamefont
  {Sebastian}}, \bibinfo {author} {\bibfnamefont {P.}~\bibnamefont {Tanedo}},
  \bibinfo {author} {\bibfnamefont {P.~A.}\ \bibnamefont {Goddard}}, \bibinfo
  {author} {\bibfnamefont {S.-C.}\ \bibnamefont {Lee}}, \bibinfo {author}
  {\bibfnamefont {A.}~\bibnamefont {Wilson}}, \bibinfo {author} {\bibfnamefont
  {S.}~\bibnamefont {Kim}}, \bibinfo {author} {\bibfnamefont {S.}~\bibnamefont
  {Cox}}, \bibinfo {author} {\bibfnamefont {R.~D.}\ \bibnamefont {McDonald}},
  \bibinfo {author} {\bibfnamefont {S.}~\bibnamefont {Hill}}, \bibinfo {author}
  {\bibfnamefont {N.}~\bibnamefont {Harrison}}, \bibinfo {author}
  {\bibfnamefont {C.~D.}\ \bibnamefont {Batista}}, \ and\ \bibinfo {author}
  {\bibfnamefont {I.~R.}\ \bibnamefont {Fisher}},\ }\href {\doibase
  10.1103/PhysRevB.74.180401} {\bibfield  {journal} {\bibinfo  {journal} {Phys.
  Rev. B}\ }\textbf {\bibinfo {volume} {74}},\ \bibinfo {pages} {180401(R)}
  (\bibinfo {year} {2006})}\BibitemShut {NoStop}%
\bibitem [{\citenamefont {Donakowski}\ \emph {et~al.}(2012)\citenamefont
  {Donakowski}, \citenamefont {Gautier}, \citenamefont {Yeon}, \citenamefont
  {Moore}, \citenamefont {Nino}, \citenamefont {Halasyamani},\ and\
  \citenamefont {Poeppelmeier}}]{Donakowski2012}%
  \BibitemOpen
  \bibfield  {author} {\bibinfo {author} {\bibfnamefont {M.~D.}\ \bibnamefont
  {Donakowski}}, \bibinfo {author} {\bibfnamefont {R.}~\bibnamefont {Gautier}},
  \bibinfo {author} {\bibfnamefont {J.}~\bibnamefont {Yeon}}, \bibinfo {author}
  {\bibfnamefont {D.~T.}\ \bibnamefont {Moore}}, \bibinfo {author}
  {\bibfnamefont {J.~C.}\ \bibnamefont {Nino}}, \bibinfo {author}
  {\bibfnamefont {P.~S.}\ \bibnamefont {Halasyamani}}, \ and\ \bibinfo {author}
  {\bibfnamefont {K.~R.}\ \bibnamefont {Poeppelmeier}},\ }\href
  {https://doi.org/10.1021/ja210984k} {\bibfield  {journal} {\bibinfo
  {journal} {J. Am. Chem. Soc.}\ }\textbf {\bibinfo {volume} {134}},\ \bibinfo
  {pages} {7679} (\bibinfo {year} {2012})}\BibitemShut {NoStop}%
\bibitem [{\citenamefont {Moriya}(1960)}]{Moriya_DMI_1960}%
  \BibitemOpen
  \bibfield  {author} {\bibinfo {author} {\bibfnamefont {T.}~\bibnamefont
  {Moriya}},\ }\href {\doibase 10.1103/PhysRev.120.91} {\bibfield  {journal}
  {\bibinfo  {journal} {Phys. Rev.}\ }\textbf {\bibinfo {volume} {120}},\
  \bibinfo {pages} {91} (\bibinfo {year} {1960})}\BibitemShut {NoStop}%
\bibitem [{\citenamefont {Kofu}\ \emph
  {et~al.}(2009{\natexlab{a}})\citenamefont {Kofu}, \citenamefont {Ueda},
  \citenamefont {Nojiri}, \citenamefont {Oshima}, \citenamefont {Zenmoto},
  \citenamefont {Rule}, \citenamefont {Gerischer}, \citenamefont {Lake},
  \citenamefont {Batista}, \citenamefont {Ueda},\ and\ \citenamefont
  {Lee}}]{Kofu2009}%
  \BibitemOpen
  \bibfield  {author} {\bibinfo {author} {\bibfnamefont {M.}~\bibnamefont
  {Kofu}}, \bibinfo {author} {\bibfnamefont {H.}~\bibnamefont {Ueda}}, \bibinfo
  {author} {\bibfnamefont {H.}~\bibnamefont {Nojiri}}, \bibinfo {author}
  {\bibfnamefont {Y.}~\bibnamefont {Oshima}}, \bibinfo {author} {\bibfnamefont
  {T.}~\bibnamefont {Zenmoto}}, \bibinfo {author} {\bibfnamefont {K.~C.}\
  \bibnamefont {Rule}}, \bibinfo {author} {\bibfnamefont {S.}~\bibnamefont
  {Gerischer}}, \bibinfo {author} {\bibfnamefont {B.}~\bibnamefont {Lake}},
  \bibinfo {author} {\bibfnamefont {C.~D.}\ \bibnamefont {Batista}}, \bibinfo
  {author} {\bibfnamefont {Y.}~\bibnamefont {Ueda}}, \ and\ \bibinfo {author}
  {\bibfnamefont {S.-H.}\ \bibnamefont {Lee}},\ }\href {\doibase
  10.1103/PhysRevLett.102.177204} {\bibfield  {journal} {\bibinfo  {journal}
  {Phys. Rev. Lett.}\ }\textbf {\bibinfo {volume} {102}},\ \bibinfo {pages}
  {177204} (\bibinfo {year} {2009}{\natexlab{a}})}\BibitemShut {NoStop}%
\bibitem [{\citenamefont {Aczel}\ \emph
  {et~al.}(2009{\natexlab{a}})\citenamefont {Aczel}, \citenamefont {Kohama},
  \citenamefont {Marcenat}, \citenamefont {Weickert}, \citenamefont {Jaime},
  \citenamefont {Ayala-Valenzuela}, \citenamefont {McDonald}, \citenamefont
  {Selesnic}, \citenamefont {Dabkowska},\ and\ \citenamefont
  {Luke}}]{Aczel2009a}%
  \BibitemOpen
  \bibfield  {author} {\bibinfo {author} {\bibfnamefont {A.~A.}\ \bibnamefont
  {Aczel}}, \bibinfo {author} {\bibfnamefont {Y.}~\bibnamefont {Kohama}},
  \bibinfo {author} {\bibfnamefont {C.}~\bibnamefont {Marcenat}}, \bibinfo
  {author} {\bibfnamefont {F.}~\bibnamefont {Weickert}}, \bibinfo {author}
  {\bibfnamefont {M.}~\bibnamefont {Jaime}}, \bibinfo {author} {\bibfnamefont
  {O.~E.}\ \bibnamefont {Ayala-Valenzuela}}, \bibinfo {author} {\bibfnamefont
  {R.~D.}\ \bibnamefont {McDonald}}, \bibinfo {author} {\bibfnamefont {S.~D.}\
  \bibnamefont {Selesnic}}, \bibinfo {author} {\bibfnamefont {H.~A.}\
  \bibnamefont {Dabkowska}}, \ and\ \bibinfo {author} {\bibfnamefont {G.~M.}\
  \bibnamefont {Luke}},\ }\href {\doibase 10.1103/PhysRevLett.103.207203}
  {\bibfield  {journal} {\bibinfo  {journal} {Phys. Rev. Lett.}\ }\textbf
  {\bibinfo {volume} {103}},\ \bibinfo {pages} {207203} (\bibinfo {year}
  {2009}{\natexlab{a}})}\BibitemShut {NoStop}%
\bibitem [{\citenamefont {Jaime}\ \emph {et~al.}(2004)\citenamefont {Jaime},
  \citenamefont {Correa}, \citenamefont {Harrison}, \citenamefont {Batista},
  \citenamefont {Kawashima}, \citenamefont {Kazuma}, \citenamefont {Jorge},
  \citenamefont {Stern}, \citenamefont {Heinmaa}, \citenamefont {Zvyagin},
  \citenamefont {Sasago},\ and\ \citenamefont {Uchinokura}}]{Jaime2004}%
  \BibitemOpen
  \bibfield  {author} {\bibinfo {author} {\bibfnamefont {M.}~\bibnamefont
  {Jaime}}, \bibinfo {author} {\bibfnamefont {V.~F.}\ \bibnamefont {Correa}},
  \bibinfo {author} {\bibfnamefont {N.}~\bibnamefont {Harrison}}, \bibinfo
  {author} {\bibfnamefont {C.~D.}\ \bibnamefont {Batista}}, \bibinfo {author}
  {\bibfnamefont {N.}~\bibnamefont {Kawashima}}, \bibinfo {author}
  {\bibfnamefont {Y.}~\bibnamefont {Kazuma}}, \bibinfo {author} {\bibfnamefont
  {G.~A.}\ \bibnamefont {Jorge}}, \bibinfo {author} {\bibfnamefont
  {R.}~\bibnamefont {Stern}}, \bibinfo {author} {\bibfnamefont
  {I.}~\bibnamefont {Heinmaa}}, \bibinfo {author} {\bibfnamefont {S.~A.}\
  \bibnamefont {Zvyagin}}, \bibinfo {author} {\bibfnamefont {Y.}~\bibnamefont
  {Sasago}}, \ and\ \bibinfo {author} {\bibfnamefont {K.}~\bibnamefont
  {Uchinokura}},\ }\href {\doibase 10.1103/PhysRevLett.93.087203} {\bibfield
  {journal} {\bibinfo  {journal} {Phys. Rev. Lett.}\ }\textbf {\bibinfo
  {volume} {93}},\ \bibinfo {pages} {087203} (\bibinfo {year}
  {2004})}\BibitemShut {NoStop}%
\bibitem [{\citenamefont {Aczel}\ \emph
  {et~al.}(2009{\natexlab{b}})\citenamefont {Aczel}, \citenamefont {Kohama},
  \citenamefont {Jaime}, \citenamefont {Ninios}, \citenamefont {Chan},
  \citenamefont {Balicas}, \citenamefont {Dabkowska},\ and\ \citenamefont
  {Luke}}]{Aczel2009}%
  \BibitemOpen
  \bibfield  {author} {\bibinfo {author} {\bibfnamefont {A.~A.}\ \bibnamefont
  {Aczel}}, \bibinfo {author} {\bibfnamefont {Y.}~\bibnamefont {Kohama}},
  \bibinfo {author} {\bibfnamefont {M.}~\bibnamefont {Jaime}}, \bibinfo
  {author} {\bibfnamefont {K.}~\bibnamefont {Ninios}}, \bibinfo {author}
  {\bibfnamefont {H.~B.}\ \bibnamefont {Chan}}, \bibinfo {author}
  {\bibfnamefont {L.}~\bibnamefont {Balicas}}, \bibinfo {author} {\bibfnamefont
  {H.~A.}\ \bibnamefont {Dabkowska}}, \ and\ \bibinfo {author} {\bibfnamefont
  {G.~M.}\ \bibnamefont {Luke}},\ }\href {\doibase 10.1103/PhysRevB.79.100409}
  {\bibfield  {journal} {\bibinfo  {journal} {Phys. Rev. B}\ }\textbf {\bibinfo
  {volume} {79}},\ \bibinfo {pages} {100409(R)} (\bibinfo {year}
  {2009}{\natexlab{b}})}\BibitemShut {NoStop}%
\bibitem [{\citenamefont {Nojiri}\ \emph {et~al.}(1999)\citenamefont {Nojiri},
  \citenamefont {Kageyama}, \citenamefont {Onizuka}, \citenamefont {Ueda},\
  and\ \citenamefont {Motokawa}}]{SrCuBO_1}%
  \BibitemOpen
  \bibfield  {author} {\bibinfo {author} {\bibfnamefont {H.}~\bibnamefont
  {Nojiri}}, \bibinfo {author} {\bibfnamefont {H.}~\bibnamefont {Kageyama}},
  \bibinfo {author} {\bibfnamefont {K.}~\bibnamefont {Onizuka}}, \bibinfo
  {author} {\bibfnamefont {Y.}~\bibnamefont {Ueda}}, \ and\ \bibinfo {author}
  {\bibfnamefont {M.}~\bibnamefont {Motokawa}},\ }\href {\doibase
  10.1143/JPSJ.68.2906} {\bibfield  {journal} {\bibinfo  {journal} {Journal of
  the Physical Society of Japan}\ }\textbf {\bibinfo {volume} {68}},\ \bibinfo
  {pages} {2906} (\bibinfo {year} {1999})}\BibitemShut {NoStop}%
\bibitem [{\citenamefont {R{\~o\~o}m}\ \emph {et~al.}(2004)\citenamefont
  {R{\~o\~o}m}, \citenamefont {H\"uvonen}, \citenamefont {Nagel}, \citenamefont
  {Hwang}, \citenamefont {Timusk},\ and\ \citenamefont {Kageyama}}]{SrCuBO_2}%
  \BibitemOpen
  \bibfield  {author} {\bibinfo {author} {\bibfnamefont {T.}~\bibnamefont
  {R{\~o\~o}m}}, \bibinfo {author} {\bibfnamefont {D.}~\bibnamefont
  {H\"uvonen}}, \bibinfo {author} {\bibfnamefont {U.}~\bibnamefont {Nagel}},
  \bibinfo {author} {\bibfnamefont {J.}~\bibnamefont {Hwang}}, \bibinfo
  {author} {\bibfnamefont {T.}~\bibnamefont {Timusk}}, \ and\ \bibinfo {author}
  {\bibfnamefont {H.}~\bibnamefont {Kageyama}},\ }\href {\doibase
  10.1103/PhysRevB.70.144417} {\bibfield  {journal} {\bibinfo  {journal} {Phys.
  Rev. B}\ }\textbf {\bibinfo {volume} {70}},\ \bibinfo {pages} {144417}
  (\bibinfo {year} {2004})}\BibitemShut {NoStop}%
\bibitem [{\citenamefont {Wang}\ \emph {et~al.}(2011)\citenamefont {Wang},
  \citenamefont {Schmidt}, \citenamefont {Goncharov}, \citenamefont {Skourski},
  \citenamefont {Wosnitza}, \citenamefont {Berger}, \citenamefont {Krug~von
  Nidda}, \citenamefont {Loidl},\ and\ \citenamefont {Deisenhofer}}]{CuTeO}%
  \BibitemOpen
  \bibfield  {author} {\bibinfo {author} {\bibfnamefont {Z.}~\bibnamefont
  {Wang}}, \bibinfo {author} {\bibfnamefont {M.}~\bibnamefont {Schmidt}},
  \bibinfo {author} {\bibfnamefont {Y.}~\bibnamefont {Goncharov}}, \bibinfo
  {author} {\bibfnamefont {Y.}~\bibnamefont {Skourski}}, \bibinfo {author}
  {\bibfnamefont {J.}~\bibnamefont {Wosnitza}}, \bibinfo {author}
  {\bibfnamefont {H.}~\bibnamefont {Berger}}, \bibinfo {author} {\bibfnamefont
  {H.-A.}\ \bibnamefont {Krug~von Nidda}}, \bibinfo {author} {\bibfnamefont
  {A.}~\bibnamefont {Loidl}}, \ and\ \bibinfo {author} {\bibfnamefont
  {J.}~\bibnamefont {Deisenhofer}},\ }\href {\doibase 10.1143/JPSJ.80.124707}
  {\bibfield  {journal} {\bibinfo  {journal} {Journal of the Physical Society
  of Japan}\ }\textbf {\bibinfo {volume} {80}},\ \bibinfo {pages} {124707}
  (\bibinfo {year} {2011})}\BibitemShut {NoStop}%
\bibitem [{\citenamefont {Kamenskyi}\ \emph {et~al.}(2013)\citenamefont
  {Kamenskyi}, \citenamefont {Wosnitza}, \citenamefont {Krzystek},
  \citenamefont {Aczel}, \citenamefont {Dabkowska}, \citenamefont {Dabkowski},
  \citenamefont {Luke},\ and\ \citenamefont {Zvyagin}}]{BaCrO}%
  \BibitemOpen
  \bibfield  {author} {\bibinfo {author} {\bibfnamefont {D.}~\bibnamefont
  {Kamenskyi}}, \bibinfo {author} {\bibfnamefont {J.}~\bibnamefont {Wosnitza}},
  \bibinfo {author} {\bibfnamefont {J.}~\bibnamefont {Krzystek}}, \bibinfo
  {author} {\bibfnamefont {A.~A.}\ \bibnamefont {Aczel}}, \bibinfo {author}
  {\bibfnamefont {H.~A.}\ \bibnamefont {Dabkowska}}, \bibinfo {author}
  {\bibfnamefont {A.~B.}\ \bibnamefont {Dabkowski}}, \bibinfo {author}
  {\bibfnamefont {G.~M.}\ \bibnamefont {Luke}}, \ and\ \bibinfo {author}
  {\bibfnamefont {S.~A.}\ \bibnamefont {Zvyagin}},\ }\href {\doibase
  10.1007/s10909-012-0722-4} {\bibfield  {journal} {\bibinfo  {journal} {J. Low
  Temp. Phys.}\ }\textbf {\bibinfo {volume} {170}},\ \bibinfo {pages} {231}
  (\bibinfo {year} {2013})}\BibitemShut {NoStop}%
\bibitem [{\citenamefont {Wang}\ \emph {et~al.}(2014)\citenamefont {Wang},
  \citenamefont {Kamenskyi}, \citenamefont {C\'epas}, \citenamefont {Schmidt},
  \citenamefont {Quintero-Castro}, \citenamefont {Islam}, \citenamefont {Lake},
  \citenamefont {Aczel}, \citenamefont {Dabkowska}, \citenamefont {Dabkowski},
  \citenamefont {Luke}, \citenamefont {Wan}, \citenamefont {Loidl},
  \citenamefont {Ozerov}, \citenamefont {Wosnitza}, \citenamefont {Zvyagin},\
  and\ \citenamefont {Deisenhofer}}]{BaCrO_SrCrO}%
  \BibitemOpen
  \bibfield  {author} {\bibinfo {author} {\bibfnamefont {Z.}~\bibnamefont
  {Wang}}, \bibinfo {author} {\bibfnamefont {D.}~\bibnamefont {Kamenskyi}},
  \bibinfo {author} {\bibfnamefont {O.}~\bibnamefont {C\'epas}}, \bibinfo
  {author} {\bibfnamefont {M.}~\bibnamefont {Schmidt}}, \bibinfo {author}
  {\bibfnamefont {D.~L.}\ \bibnamefont {Quintero-Castro}}, \bibinfo {author}
  {\bibfnamefont {A.~T. M.~N.}\ \bibnamefont {Islam}}, \bibinfo {author}
  {\bibfnamefont {B.}~\bibnamefont {Lake}}, \bibinfo {author} {\bibfnamefont
  {A.~A.}\ \bibnamefont {Aczel}}, \bibinfo {author} {\bibfnamefont {H.~A.}\
  \bibnamefont {Dabkowska}}, \bibinfo {author} {\bibfnamefont {A.~B.}\
  \bibnamefont {Dabkowski}}, \bibinfo {author} {\bibfnamefont {G.~M.}\
  \bibnamefont {Luke}}, \bibinfo {author} {\bibfnamefont {Y.}~\bibnamefont
  {Wan}}, \bibinfo {author} {\bibfnamefont {A.}~\bibnamefont {Loidl}}, \bibinfo
  {author} {\bibfnamefont {M.}~\bibnamefont {Ozerov}}, \bibinfo {author}
  {\bibfnamefont {J.}~\bibnamefont {Wosnitza}}, \bibinfo {author}
  {\bibfnamefont {S.~A.}\ \bibnamefont {Zvyagin}}, \ and\ \bibinfo {author}
  {\bibfnamefont {J.}~\bibnamefont {Deisenhofer}},\ }\href
  {https://link.aps.org/doi/10.1103/PhysRevB.89.174406} {\bibfield  {journal}
  {\bibinfo  {journal} {Phys. Rev. B}\ }\textbf {\bibinfo {volume} {89}},\
  \bibinfo {pages} {174406} (\bibinfo {year} {2014})}\BibitemShut {NoStop}%
\bibitem [{\citenamefont {Wang}\ \emph {et~al.}(2012)\citenamefont {Wang},
  \citenamefont {Schmidt}, \citenamefont {G\"unther}, \citenamefont {Mayr},
  \citenamefont {Wan}, \citenamefont {Lee}, \citenamefont {Ueda}, \citenamefont
  {Ueda}, \citenamefont {Loidl},\ and\ \citenamefont {Deisenhofer}}]{SrCrO}%
  \BibitemOpen
  \bibfield  {author} {\bibinfo {author} {\bibfnamefont {Z.}~\bibnamefont
  {Wang}}, \bibinfo {author} {\bibfnamefont {M.}~\bibnamefont {Schmidt}},
  \bibinfo {author} {\bibfnamefont {A.}~\bibnamefont {G\"unther}}, \bibinfo
  {author} {\bibfnamefont {F.}~\bibnamefont {Mayr}}, \bibinfo {author}
  {\bibfnamefont {Y.}~\bibnamefont {Wan}}, \bibinfo {author} {\bibfnamefont
  {S.-H.}\ \bibnamefont {Lee}}, \bibinfo {author} {\bibfnamefont
  {H.}~\bibnamefont {Ueda}}, \bibinfo {author} {\bibfnamefont {Y.}~\bibnamefont
  {Ueda}}, \bibinfo {author} {\bibfnamefont {A.}~\bibnamefont {Loidl}}, \ and\
  \bibinfo {author} {\bibfnamefont {J.}~\bibnamefont {Deisenhofer}},\ }\href
  {https://link.aps.org/doi/10.1103/PhysRevB.85.224304} {\bibfield  {journal}
  {\bibinfo  {journal} {Phys. Rev. B}\ }\textbf {\bibinfo {volume} {85}},\
  \bibinfo {pages} {224304} (\bibinfo {year} {2012})}\BibitemShut {NoStop}%
\bibitem [{SI()}]{SI}%
  \BibitemOpen
  \href@noop {} {}\bibinfo {note} {See supplemental information at XXXXXXXX for
  experimental details and further analysis of magnetometry, $\mu^+$SR, ESR and
  DFT.}\BibitemShut {Stop}%
\bibitem [{\citenamefont {Manson}\ \emph
  {et~al.}(2009{\natexlab{a}})\citenamefont {Manson}, \citenamefont
  {Schlueter}, \citenamefont {Funk}, \citenamefont {Southerland}, \citenamefont
  {Twamley}, \citenamefont {Lancaster}, \citenamefont {Blundell}, \citenamefont
  {Baker}, \citenamefont {Pratt}, \citenamefont {Singleton}, \citenamefont
  {McDonald}, \citenamefont {Goddard}, \citenamefont {Sengupta}, \citenamefont
  {Batista}, \citenamefont {Ding}, \citenamefont {Lee}, \citenamefont
  {Whangbo}, \citenamefont {Franke}, \citenamefont {Cox}, \citenamefont
  {Baines},\ and\ \citenamefont {Trial}}]{Manson2009a}%
  \BibitemOpen
  \bibfield  {author} {\bibinfo {author} {\bibfnamefont {J.~L.}\ \bibnamefont
  {Manson}}, \bibinfo {author} {\bibfnamefont {J.~A.}\ \bibnamefont
  {Schlueter}}, \bibinfo {author} {\bibfnamefont {K.~A.}\ \bibnamefont {Funk}},
  \bibinfo {author} {\bibfnamefont {H.~I.}\ \bibnamefont {Southerland}},
  \bibinfo {author} {\bibfnamefont {B.}~\bibnamefont {Twamley}}, \bibinfo
  {author} {\bibfnamefont {T.}~\bibnamefont {Lancaster}}, \bibinfo {author}
  {\bibfnamefont {S.~J.}\ \bibnamefont {Blundell}}, \bibinfo {author}
  {\bibfnamefont {P.~J.}\ \bibnamefont {Baker}}, \bibinfo {author}
  {\bibfnamefont {F.~L.}\ \bibnamefont {Pratt}}, \bibinfo {author}
  {\bibfnamefont {J.}~\bibnamefont {Singleton}}, \bibinfo {author}
  {\bibfnamefont {R.~D.}\ \bibnamefont {McDonald}}, \bibinfo {author}
  {\bibfnamefont {P.~A.}\ \bibnamefont {Goddard}}, \bibinfo {author}
  {\bibfnamefont {P.}~\bibnamefont {Sengupta}}, \bibinfo {author}
  {\bibfnamefont {C.~D.}\ \bibnamefont {Batista}}, \bibinfo {author}
  {\bibfnamefont {L.}~\bibnamefont {Ding}}, \bibinfo {author} {\bibfnamefont
  {C.}~\bibnamefont {Lee}}, \bibinfo {author} {\bibfnamefont {M.-H.}\
  \bibnamefont {Whangbo}}, \bibinfo {author} {\bibfnamefont {I.}~\bibnamefont
  {Franke}}, \bibinfo {author} {\bibfnamefont {S.}~\bibnamefont {Cox}},
  \bibinfo {author} {\bibfnamefont {C.}~\bibnamefont {Baines}}, \ and\ \bibinfo
  {author} {\bibfnamefont {D.}~\bibnamefont {Trial}},\ }\href {\doibase
  10.1021/ja808761d} {\bibfield  {journal} {\bibinfo  {journal} {J. Am. Chem.
  Soc.}\ }\textbf {\bibinfo {volume} {131}},\ \bibinfo {pages} {6733} (\bibinfo
  {year} {2009}{\natexlab{a}})}\BibitemShut {NoStop}%
\bibitem [{\citenamefont {Huddart}\ \emph {et~al.}(2019)\citenamefont
  {Huddart}, \citenamefont {Brambleby}, \citenamefont {Lancaster},
  \citenamefont {Goddard}, \citenamefont {Xiao}, \citenamefont {Blundell},
  \citenamefont {Pratt}, \citenamefont {Singleton}, \citenamefont {Macchi},
  \citenamefont {Scatena}, \citenamefont {Barton},\ and\ \citenamefont
  {Manson}}]{Huddart2019}%
  \BibitemOpen
  \bibfield  {author} {\bibinfo {author} {\bibfnamefont {B.~M.}\ \bibnamefont
  {Huddart}}, \bibinfo {author} {\bibfnamefont {J.}~\bibnamefont {Brambleby}},
  \bibinfo {author} {\bibfnamefont {T.}~\bibnamefont {Lancaster}}, \bibinfo
  {author} {\bibfnamefont {P.~A.}\ \bibnamefont {Goddard}}, \bibinfo {author}
  {\bibfnamefont {F.}~\bibnamefont {Xiao}}, \bibinfo {author} {\bibfnamefont
  {S.~J.}\ \bibnamefont {Blundell}}, \bibinfo {author} {\bibfnamefont {F.~L.}\
  \bibnamefont {Pratt}}, \bibinfo {author} {\bibfnamefont {J.}~\bibnamefont
  {Singleton}}, \bibinfo {author} {\bibfnamefont {P.}~\bibnamefont {Macchi}},
  \bibinfo {author} {\bibfnamefont {R.}~\bibnamefont {Scatena}}, \bibinfo
  {author} {\bibfnamefont {A.~M.}\ \bibnamefont {Barton}}, \ and\ \bibinfo
  {author} {\bibfnamefont {J.~L.}\ \bibnamefont {Manson}},\ }\href {\doibase
  10.1039/C8CP07160H} {\bibfield  {journal} {\bibinfo  {journal} {Phys. Chem.
  Chem. Phys.}\ }\textbf {\bibinfo {volume} {21}},\ \bibinfo {pages} {1014}
  (\bibinfo {year} {2019})}\BibitemShut {NoStop}%
\bibitem [{\citenamefont {Lanza}\ \emph {et~al.}(2014)\citenamefont {Lanza},
  \citenamefont {Fiolka}, \citenamefont {Fisch}, \citenamefont {Casati},
  \citenamefont {Skoulatos}, \citenamefont {Rüegg}, \citenamefont {Krämer},\
  and\ \citenamefont {Macchi}}]{CuF2H2Opyz}%
  \BibitemOpen
  \bibfield  {author} {\bibinfo {author} {\bibfnamefont {A.}~\bibnamefont
  {Lanza}}, \bibinfo {author} {\bibfnamefont {C.}~\bibnamefont {Fiolka}},
  \bibinfo {author} {\bibfnamefont {M.}~\bibnamefont {Fisch}}, \bibinfo
  {author} {\bibfnamefont {N.}~\bibnamefont {Casati}}, \bibinfo {author}
  {\bibfnamefont {M.}~\bibnamefont {Skoulatos}}, \bibinfo {author}
  {\bibfnamefont {C.}~\bibnamefont {Rüegg}}, \bibinfo {author} {\bibfnamefont
  {K.~W.}\ \bibnamefont {Krämer}}, \ and\ \bibinfo {author} {\bibfnamefont
  {P.}~\bibnamefont {Macchi}},\ }\href {\doibase 10.1039/C4CC06696K} {\bibfield
   {journal} {\bibinfo  {journal} {Chem. Commun.}\ }\textbf {\bibinfo {volume}
  {50}},\ \bibinfo {pages} {14504} (\bibinfo {year} {2014})}\BibitemShut
  {NoStop}%
\bibitem [{\citenamefont {Kravchina}\ \emph {et~al.}(2011)\citenamefont
  {Kravchina}, \citenamefont {Kaplienko}, \citenamefont {Nikolova},
  \citenamefont {Anders}, \citenamefont {Ziolkovskii}, \citenamefont
  {Orendachova},\ and\ \citenamefont {Kajnakova}}]{Kravchina2011}%
  \BibitemOpen
  \bibfield  {author} {\bibinfo {author} {\bibfnamefont {O.~V.}\ \bibnamefont
  {Kravchina}}, \bibinfo {author} {\bibfnamefont {A.~I.}\ \bibnamefont
  {Kaplienko}}, \bibinfo {author} {\bibfnamefont {E.~P.}\ \bibnamefont
  {Nikolova}}, \bibinfo {author} {\bibfnamefont {A.~G.}\ \bibnamefont
  {Anders}}, \bibinfo {author} {\bibfnamefont {D.~V.}\ \bibnamefont
  {Ziolkovskii}}, \bibinfo {author} {\bibfnamefont {A.}~\bibnamefont
  {Orendachova}}, \ and\ \bibinfo {author} {\bibfnamefont {M.}~\bibnamefont
  {Kajnakova}},\ }\href {\doibase 10.1134/S1990793111020205} {\bibfield
  {journal} {\bibinfo  {journal} {Russian Journal of Physical Chemistry B}\
  }\textbf {\bibinfo {volume} {5}},\ \bibinfo {pages} {209} (\bibinfo {year}
  {2011})}\BibitemShut {NoStop}%
\bibitem [{\citenamefont {Feyerherm}\ and\ \citenamefont
  {Abens}(2000)}]{Feyerherm2000}%
  \BibitemOpen
  \bibfield  {author} {\bibinfo {author} {\bibfnamefont {R.}~\bibnamefont
  {Feyerherm}}\ and\ \bibinfo {author} {\bibfnamefont {S.}~\bibnamefont
  {Abens}},\ }\href
  {http://iopscience.iop.org/article/10.1088/0953-8984/12/39/312/pdf}
  {\bibfield  {journal} {\bibinfo  {journal} {J. Phys. Condens. Matter}\
  }\textbf {\bibinfo {volume} {12}},\ \bibinfo {pages} {8495} (\bibinfo {year}
  {2000})}\BibitemShut {NoStop}%
\bibitem [{\citenamefont {Liu}\ \emph {et~al.}(2019)\citenamefont {Liu},
  \citenamefont {Kittaka}, \citenamefont {Johnson}, \citenamefont {Lancaster},
  \citenamefont {Singleton}, \citenamefont {Sakakibara}, \citenamefont
  {Kohama}, \citenamefont {van Tol}, \citenamefont {Ardavan}, \citenamefont
  {Williams}, \citenamefont {Blundell}, \citenamefont {Manson}, \citenamefont
  {Manson},\ and\ \citenamefont {Goddard}}]{Liu2019}%
  \BibitemOpen
  \bibfield  {author} {\bibinfo {author} {\bibfnamefont {J.}~\bibnamefont
  {Liu}}, \bibinfo {author} {\bibfnamefont {S.}~\bibnamefont {Kittaka}},
  \bibinfo {author} {\bibfnamefont {R.~D.}\ \bibnamefont {Johnson}}, \bibinfo
  {author} {\bibfnamefont {T.}~\bibnamefont {Lancaster}}, \bibinfo {author}
  {\bibfnamefont {J.}~\bibnamefont {Singleton}}, \bibinfo {author}
  {\bibfnamefont {T.}~\bibnamefont {Sakakibara}}, \bibinfo {author}
  {\bibfnamefont {Y.}~\bibnamefont {Kohama}}, \bibinfo {author} {\bibfnamefont
  {J.}~\bibnamefont {van Tol}}, \bibinfo {author} {\bibfnamefont
  {A.}~\bibnamefont {Ardavan}}, \bibinfo {author} {\bibfnamefont {B.~H.}\
  \bibnamefont {Williams}}, \bibinfo {author} {\bibfnamefont {S.~J.}\
  \bibnamefont {Blundell}}, \bibinfo {author} {\bibfnamefont {Z.~E.}\
  \bibnamefont {Manson}}, \bibinfo {author} {\bibfnamefont {J.~L.}\
  \bibnamefont {Manson}}, \ and\ \bibinfo {author} {\bibfnamefont {P.~A.}\
  \bibnamefont {Goddard}},\ }\href {\doibase 10.1103/PhysRevLett.122.057207}
  {\bibfield  {journal} {\bibinfo  {journal} {Phys. Rev. Lett.}\ }\textbf
  {\bibinfo {volume} {122}},\ \bibinfo {pages} {057207} (\bibinfo {year}
  {2019})}\BibitemShut {NoStop}%
\bibitem [{\citenamefont {Athas}\ \emph {et~al.}(1993)\citenamefont {Athas},
  \citenamefont {Brooks}, \citenamefont {Klepper}, \citenamefont {Uji},\ and\
  \citenamefont {Tokumoto}}]{Athas1993}%
  \BibitemOpen
  \bibfield  {author} {\bibinfo {author} {\bibfnamefont {G.~J.}\ \bibnamefont
  {Athas}}, \bibinfo {author} {\bibfnamefont {J.~S.}\ \bibnamefont {Brooks}},
  \bibinfo {author} {\bibfnamefont {S.~J.}\ \bibnamefont {Klepper}}, \bibinfo
  {author} {\bibfnamefont {S.}~\bibnamefont {Uji}}, \ and\ \bibinfo {author}
  {\bibfnamefont {M.}~\bibnamefont {Tokumoto}},\ }\href
  {https://doi.org/10.1063/1.1144336} {\bibfield  {journal} {\bibinfo
  {journal} {Rev. Sci. Instrum.}\ }\textbf {\bibinfo {volume} {64}},\ \bibinfo
  {pages} {3248} (\bibinfo {year} {1993})}\BibitemShut {NoStop}%
\bibitem [{\citenamefont {Coffey}\ \emph {et~al.}(2000)\citenamefont {Coffey},
  \citenamefont {Bayindir}, \citenamefont {DeCarolis}, \citenamefont {Bennett},
  \citenamefont {Esper},\ and\ \citenamefont {Agosta}}]{Coffey2000}%
  \BibitemOpen
  \bibfield  {author} {\bibinfo {author} {\bibfnamefont {T.}~\bibnamefont
  {Coffey}}, \bibinfo {author} {\bibfnamefont {Z.}~\bibnamefont {Bayindir}},
  \bibinfo {author} {\bibfnamefont {J.~F.}\ \bibnamefont {DeCarolis}}, \bibinfo
  {author} {\bibfnamefont {M.}~\bibnamefont {Bennett}}, \bibinfo {author}
  {\bibfnamefont {G.}~\bibnamefont {Esper}}, \ and\ \bibinfo {author}
  {\bibfnamefont {C.~C.}\ \bibnamefont {Agosta}},\ }\href
  {https://aip.scitation.org/doi/abs/10.1063/1.1321301} {\bibfield  {journal}
  {\bibinfo  {journal} {Rev. Sci. Instrum.}\ }\textbf {\bibinfo {volume}
  {71}},\ \bibinfo {pages} {4600} (\bibinfo {year} {2000})}\BibitemShut
  {NoStop}%
\bibitem [{\citenamefont {Brambleby}\ \emph {et~al.}(2017)\citenamefont
  {Brambleby}, \citenamefont {Goddard}, \citenamefont {Singleton},
  \citenamefont {Jaime}, \citenamefont {Lancaster}, \citenamefont {Huang},
  \citenamefont {Wosnitza}, \citenamefont {Topping}, \citenamefont {Carreiro},
  \citenamefont {Tran}, \citenamefont {Manson},\ and\ \citenamefont
  {Manson}}]{Brambleby2017b}%
  \BibitemOpen
  \bibfield  {author} {\bibinfo {author} {\bibfnamefont {J.}~\bibnamefont
  {Brambleby}}, \bibinfo {author} {\bibfnamefont {P.~A.}\ \bibnamefont
  {Goddard}}, \bibinfo {author} {\bibfnamefont {J.}~\bibnamefont {Singleton}},
  \bibinfo {author} {\bibfnamefont {M.}~\bibnamefont {Jaime}}, \bibinfo
  {author} {\bibfnamefont {T.}~\bibnamefont {Lancaster}}, \bibinfo {author}
  {\bibfnamefont {L.}~\bibnamefont {Huang}}, \bibinfo {author} {\bibfnamefont
  {J.}~\bibnamefont {Wosnitza}}, \bibinfo {author} {\bibfnamefont {C.~V.}\
  \bibnamefont {Topping}}, \bibinfo {author} {\bibfnamefont {K.~E.}\
  \bibnamefont {Carreiro}}, \bibinfo {author} {\bibfnamefont {H.~E.}\
  \bibnamefont {Tran}}, \bibinfo {author} {\bibfnamefont {Z.~E.}\ \bibnamefont
  {Manson}}, \ and\ \bibinfo {author} {\bibfnamefont {J.~L.}\ \bibnamefont
  {Manson}},\ }\href {\doibase 10.1103/PhysRevB.95.024404} {\bibfield
  {journal} {\bibinfo  {journal} {Phys. Rev. B}\ }\textbf {\bibinfo {volume}
  {95}},\ \bibinfo {pages} {024404} (\bibinfo {year} {2017})}\BibitemShut
  {NoStop}%
\bibitem [{\citenamefont {Ghannadzadeh}\ \emph {et~al.}(2011)\citenamefont
  {Ghannadzadeh}, \citenamefont {Coak}, \citenamefont {Franke}, \citenamefont
  {Goddard}, \citenamefont {Singleton},\ and\ \citenamefont
  {Manson}}]{Ghannadzadeh2011}%
  \BibitemOpen
  \bibfield  {author} {\bibinfo {author} {\bibfnamefont {S.}~\bibnamefont
  {Ghannadzadeh}}, \bibinfo {author} {\bibfnamefont {M.}~\bibnamefont {Coak}},
  \bibinfo {author} {\bibfnamefont {I.}~\bibnamefont {Franke}}, \bibinfo
  {author} {\bibfnamefont {P.~A.}\ \bibnamefont {Goddard}}, \bibinfo {author}
  {\bibfnamefont {J.}~\bibnamefont {Singleton}}, \ and\ \bibinfo {author}
  {\bibfnamefont {J.~L.}\ \bibnamefont {Manson}},\ }\href {\doibase
  10.1063/1.3653395} {\bibfield  {journal} {\bibinfo  {journal} {Review of
  Scientific Instruments}\ }\textbf {\bibinfo {volume} {82}},\ \bibinfo {pages}
  {113902} (\bibinfo {year} {2011})}\BibitemShut {NoStop}%
\bibitem [{\citenamefont {Manson}\ \emph {et~al.}(2018)\citenamefont {Manson},
  \citenamefont {Brambleby}, \citenamefont {Goddard}, \citenamefont {Spurgeon},
  \citenamefont {Villa}, \citenamefont {Liu}, \citenamefont {Ghannadzadeh},
  \citenamefont {Foronda}, \citenamefont {Singleton}, \citenamefont
  {Lancaster}, \citenamefont {Clark}, \citenamefont {Thomas}, \citenamefont
  {Xiao}, \citenamefont {Williams}, \citenamefont {Pratt}, \citenamefont
  {Blundell}, \citenamefont {Topping}, \citenamefont {Baines}, \citenamefont
  {Campana},\ and\ \citenamefont {Noll}}]{Manson2018}%
  \BibitemOpen
  \bibfield  {author} {\bibinfo {author} {\bibfnamefont {J.~L.}\ \bibnamefont
  {Manson}}, \bibinfo {author} {\bibfnamefont {J.}~\bibnamefont {Brambleby}},
  \bibinfo {author} {\bibfnamefont {P.~A.}\ \bibnamefont {Goddard}}, \bibinfo
  {author} {\bibfnamefont {P.~M.}\ \bibnamefont {Spurgeon}}, \bibinfo {author}
  {\bibfnamefont {J.~A.}\ \bibnamefont {Villa}}, \bibinfo {author}
  {\bibfnamefont {J.}~\bibnamefont {Liu}}, \bibinfo {author} {\bibfnamefont
  {S.}~\bibnamefont {Ghannadzadeh}}, \bibinfo {author} {\bibfnamefont
  {F.}~\bibnamefont {Foronda}}, \bibinfo {author} {\bibfnamefont
  {J.}~\bibnamefont {Singleton}}, \bibinfo {author} {\bibfnamefont
  {T.}~\bibnamefont {Lancaster}}, \bibinfo {author} {\bibfnamefont {S.~J.}\
  \bibnamefont {Clark}}, \bibinfo {author} {\bibfnamefont {I.~O.}\ \bibnamefont
  {Thomas}}, \bibinfo {author} {\bibfnamefont {F.}~\bibnamefont {Xiao}},
  \bibinfo {author} {\bibfnamefont {R.~C.}\ \bibnamefont {Williams}}, \bibinfo
  {author} {\bibfnamefont {F.~L.}\ \bibnamefont {Pratt}}, \bibinfo {author}
  {\bibfnamefont {S.~J.}\ \bibnamefont {Blundell}}, \bibinfo {author}
  {\bibfnamefont {C.~V.}\ \bibnamefont {Topping}}, \bibinfo {author}
  {\bibfnamefont {C.}~\bibnamefont {Baines}}, \bibinfo {author} {\bibfnamefont
  {C.}~\bibnamefont {Campana}}, \ and\ \bibinfo {author} {\bibfnamefont
  {B.}~\bibnamefont {Noll}},\ }\href {\doibase 10.1038/s41598-018-23054-6}
  {\bibfield  {journal} {\bibinfo  {journal} {Scientific Reports}\ }\textbf
  {\bibinfo {volume} {8}},\ \bibinfo {pages} {4745} (\bibinfo {year}
  {2018})}\BibitemShut {NoStop}%
\bibitem [{\citenamefont {Nakajima}\ \emph {et~al.}(2006)\citenamefont
  {Nakajima}, \citenamefont {Mitamura},\ and\ \citenamefont
  {Ueda}}]{Nakajima2006}%
  \BibitemOpen
  \bibfield  {author} {\bibinfo {author} {\bibfnamefont {T.}~\bibnamefont
  {Nakajima}}, \bibinfo {author} {\bibfnamefont {H.}~\bibnamefont {Mitamura}},
  \ and\ \bibinfo {author} {\bibfnamefont {Y.}~\bibnamefont {Ueda}},\ }\href
  {\doibase 10.1143/JPSJ.75.054706} {\bibfield  {journal} {\bibinfo  {journal}
  {Journal of the Physical Society of Japan}\ }\textbf {\bibinfo {volume}
  {75}},\ \bibinfo {pages} {054706} (\bibinfo {year} {2006})}\BibitemShut
  {NoStop}%
\bibitem [{\citenamefont {Zvyagin}\ \emph {et~al.}(2015)\citenamefont
  {Zvyagin}, \citenamefont {Ozerov}, \citenamefont {Kamenskyi}, \citenamefont
  {Wosnitza}, \citenamefont {Krzystek}, \citenamefont {Yoshizawa},
  \citenamefont {Hagiwara}, \citenamefont {Hu}, \citenamefont {Ryu},
  \citenamefont {Petrovic},\ and\ \citenamefont {Zhitomirsky}}]{CsCuBr}%
  \BibitemOpen
  \bibfield  {author} {\bibinfo {author} {\bibfnamefont {S.~A.}\ \bibnamefont
  {Zvyagin}}, \bibinfo {author} {\bibfnamefont {M.}~\bibnamefont {Ozerov}},
  \bibinfo {author} {\bibfnamefont {D.}~\bibnamefont {Kamenskyi}}, \bibinfo
  {author} {\bibfnamefont {J.}~\bibnamefont {Wosnitza}}, \bibinfo {author}
  {\bibfnamefont {J.}~\bibnamefont {Krzystek}}, \bibinfo {author}
  {\bibfnamefont {D.}~\bibnamefont {Yoshizawa}}, \bibinfo {author}
  {\bibfnamefont {M.}~\bibnamefont {Hagiwara}}, \bibinfo {author}
  {\bibfnamefont {R.}~\bibnamefont {Hu}}, \bibinfo {author} {\bibfnamefont
  {H.}~\bibnamefont {Ryu}}, \bibinfo {author} {\bibfnamefont {C.}~\bibnamefont
  {Petrovic}}, \ and\ \bibinfo {author} {\bibfnamefont {M.~E.}\ \bibnamefont
  {Zhitomirsky}},\ }\href {https://doi.org/10.1088/1367-2630/17/11/113059}
  {\bibfield  {journal} {\bibinfo  {journal} {New Journal of Physics}\ }\textbf
  {\bibinfo {volume} {17}},\ \bibinfo {pages} {113059} (\bibinfo {year}
  {2015})}\BibitemShut {NoStop}%
\bibitem [{\citenamefont {Ceulemans}\ \emph {et~al.}(2000)\citenamefont
  {Ceulemans}, \citenamefont {Chibotaru}, \citenamefont {Heylen}, \citenamefont
  {Pierloot},\ and\ \citenamefont {Vanquickenborne}}]{RFA}%
  \BibitemOpen
  \bibfield  {author} {\bibinfo {author} {\bibfnamefont {A.}~\bibnamefont
  {Ceulemans}}, \bibinfo {author} {\bibfnamefont {L.~F.}\ \bibnamefont
  {Chibotaru}}, \bibinfo {author} {\bibfnamefont {G.~A.}\ \bibnamefont
  {Heylen}}, \bibinfo {author} {\bibfnamefont {K.}~\bibnamefont {Pierloot}}, \
  and\ \bibinfo {author} {\bibfnamefont {L.~G.}\ \bibnamefont
  {Vanquickenborne}},\ }\href {\doibase 10.1021/cr960129k} {\bibfield
  {journal} {\bibinfo  {journal} {Chemical Reviews}\ }\textbf {\bibinfo
  {volume} {100}},\ \bibinfo {pages} {787} (\bibinfo {year}
  {2000})}\BibitemShut {NoStop}%
\bibitem [{\citenamefont {Kofu}\ \emph
  {et~al.}(2009{\natexlab{b}})\citenamefont {Kofu}, \citenamefont {Kim},
  \citenamefont {Ji}, \citenamefont {Lee}, \citenamefont {Ueda}, \citenamefont
  {Qiu}, \citenamefont {Kang}, \citenamefont {Green},\ and\ \citenamefont
  {Ueda}}]{Kofu}%
  \BibitemOpen
  \bibfield  {author} {\bibinfo {author} {\bibfnamefont {M.}~\bibnamefont
  {Kofu}}, \bibinfo {author} {\bibfnamefont {J.-H.}\ \bibnamefont {Kim}},
  \bibinfo {author} {\bibfnamefont {S.}~\bibnamefont {Ji}}, \bibinfo {author}
  {\bibfnamefont {S.-H.}\ \bibnamefont {Lee}}, \bibinfo {author} {\bibfnamefont
  {H.}~\bibnamefont {Ueda}}, \bibinfo {author} {\bibfnamefont {Y.}~\bibnamefont
  {Qiu}}, \bibinfo {author} {\bibfnamefont {H.-J.}\ \bibnamefont {Kang}},
  \bibinfo {author} {\bibfnamefont {M.~A.}\ \bibnamefont {Green}}, \ and\
  \bibinfo {author} {\bibfnamefont {Y.}~\bibnamefont {Ueda}},\ }\href {\doibase
  10.1103/PhysRevLett.102.037206} {\bibfield  {journal} {\bibinfo  {journal}
  {Phys. Rev. Lett.}\ }\textbf {\bibinfo {volume} {102}},\ \bibinfo {pages}
  {037206} (\bibinfo {year} {2009}{\natexlab{b}})}\BibitemShut {NoStop}%
\bibitem [{\citenamefont {Matsumoto}\ \emph {et~al.}(2008)\citenamefont
  {Matsumoto}, \citenamefont {Shoji},\ and\ \citenamefont
  {Koga}}]{Matsumoto2008}%
  \BibitemOpen
  \bibfield  {author} {\bibinfo {author} {\bibfnamefont {M.}~\bibnamefont
  {Matsumoto}}, \bibinfo {author} {\bibfnamefont {T.}~\bibnamefont {Shoji}}, \
  and\ \bibinfo {author} {\bibfnamefont {M.}~\bibnamefont {Koga}},\ }\href
  {\doibase 10.1143/JPSJ.77.074712} {\bibfield  {journal} {\bibinfo  {journal}
  {J. Phys. Soc. Japan}\ }\textbf {\bibinfo {volume} {77}},\ \bibinfo {pages}
  {1} (\bibinfo {year} {2008})}\BibitemShut {NoStop}%
\bibitem [{\citenamefont {Butler}\ \emph {et~al.}(1976)\citenamefont {Butler},
  \citenamefont {Walker},\ and\ \citenamefont {Soos}}]{butler}%
  \BibitemOpen
  \bibfield  {author} {\bibinfo {author} {\bibfnamefont {M.~A.}\ \bibnamefont
  {Butler}}, \bibinfo {author} {\bibfnamefont {L.~R.}\ \bibnamefont {Walker}},
  \ and\ \bibinfo {author} {\bibfnamefont {Z.~G.}\ \bibnamefont {Soos}},\
  }\href {\doibase 10.1063/1.432709} {\bibfield  {journal} {\bibinfo  {journal}
  {The Journal of Chemical Physics}\ }\textbf {\bibinfo {volume} {64}},\
  \bibinfo {pages} {3592} (\bibinfo {year} {1976})}\BibitemShut {NoStop}%
\bibitem [{\citenamefont {Hayano}\ \emph {et~al.}(1979)\citenamefont {Hayano},
  \citenamefont {Uemura}, \citenamefont {Imazato}, \citenamefont {Nishida},
  \citenamefont {Yamazaki},\ and\ \citenamefont {Kubo}}]{Hayano1979}%
  \BibitemOpen
  \bibfield  {author} {\bibinfo {author} {\bibfnamefont {R.~S.}\ \bibnamefont
  {Hayano}}, \bibinfo {author} {\bibfnamefont {Y.~J.}\ \bibnamefont {Uemura}},
  \bibinfo {author} {\bibfnamefont {J.}~\bibnamefont {Imazato}}, \bibinfo
  {author} {\bibfnamefont {N.}~\bibnamefont {Nishida}}, \bibinfo {author}
  {\bibfnamefont {T.}~\bibnamefont {Yamazaki}}, \ and\ \bibinfo {author}
  {\bibfnamefont {R.}~\bibnamefont {Kubo}},\ }\href {\doibase
  10.1103/PhysRevB.20.850} {\bibfield  {journal} {\bibinfo  {journal} {Phys.
  Rev. B}\ }\textbf {\bibinfo {volume} {20}},\ \bibinfo {pages} {850} (\bibinfo
  {year} {1979})}\BibitemShut {NoStop}%
\bibitem [{\citenamefont {Pratt}\ \emph {et~al.}(2006)\citenamefont {Pratt},
  \citenamefont {Blundell}, \citenamefont {Lancaster}, \citenamefont {Baines},\
  and\ \citenamefont {Takagi}}]{Pratt2006}%
  \BibitemOpen
  \bibfield  {author} {\bibinfo {author} {\bibfnamefont {F.~L.}\ \bibnamefont
  {Pratt}}, \bibinfo {author} {\bibfnamefont {S.~J.}\ \bibnamefont {Blundell}},
  \bibinfo {author} {\bibfnamefont {T.}~\bibnamefont {Lancaster}}, \bibinfo
  {author} {\bibfnamefont {C.}~\bibnamefont {Baines}}, \ and\ \bibinfo {author}
  {\bibfnamefont {S.}~\bibnamefont {Takagi}},\ }\href {\doibase
  10.1103/PhysRevLett.96.247203} {\bibfield  {journal} {\bibinfo  {journal}
  {Phys. Rev. Lett.}\ }\textbf {\bibinfo {volume} {96}},\ \bibinfo {pages}
  {247203} (\bibinfo {year} {2006})}\BibitemShut {NoStop}%
\bibitem [{\citenamefont {Xiao}\ \emph {et~al.}(2015)\citenamefont {Xiao},
  \citenamefont {M\"oller}, \citenamefont {Lancaster}, \citenamefont
  {Williams}, \citenamefont {Pratt}, \citenamefont {Blundell}, \citenamefont
  {Ceresoli}, \citenamefont {Barton},\ and\ \citenamefont {Manson}}]{Xiao2015}%
  \BibitemOpen
  \bibfield  {author} {\bibinfo {author} {\bibfnamefont {F.}~\bibnamefont
  {Xiao}}, \bibinfo {author} {\bibfnamefont {J.~S.}\ \bibnamefont {M\"oller}},
  \bibinfo {author} {\bibfnamefont {T.}~\bibnamefont {Lancaster}}, \bibinfo
  {author} {\bibfnamefont {R.~C.}\ \bibnamefont {Williams}}, \bibinfo {author}
  {\bibfnamefont {F.~L.}\ \bibnamefont {Pratt}}, \bibinfo {author}
  {\bibfnamefont {S.~J.}\ \bibnamefont {Blundell}}, \bibinfo {author}
  {\bibfnamefont {D.}~\bibnamefont {Ceresoli}}, \bibinfo {author}
  {\bibfnamefont {A.~M.}\ \bibnamefont {Barton}}, \ and\ \bibinfo {author}
  {\bibfnamefont {J.~L.}\ \bibnamefont {Manson}},\ }\href {\doibase
  10.1103/PhysRevB.91.144417} {\bibfield  {journal} {\bibinfo  {journal} {Phys.
  Rev. B}\ }\textbf {\bibinfo {volume} {91}},\ \bibinfo {pages} {144417}
  (\bibinfo {year} {2015})}\BibitemShut {NoStop}%
\bibitem [{\citenamefont {Clark}\ \emph {et~al.}(2005)\citenamefont {Clark},
  \citenamefont {Segall}, \citenamefont {Pickard}, \citenamefont {Hasnip},
  \citenamefont {Probert}, \citenamefont {Refson},\ and\ \citenamefont
  {Payne}}]{CASTEP}%
  \BibitemOpen
  \bibfield  {author} {\bibinfo {author} {\bibfnamefont {S.~J.}\ \bibnamefont
  {Clark}}, \bibinfo {author} {\bibfnamefont {M.~D.}\ \bibnamefont {Segall}},
  \bibinfo {author} {\bibfnamefont {C.~J.}\ \bibnamefont {Pickard}}, \bibinfo
  {author} {\bibfnamefont {P.~J.}\ \bibnamefont {Hasnip}}, \bibinfo {author}
  {\bibfnamefont {M.~I.}\ \bibnamefont {Probert}}, \bibinfo {author}
  {\bibfnamefont {K.}~\bibnamefont {Refson}}, \ and\ \bibinfo {author}
  {\bibfnamefont {M.~C.}\ \bibnamefont {Payne}},\ }\href {\doibase
  10.1524/zkri.220.5.567.65075} {\bibfield  {journal} {\bibinfo  {journal}
  {Zeitschrift fur Krist.}\ }\textbf {\bibinfo {volume} {220}},\ \bibinfo
  {pages} {567} (\bibinfo {year} {2005})}\BibitemShut {NoStop}%
\bibitem [{\citenamefont {Thomas}\ \emph {et~al.}(2017)\citenamefont {Thomas},
  \citenamefont {Clark},\ and\ \citenamefont {Lancaster}}]{IOThomas}%
  \BibitemOpen
  \bibfield  {author} {\bibinfo {author} {\bibfnamefont {I.~O.}\ \bibnamefont
  {Thomas}}, \bibinfo {author} {\bibfnamefont {S.~J.}\ \bibnamefont {Clark}}, \
  and\ \bibinfo {author} {\bibfnamefont {T.}~\bibnamefont {Lancaster}},\ }\href
  {\doibase 10.1103/PhysRevB.96.094403} {\bibfield  {journal} {\bibinfo
  {journal} {Phys. Rev. B}\ }\textbf {\bibinfo {volume} {96}},\ \bibinfo
  {pages} {094403} (\bibinfo {year} {2017})}\BibitemShut {NoStop}%
\bibitem [{\citenamefont {Tsirlin}\ and\ \citenamefont
  {Rosner}(2009)}]{tsirlin}%
  \BibitemOpen
  \bibfield  {author} {\bibinfo {author} {\bibfnamefont {A.~A.}\ \bibnamefont
  {Tsirlin}}\ and\ \bibinfo {author} {\bibfnamefont {H.}~\bibnamefont
  {Rosner}},\ }\href {\doibase 10.1103/PhysRevB.79.214417} {\bibfield
  {journal} {\bibinfo  {journal} {Phys. Rev. B}\ }\textbf {\bibinfo {volume}
  {79}},\ \bibinfo {pages} {214417} (\bibinfo {year} {2009})}\BibitemShut
  {NoStop}%
\bibitem [{\citenamefont {Kenny}\ \emph {et~al.}(2021)\citenamefont {Kenny},
  \citenamefont {Jacko},\ and\ \citenamefont {Powell}}]{powell}%
  \BibitemOpen
  \bibfield  {author} {\bibinfo {author} {\bibfnamefont {E.~P.}\ \bibnamefont
  {Kenny}}, \bibinfo {author} {\bibfnamefont {A.~C.}\ \bibnamefont {Jacko}}, \
  and\ \bibinfo {author} {\bibfnamefont {B.~J.}\ \bibnamefont {Powell}},\
  }\href {\doibase 10.1021/acs.inorgchem.1c00532} {\bibfield  {journal}
  {\bibinfo  {journal} {Inorganic Chemistry}\ }\textbf {\bibinfo {volume}
  {60}},\ \bibinfo {pages} {11907} (\bibinfo {year} {2021})}\BibitemShut
  {NoStop}%
\bibitem [{\citenamefont {Tachiki}\ and\ \citenamefont
  {Yamada}(1970)}]{Tachiki1970}%
  \BibitemOpen
  \bibfield  {author} {\bibinfo {author} {\bibfnamefont {M.}~\bibnamefont
  {Tachiki}}\ and\ \bibinfo {author} {\bibfnamefont {T.}~\bibnamefont
  {Yamada}},\ }\href {https://doi.org/10.1143/JPSJ.28.1413} {\bibfield
  {journal} {\bibinfo  {journal} {J. Phys. Soc. Japan}\ }\textbf {\bibinfo
  {volume} {28}},\ \bibinfo {pages} {1413} (\bibinfo {year}
  {1970})}\BibitemShut {NoStop}%
\bibitem [{\citenamefont {Ghannadzadeh}\ \emph {et~al.}(2013)\citenamefont
  {Ghannadzadeh}, \citenamefont {M\"oller}, \citenamefont {Goddard},
  \citenamefont {Lancaster}, \citenamefont {Xiao}, \citenamefont {Blundell},
  \citenamefont {Maisuradze}, \citenamefont {Khasanov}, \citenamefont {Manson},
  \citenamefont {Tozer}, \citenamefont {Graf},\ and\ \citenamefont
  {Schlueter}}]{Ghannadzadeh2013}%
  \BibitemOpen
  \bibfield  {author} {\bibinfo {author} {\bibfnamefont {S.}~\bibnamefont
  {Ghannadzadeh}}, \bibinfo {author} {\bibfnamefont {J.~S.}\ \bibnamefont
  {M\"oller}}, \bibinfo {author} {\bibfnamefont {P.~A.}\ \bibnamefont
  {Goddard}}, \bibinfo {author} {\bibfnamefont {T.}~\bibnamefont {Lancaster}},
  \bibinfo {author} {\bibfnamefont {F.}~\bibnamefont {Xiao}}, \bibinfo {author}
  {\bibfnamefont {S.~J.}\ \bibnamefont {Blundell}}, \bibinfo {author}
  {\bibfnamefont {A.}~\bibnamefont {Maisuradze}}, \bibinfo {author}
  {\bibfnamefont {R.}~\bibnamefont {Khasanov}}, \bibinfo {author}
  {\bibfnamefont {J.~L.}\ \bibnamefont {Manson}}, \bibinfo {author}
  {\bibfnamefont {S.~W.}\ \bibnamefont {Tozer}}, \bibinfo {author}
  {\bibfnamefont {D.}~\bibnamefont {Graf}}, \ and\ \bibinfo {author}
  {\bibfnamefont {J.~A.}\ \bibnamefont {Schlueter}},\ }\href {\doibase
  10.1103/PhysRevB.87.241102} {\bibfield  {journal} {\bibinfo  {journal} {Phys.
  Rev. B}\ }\textbf {\bibinfo {volume} {87}},\ \bibinfo {pages} {241102(R)}
  (\bibinfo {year} {2013})}\BibitemShut {NoStop}%
\bibitem [{\citenamefont {Wang}\ \emph {et~al.}(2013)\citenamefont {Wang},
  \citenamefont {Jain}, \citenamefont {Choi}, \citenamefont {van Tol},
  \citenamefont {Cheetham}, \citenamefont {Kroto}, \citenamefont {Koo},
  \citenamefont {Zhou}, \citenamefont {Hwang}, \citenamefont {Choi},
  \citenamefont {Whangbo},\ and\ \citenamefont {Dalal}}]{Wang2013}%
  \BibitemOpen
  \bibfield  {author} {\bibinfo {author} {\bibfnamefont {Z.}~\bibnamefont
  {Wang}}, \bibinfo {author} {\bibfnamefont {P.}~\bibnamefont {Jain}}, \bibinfo
  {author} {\bibfnamefont {K.-Y.}\ \bibnamefont {Choi}}, \bibinfo {author}
  {\bibfnamefont {J.}~\bibnamefont {van Tol}}, \bibinfo {author} {\bibfnamefont
  {A.~K.}\ \bibnamefont {Cheetham}}, \bibinfo {author} {\bibfnamefont {H.~W.}\
  \bibnamefont {Kroto}}, \bibinfo {author} {\bibfnamefont {H.-J.}\ \bibnamefont
  {Koo}}, \bibinfo {author} {\bibfnamefont {H.}~\bibnamefont {Zhou}}, \bibinfo
  {author} {\bibfnamefont {J.}~\bibnamefont {Hwang}}, \bibinfo {author}
  {\bibfnamefont {E.~S.}\ \bibnamefont {Choi}}, \bibinfo {author}
  {\bibfnamefont {M.-H.}\ \bibnamefont {Whangbo}}, \ and\ \bibinfo {author}
  {\bibfnamefont {N.~S.}\ \bibnamefont {Dalal}},\ }\href {\doibase
  10.1103/PhysRevB.87.224406} {\bibfield  {journal} {\bibinfo  {journal} {Phys.
  Rev. B}\ }\textbf {\bibinfo {volume} {87}},\ \bibinfo {pages} {224406}
  (\bibinfo {year} {2013})}\BibitemShut {NoStop}%
\bibitem [{\citenamefont {Manson}\ \emph
  {et~al.}(2009{\natexlab{b}})\citenamefont {Manson}, \citenamefont
  {Schlueter}, \citenamefont {Funk}, \citenamefont {Southerland}, \citenamefont
  {Twamley}, \citenamefont {Lancaster}, \citenamefont {Blundell}, \citenamefont
  {Baker}, \citenamefont {Pratt}, \citenamefont {Singleton}, \citenamefont
  {McDonald}, \citenamefont {Goddard}, \citenamefont {Sengupta}, \citenamefont
  {Batista}, \citenamefont {Ding}, \citenamefont {Lee}, \citenamefont
  {Whangbo}, \citenamefont {Franke}, \citenamefont {Cox}, \citenamefont
  {Baines},\ and\ \citenamefont {Trial}}]{Manson2009b}%
  \BibitemOpen
  \bibfield  {author} {\bibinfo {author} {\bibfnamefont {J.~L.}\ \bibnamefont
  {Manson}}, \bibinfo {author} {\bibfnamefont {J.~A.}\ \bibnamefont
  {Schlueter}}, \bibinfo {author} {\bibfnamefont {K.~A.}\ \bibnamefont {Funk}},
  \bibinfo {author} {\bibfnamefont {H.~I.}\ \bibnamefont {Southerland}},
  \bibinfo {author} {\bibfnamefont {B.}~\bibnamefont {Twamley}}, \bibinfo
  {author} {\bibfnamefont {T.}~\bibnamefont {Lancaster}}, \bibinfo {author}
  {\bibfnamefont {S.~J.}\ \bibnamefont {Blundell}}, \bibinfo {author}
  {\bibfnamefont {P.~J.}\ \bibnamefont {Baker}}, \bibinfo {author}
  {\bibfnamefont {F.~L.}\ \bibnamefont {Pratt}}, \bibinfo {author}
  {\bibfnamefont {J.}~\bibnamefont {Singleton}}, \bibinfo {author}
  {\bibfnamefont {R.~D.}\ \bibnamefont {McDonald}}, \bibinfo {author}
  {\bibfnamefont {P.~A.}\ \bibnamefont {Goddard}}, \bibinfo {author}
  {\bibfnamefont {P.}~\bibnamefont {Sengupta}}, \bibinfo {author}
  {\bibfnamefont {C.~D.}\ \bibnamefont {Batista}}, \bibinfo {author}
  {\bibfnamefont {L.}~\bibnamefont {Ding}}, \bibinfo {author} {\bibfnamefont
  {C.}~\bibnamefont {Lee}}, \bibinfo {author} {\bibfnamefont {M.~H.}\
  \bibnamefont {Whangbo}}, \bibinfo {author} {\bibfnamefont {I.}~\bibnamefont
  {Franke}}, \bibinfo {author} {\bibfnamefont {S.}~\bibnamefont {Cox}},
  \bibinfo {author} {\bibfnamefont {C.}~\bibnamefont {Baines}}, \ and\ \bibinfo
  {author} {\bibfnamefont {D.}~\bibnamefont {Trial}},\ }\href {\doibase
  10.1021/ja808761d} {\bibfield  {journal} {\bibinfo  {journal} {J. Am. Chem.
  Soc.}\ }\textbf {\bibinfo {volume} {131}},\ \bibinfo {pages} {6733} (\bibinfo
  {year} {2009}{\natexlab{b}})}\BibitemShut {NoStop}%
\bibitem [{\citenamefont {Brambleby}\ \emph {et~al.}(2015)\citenamefont
  {Brambleby}, \citenamefont {Goddard}, \citenamefont {Johnson}, \citenamefont
  {Liu}, \citenamefont {Kaminski}, \citenamefont {Ardavan}, \citenamefont
  {Steele}, \citenamefont {Blundell}, \citenamefont {Lancaster}, \citenamefont
  {Manuel}, \citenamefont {Baker}, \citenamefont {Singleton}, \citenamefont
  {Schwalbe}, \citenamefont {Spurgeon}, \citenamefont {Tran}, \citenamefont
  {Peterson}, \citenamefont {Corbey},\ and\ \citenamefont
  {Manson}}]{Brambleby2015}%
  \BibitemOpen
  \bibfield  {author} {\bibinfo {author} {\bibfnamefont {J.}~\bibnamefont
  {Brambleby}}, \bibinfo {author} {\bibfnamefont {P.~A.}\ \bibnamefont
  {Goddard}}, \bibinfo {author} {\bibfnamefont {R.~D.}\ \bibnamefont
  {Johnson}}, \bibinfo {author} {\bibfnamefont {J.}~\bibnamefont {Liu}},
  \bibinfo {author} {\bibfnamefont {D.}~\bibnamefont {Kaminski}}, \bibinfo
  {author} {\bibfnamefont {A.}~\bibnamefont {Ardavan}}, \bibinfo {author}
  {\bibfnamefont {A.~J.}\ \bibnamefont {Steele}}, \bibinfo {author}
  {\bibfnamefont {S.~J.}\ \bibnamefont {Blundell}}, \bibinfo {author}
  {\bibfnamefont {T.}~\bibnamefont {Lancaster}}, \bibinfo {author}
  {\bibfnamefont {P.}~\bibnamefont {Manuel}}, \bibinfo {author} {\bibfnamefont
  {P.~J.}\ \bibnamefont {Baker}}, \bibinfo {author} {\bibfnamefont
  {J.}~\bibnamefont {Singleton}}, \bibinfo {author} {\bibfnamefont {S.~G.}\
  \bibnamefont {Schwalbe}}, \bibinfo {author} {\bibfnamefont {P.~M.}\
  \bibnamefont {Spurgeon}}, \bibinfo {author} {\bibfnamefont {H.~E.}\
  \bibnamefont {Tran}}, \bibinfo {author} {\bibfnamefont {P.~K.}\ \bibnamefont
  {Peterson}}, \bibinfo {author} {\bibfnamefont {J.~F.}\ \bibnamefont
  {Corbey}}, \ and\ \bibinfo {author} {\bibfnamefont {J.~L.}\ \bibnamefont
  {Manson}},\ }\href {\doibase 10.1103/PhysRevB.92.134406} {\bibfield
  {journal} {\bibinfo  {journal} {Phys. Rev. B}\ }\textbf {\bibinfo {volume}
  {92}},\ \bibinfo {pages} {134406} (\bibinfo {year} {2015})}\BibitemShut
  {NoStop}%
\bibitem [{\citenamefont {Waki}\ \emph {et~al.}(2005)\citenamefont {Waki},
  \citenamefont {Kato}, \citenamefont {Itoh}, \citenamefont {Michioka},
  \citenamefont {Yoshimura},\ and\ \citenamefont {Goto}}]{Waki2005Pb2V3O9}%
  \BibitemOpen
  \bibfield  {author} {\bibinfo {author} {\bibfnamefont {T.}~\bibnamefont
  {Waki}}, \bibinfo {author} {\bibfnamefont {M.}~\bibnamefont {Kato}}, \bibinfo
  {author} {\bibfnamefont {Y.}~\bibnamefont {Itoh}}, \bibinfo {author}
  {\bibfnamefont {C.}~\bibnamefont {Michioka}}, \bibinfo {author}
  {\bibfnamefont {K.}~\bibnamefont {Yoshimura}}, \ and\ \bibinfo {author}
  {\bibfnamefont {T.}~\bibnamefont {Goto}},\ }\href {\doibase
  https://doi.org/10.1016/j.jpcs.2005.05.025} {\bibfield  {journal} {\bibinfo
  {journal} {Journal of Physics and Chemistry of Solids}\ }\textbf {\bibinfo
  {volume} {66}},\ \bibinfo {pages} {1432} (\bibinfo {year}
  {2005})}\BibitemShut {NoStop}%
\bibitem [{\citenamefont {Mentré}\ \emph {et~al.}(1999)\citenamefont
  {Mentré}, \citenamefont {Dhaussy}, \citenamefont {Abraham}, \citenamefont
  {Suard},\ and\ \citenamefont {Steinfink}}]{Mentre1999Pb2V3O9}%
  \BibitemOpen
  \bibfield  {author} {\bibinfo {author} {\bibfnamefont {O.}~\bibnamefont
  {Mentré}}, \bibinfo {author} {\bibfnamefont {A.~C.}\ \bibnamefont
  {Dhaussy}}, \bibinfo {author} {\bibfnamefont {F.}~\bibnamefont {Abraham}},
  \bibinfo {author} {\bibfnamefont {E.}~\bibnamefont {Suard}}, \ and\ \bibinfo
  {author} {\bibfnamefont {H.}~\bibnamefont {Steinfink}},\ }\href {\doibase
  10.1021/cm990073l} {\bibfield  {journal} {\bibinfo  {journal} {Chemistry of
  Materials}\ }\textbf {\bibinfo {volume} {11}},\ \bibinfo {pages} {2408}
  (\bibinfo {year} {1999})}\BibitemShut {NoStop}%
\bibitem [{\citenamefont {Dalal}(2017)}]{DalalJT}%
  \BibitemOpen
  \bibfield  {author} {\bibinfo {author} {\bibfnamefont {M.}~\bibnamefont
  {Dalal}},\ }\href
  {https://www.dalalinstitute.com/wp-content/uploads/Books/A-Textbook-of-Inorganic-Chemistry-Volume-1/ATOICV1-8-7-Jahn-Teller-Effect.pdf}
  {\emph {\bibinfo {title} {A Textbook of Inorganic Chemistry -- Volume 1}}}\
  (\bibinfo  {publisher} {Amazon Digital Services LLC - KDP Print US},\
  \bibinfo {year} {2017})\ pp.\ \bibinfo {pages} {312--321}\BibitemShut
  {NoStop}%
\bibitem [{\citenamefont {Butcher}\ \emph {et~al.}(2008)\citenamefont
  {Butcher}, \citenamefont {Landee}, \citenamefont {Turnbull},\ and\
  \citenamefont {Xiao}}]{Butcher2008}%
  \BibitemOpen
  \bibfield  {author} {\bibinfo {author} {\bibfnamefont {R.~T.}\ \bibnamefont
  {Butcher}}, \bibinfo {author} {\bibfnamefont {C.~P.}\ \bibnamefont {Landee}},
  \bibinfo {author} {\bibfnamefont {M.~M.}\ \bibnamefont {Turnbull}}, \ and\
  \bibinfo {author} {\bibfnamefont {F.}~\bibnamefont {Xiao}},\ }\href {\doibase
  10.1016/j.ica.2008.03.090} {\bibfield  {journal} {\bibinfo  {journal}
  {Inorganica Chim. Acta}\ }\textbf {\bibinfo {volume} {361}},\ \bibinfo
  {pages} {3654} (\bibinfo {year} {2008})}\BibitemShut {NoStop}%
\bibitem [{\citenamefont {Lapidus}\ \emph {et~al.}(2013)\citenamefont
  {Lapidus}, \citenamefont {Manson}, \citenamefont {Liu}, \citenamefont
  {Smith}, \citenamefont {Goddard}, \citenamefont {Bendix}, \citenamefont
  {Topping}, \citenamefont {Singleton}, \citenamefont {Dunmars}, \citenamefont
  {Mitchell},\ and\ \citenamefont {Schlueter}}]{Lapidus2013}%
  \BibitemOpen
  \bibfield  {author} {\bibinfo {author} {\bibfnamefont {S.~H.}\ \bibnamefont
  {Lapidus}}, \bibinfo {author} {\bibfnamefont {J.~L.}\ \bibnamefont {Manson}},
  \bibinfo {author} {\bibfnamefont {J.}~\bibnamefont {Liu}}, \bibinfo {author}
  {\bibfnamefont {M.~J.}\ \bibnamefont {Smith}}, \bibinfo {author}
  {\bibfnamefont {P.}~\bibnamefont {Goddard}}, \bibinfo {author} {\bibfnamefont
  {J.}~\bibnamefont {Bendix}}, \bibinfo {author} {\bibfnamefont {C.~V.}\
  \bibnamefont {Topping}}, \bibinfo {author} {\bibfnamefont {J.}~\bibnamefont
  {Singleton}}, \bibinfo {author} {\bibfnamefont {C.}~\bibnamefont {Dunmars}},
  \bibinfo {author} {\bibfnamefont {J.~F.}\ \bibnamefont {Mitchell}}, \ and\
  \bibinfo {author} {\bibfnamefont {J.~A.}\ \bibnamefont {Schlueter}},\ }\href
  {\doibase 10.1039/c3cc41394b} {\bibfield  {journal} {\bibinfo  {journal}
  {Chem. Commun.}\ }\textbf {\bibinfo {volume} {49}},\ \bibinfo {pages} {3558}
  (\bibinfo {year} {2013})}\BibitemShut {NoStop}%
\bibitem [{\citenamefont {Lancaster}\ \emph {et~al.}(2019)\citenamefont
  {Lancaster}, \citenamefont {Huddart}, \citenamefont {Williams}, \citenamefont
  {Xiao}, \citenamefont {Franke}, \citenamefont {Baker}, \citenamefont {Pratt},
  \citenamefont {Blundell}, \citenamefont {Schlueter}, \citenamefont {Mills},
  \citenamefont {Maahs},\ and\ \citenamefont {Preuss}}]{Lancaster2019}%
  \BibitemOpen
  \bibfield  {author} {\bibinfo {author} {\bibfnamefont {T.}~\bibnamefont
  {Lancaster}}, \bibinfo {author} {\bibfnamefont {B.~M.}\ \bibnamefont
  {Huddart}}, \bibinfo {author} {\bibfnamefont {R.~C.}\ \bibnamefont
  {Williams}}, \bibinfo {author} {\bibfnamefont {F.}~\bibnamefont {Xiao}},
  \bibinfo {author} {\bibfnamefont {K.~J.~A.}\ \bibnamefont {Franke}}, \bibinfo
  {author} {\bibfnamefont {P.~J.}\ \bibnamefont {Baker}}, \bibinfo {author}
  {\bibfnamefont {F.~L.}\ \bibnamefont {Pratt}}, \bibinfo {author}
  {\bibfnamefont {S.~J.}\ \bibnamefont {Blundell}}, \bibinfo {author}
  {\bibfnamefont {J.~A.}\ \bibnamefont {Schlueter}}, \bibinfo {author}
  {\bibfnamefont {M.~B.}\ \bibnamefont {Mills}}, \bibinfo {author}
  {\bibfnamefont {A.~C.}\ \bibnamefont {Maahs}}, \ and\ \bibinfo {author}
  {\bibfnamefont {K.~E.}\ \bibnamefont {Preuss}},\ }\href
  {https://doi.org/10.1088/1361-648x/ab2cb6} {\bibfield  {journal} {\bibinfo
  {journal} {Journal of Physics: Condensed Matter}\ }\textbf {\bibinfo {volume}
  {31}},\ \bibinfo {pages} {394002} (\bibinfo {year} {2019})}\BibitemShut
  {NoStop}%
\end{thebibliography}
\end{document}


\title{Anomalous magnetic exchange in a dimerized quantum-magnet \\composed of unlike spin species - supplemental information}

\

\author{S. P. M. Curley}
\affiliation{Department of Physics, University of Warwick, Gibbet Hill Road, Coventry, CV4 7AL, UK}
\author{B. M. Huddart}
\affiliation{Centre for Materials Physics, Durham University, Durham, DH1 3LE, UK}
\author{D. Kamenskyi}
\affiliation{Experimental Physics V, Center for Electronic Correlations and Magnetism,
Institute of Physics, University of Augsburg, 86135 Augsburg, Germany}
\affiliation{Molecular Photoscience Research Center, Kobe University, 657-8501 Kobe, Japan}
\author{M. J. Coak}
\affiliation{Department of Physics, University of Warwick, Gibbet Hill Road, Coventry, CV4 7AL, UK}
\author{R. C. Williams}
\affiliation{Department of Physics, University of Warwick, Gibbet Hill Road, Coventry, CV4 7AL, UK}
\author{S. Ghannadzadeh}
\affiliation{Department of Physics, Clarendon Laboratory, Oxford University, Parks Road,
Oxford, OX1 3PU, UK}
\author{A. Schneider}
\affiliation{Experimental Physics VI, Center for Electronic Correlations and Magnetism,
Institute of Physics, University of Augsburg, 86135 Augsburg, Germany}
\author{S. Okubo}
\affiliation{Molecular Photoscience Research Center, Kobe University, 657-8501 Kobe, Japan}
\author{T. Sakurai}
\affiliation{Molecular Photoscience Research Center, Kobe University, 657-8501 Kobe, Japan}
\author {H. Ohta}
\affiliation{Molecular Photoscience Research Center, Kobe University, 657-8501 Kobe, Japan}
\author{J. P. Tidey}
\affiliation{Department of Chemistry, University of Warwick, Gibbet Hill, Coventry CV4 7AL, UK}
\author{D. Graf}
\affiliation{National High Magnetic Field Laboratory, Florida State University, Tallahassee, Florida 32310, USA}
\author{S. J. Clark}
\affiliation{Centre for Materials Physics, Durham University, Durham DH1 3LE, UK}
\author{S. J. Blundell}
\affiliation{Department of Physics, Clarendon Laboratory, Oxford University, Parks Road,
Oxford, OX1 3PU, UK}
\author{F. L. Pratt}
\affiliation{ISIS Facility, Rutherford Appleton Laboratory, Chilton, Oxfordshire, OX11 0QX, UK}
\author{M. T. F. Telling}
\affiliation{ISIS Facility, Rutherford Appleton Laboratory, Chilton, Oxfordshire, OX11 0QX, UK}
\author{T. Lancaster}
\email{tom.lancaster@durham.ac.uk}
\affiliation{Centre for Materials Physics, Durham University, Durham, DH1 3LE, UK}
\author{J. L. Manson}
 \email{jmanson@ewu.edu}
\affiliation{Department of Chemistry and Biochemistry, Eastern Washington University, Cheney, Washington 99004, USA}
\author{P. A. Goddard}
 \email{p.goddard@warwick.ac.uk}
\affiliation{Department of Physics, University of Warwick, Gibbet Hill Road, Coventry, CV4 7AL, UK}

\maketitle


\renewcommand{\thefigure}{S\arabic{figure}}
\renewcommand{\thetable}{S\Roman{table}}

\section{Experimental details}

\subsection{Synthesis and structure}


A crystal of dimensions 0.48 x 0.32 x 0.14 mm was affixed to a 50~$\mu$m Mitigen micromount using Fomblin-Y perfluoroether. Data were collected using a Rigaku Oxford Diffraction XtalLAB Synergy, Dualflex diffractometer \cite{SynergyS} equipped with a HyPix-6000HE Hybrid Photon Counting area detector and employing Mo K$_{\alpha}$ radiation ($\lambda = 0.71073$\,\AA), generated by a PhotonJet, micro-focus sealed X-ray tube. Data were collected to $d_{\rm{min}} = 0.50$\,\AA\,, indexed and reduced using CrysAlisPro~\cite{CrysAlisPRO} and absorption corrected for by Gaussian integration over a multifaceted crystal model.

Structural solution was performed by direct methods using SHELXS and refinement performed using SHELXL~\cite{ShelXL}, implemented through Olex2. H-atoms were located from the difference map and restraints placed on the O---H 1,2- and the H---O---H 1,3-distances to maintain reasonable geometry while allowing geometric freedom with respect to the M-O bond. Non-H atom atomic displacements were treated anisotropically with those of hydrogen atoms were treated isotropically and riding on the donor atom. While the Flack parameter refined to essentially zero, the potentially inversion twin law has been retained for the purpose of discussion. Table~\ref{tab:xrayCuVdimer} shows the details of the data reduction and structure solution.

\noindent
\begin{table}[t]
\centering
\caption{Single crystal X-ray data and refinements details for CuVOF$_4$(H$_2$O)$_6 \cdot$H$_2$O.}
\label{tab:xrayCuVdimer}
\small
\begin{tabular}{l c} 
\hline
\hline
Parameter (units) & Fit Results (error) \\
\hline
Instrument & Rigaku Oxford Diffraction \\ 
  & Synergy S  \\
 Method & Single crystal \\
Empirical formula &	Cu F$_4$ H$_{14}$ O$_8$ V  \\
Formula weight (g mol$^{-1}$) &	332.59  \\
Temperature (K) &	150.00(10)  \\
Crystal system &	orthorhombic  \\
Space group &	$P n a 2_1$  \\
$a$ (\AA)	 & 15.65164(16)  \\
$b$ (\AA)  &	8.24054(8)  \\
$c$ (\AA)  &	7.37877(7) \\  
$\alpha$ ($^{\circ}$) &	90  \\
$\beta$ ($^{\circ}$)  &	90  \\
$\gamma$ ($^{\circ}$) &	90   \\
Volume (\AA$^{3}$) &	951.699(16)  \\
Z &	4  \\
$\rho_{\rm calc}$ (g cm$^{-3}$) &	2.321 \\
$\mu$ (mm$^{‑1}$) &	3.301  \\
Crystal size (mm$^3$) &	0.479 × 0.320 × 0.143 \\
Radiation &	Mo K$_{\alpha}$ ($\lambda$ = 0.71073) \\
2$\theta$ range for data collection ($^{\circ}$) &	7.18 to 89.88 \\
$h$-index range &	$-31 \leq h \leq 31$ \\
$k$-index range & $-14 \leq k \leq 16$ \\
$l$-index range & $-14 \leq l \leq 14$ \\
Reflections collected &	37584 \\
Independent reflections &	7743  \\
Data/restraints/parameters &	7743/22/171 \\
Goodness-of-fit on F2 &	1.069 \\
Final R indexes [$I \geq 2\sigma$ ($I$)] &	$R_1$ = 0.0226 \\    &  w$R_2$ = 0.0602 \\
Final R indexes [all data] &	$R_1$ = 0.0232\\
  & w$R_2$ = 0.0604 \\
Largest diff. peak/hole (e \AA$^{-3}$) &	0.78/-1.10  \\
Flack parameter	& 0.012(5) \\

\hline 
\end{tabular}
\end{table}
\noindent
\subsection{Magnetometry}

\subsubsection{SQUID magnetometry}

SQUID magnetometry measurements were performed using a Quantum Design MPMS-XL SQUID magnetometer. Single-crystals were orientated such that the crystallographic $a$-axis was aligned parallel and perpendicular to the applied magnetic field $H$. Samples were mounted inside a gelatin capsule and held in place with a small application of Apiezon vacuum M-grease. The capsule was then mounted inside a low magnetic background drinking straw and loaded into the sample chamber. The temperature dependence of the molar susceptibility $\chi$ was obtained in the linear limit using the relation $\chi = M/nH$ where $M$ is the measured magnetic moment and $n$ is the number of moles of the compound.

\noindent
\subsubsection{Radio frequency susceptometry}

The dynamic magnetic susceptibility (d$M$/d$H$) was measured using as radio-frequency (RF) oscillator circuit technique. The set-up involves a RF circuit and a Tunnel-Diode oscillator (TDO) and is based on an LCR circuit. A single-crystal was placed within a small detector coil with applied field orientated parallel then perpendicular to the crystallographic $a$-axis. The detector coil is inductively coupled to the TDO which measures changes in the oscillation frequency ($\Delta\omega$) of the circuit. This can be related to the real part of the dynamic susceptibility,
\noindent
\begin{equation}
    \frac{\Delta\omega}{\omega} = -\pi f \frac{\rm{d}\textit{M}}{\rm{d}\textit{H}} 
\end{equation}
\noindent
where $f$ is the filling factor \cite{Clover1970,
Coffey2000}. The detector-coil and TDO were mounted into a top-loading liquid sorption pumped $^3$He cryostat. Fields of up to 35~T were obtained by utilising a DC-field resistive magnet located in Cell-12 at the National High Magnetic Field Laboratory, Florida, US. Empty-coil field sweeps were collected as a background and subtracted from similar temperature total frequency responses in order to isolate the sample response.

\subsection{Electron-spin resonance}

Electron-spin resonance (ESR) experiments were performed over wide ranges of frequencies ($9-500$ GHz), magnetic fields (up to 15 T) and temperatures ($2-295$ K). The low-frequency ESR studies were performed on single-crystals of CuVOF$_4$(H$_2$O)$_6\cdot$H$_2$O using a commercially available X-Band  Bruker ESR spectrometer operating at 9.35 GHz at Augsburg University. High-frequency ESR experiments were done using a multi-frequency spectrometer operated in combination with a 16 T pulsed magnet (5 msec pulse) at Kobe University (Japan). Backward Wave Oscillators and Gunn Diodes were employed as sources of mm- and submm-wavelength radiation. 2,2-Diphenyl-1-picrylhydrazyl (DPPH) was used as a standard magnetic field marker with $g$-factor $g=2.0036$.

\subsection{Muon-spin relaxation}

Zero-field (ZF) and longitudinal field (LF) muon-spin relaxation ($\mu^{+}$SR) measurements \cite{steve_review} were made on a polycrystalline sample of CuVOF$_4$(H$_2$O)$_6\cdot$H$_2$O using the HiFi spectrometer at the ISIS facility, Rutherford Appleton Laboratory, UK.
In a $\mu^{+}$SR experiment \cite{steve_review} spin polarized muons
are implanted into the sample, where they precess about the total magnetic field $B$ at the muon site at a frequency $\nu=\gamma_{\mu} B / 2\pi$, where $\gamma_{\mu}$ is the muon gyromagnetic ratio ($=2\pi \times 135.5$~MHz~~T$^{-1}$).  These muons decay with an average lifetime of 2.2~$\mu$s into two neutrinos and a positron.  Due to the parity-violating nature of the weak interaction, positrons are emitted preferentially along the instantaneous muon spin direction.  Recording the direction of emitted positron therefore allows us to infer the muon spin polarization at the time of decay.  The quantity of interest is the asymmetry 
\begin{equation}
A(t)=\frac{N_{\textrm{F}}(t)-\alpha N_{\textrm{B}}(t)}{N_{\textrm{F}}(t)+\alpha N_{\textrm{B}}(t)},
\label{eq:zf_asymmetry}
\end{equation}
where $N_{\textrm{F/B}}$ is the number of positrons detected in the forward/backward detectors and $\alpha$ is an experimental calibration constant.  The asymmetry $A(t)$ is proportional to the spin polarization of the muon ensemble.

For a magnetically-ordered compound one observes oscillations in the asymmetry $A(t)$.  For a distribution of magnetic fields the spins will each precess at a different frequency, resulting in a relaxation of the muon polarization. When dynamics are present in the fast fluctuation limit \cite{dalmas,hayano}, the relaxation rate is expected to vary as $\lambda \propto  {\Delta}^{2} \tau$, where $\Delta=\sqrt{{\gamma}_{\mu}^{2}\langle(B-\langle B \rangle)^{2} \rangle }$ is the second moment of  the
local magnetic field distribution at the muon site and $\tau$ is the correlation time. In an LF $\mu^{+}$SR experiment a field is applied parallel to the direction of the initial muon spin.  Being parallel to the muon spin, the applied field does not result in precession but instead `locks in' the spin direction of the muon, since a large field component lies along the initial muon polarization direction. This allows us to probe the dynamics of the system, with time-varying magnetic fields at the muon site being able to flip muon spins.

\section{Results}

\subsection{Magnetometry}

\begin{figure}[t]
	\includegraphics[width=\columnwidth]{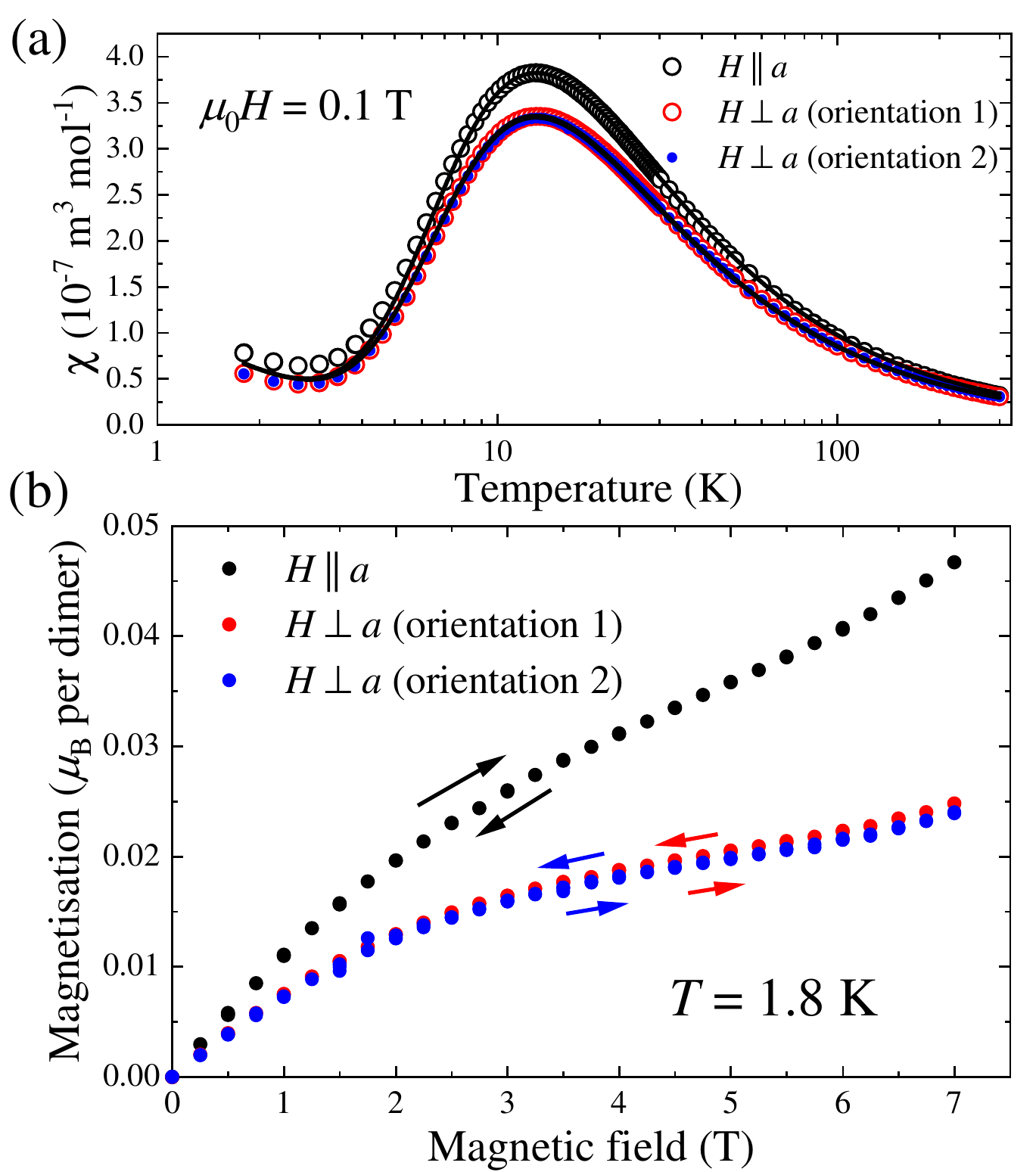}
	\caption{Orientation dependent SQUID magnetometry for CuVOF$_4$(H$_2$O)$_6\cdot$H$_2$O with field along three orthogonal crystal directions for the mangnetic susceptibility plotted as a function of temperature (a) and magnetic moment plotted a function of applied magnetic field (b). Solids lines in (a) are fits to Eq.~\ref{eq:BB_full}.}
	\label{fig:SQUID_angdep}   
\end{figure}

The results of SQUID magnetometry measurements performed on an orientated single-crystal of CuVOF$_4$(H$_2$O)$_6\cdot$H$_2$O are shown in Fig.~\ref{fig:SQUID_angdep}. The magnetic susceptibility $\chi(T)$ shows little difference between measurements made with the field along the two orthogonal directions for $H \perp a$, whilst the hump in $\chi(T)$ for $H \parallel a$ has a slightly larger amplitude. All datasets were fit to an interacting dimer model of the form,
\noindent
\begin{multline}\label{eq:BB_full}
        \chi = (1-\rho)\frac{N_{\rm A} (g \mu_{\rm B})^2}{k_{\rm B}T(3+\exp{(J_0/T)}+ nJ'/T)}\\+\rho \frac{(g_p \mu_{\rm B})^2N_{\rm A} \mu_0 S (S+1)}{3 k_{\rm B} T}
\end{multline}
where here $N_{\rm A}$ is Avogadro’s number, $\mu_{\rm B}$ is the Bohr magneton, $\mu_0$ is the permeability of free space, $J_0$
is the intradimer exchange constant, $J'$ is the
interdimer exchange constant and $n$ the number of nearest dimer neighbors. The second half of the equation models the low-temperature upturn in $\chi(T)$ as an ensemble of free $S = 1/2$ ions with an isotropic $g$-factor of $g_p = 2$ with the parameter $\rho$ capturing the fraction of the sample attributable to isotropic $S = 1/2$ ions. As in described in the main text, $\chi(T)$ datasets were fit simultaneously with only $g$-factors free to vary for each dataset (resultant fit parameters can be found in the main text).

Magnetisation $M$($H$) for field along all three orthogonal crystal directions each possess the same qualitative $M$($H$) profile. No hysteretic behaviour is seen along any direction, indicating an absence of any ferrimagnetism in the system. For $H \parallel a$, $M$($H$) rises more rapidly than for $H$ along the other two orthogonal orientations. This may indicate the presence of a DM vector orientated slightly along the $a$-axis.
\noindent
\subsection{Electron-spin resonance}

\begin{figure}[t]
	\includegraphics[width=\columnwidth]{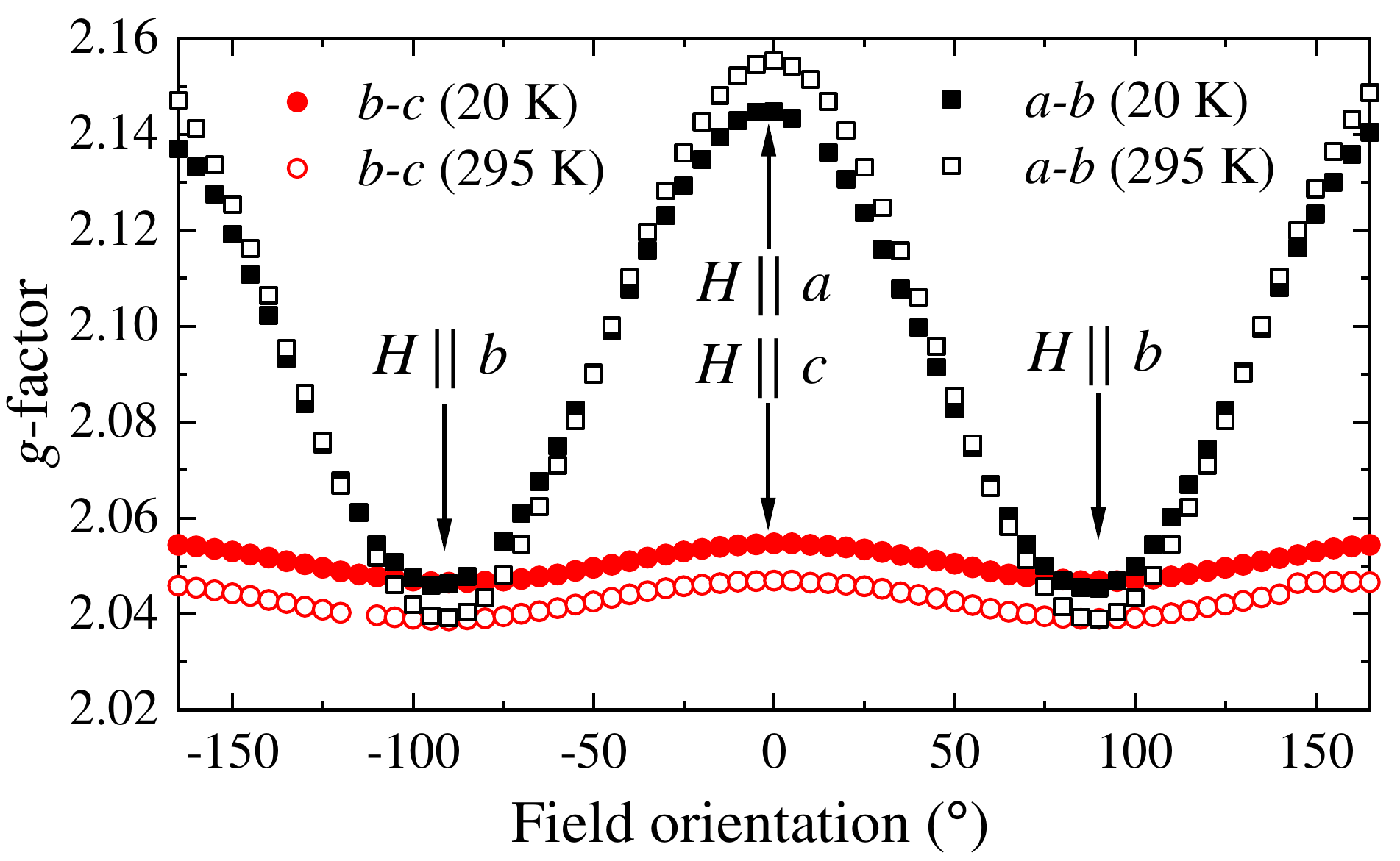}
	\caption{The angular dependence of the $g$-factor of the transitions within the excited triplet state ($S=1$) in CuVOF$_4$(H$_2$O)$_6\cdot$H$_2$O.}
	\label{fig:ESRang}   
\end{figure}

Angular dependencies of the triplet transitions $g$-factor were measured at 20 and 295 K when these transitions are dominate in the spectrum. The results are shown in Fig.~\ref{fig:ESRang}. Apart the anisotropy of the g-factor, we also observe significant anisotropy of the EPR linewidth, which becomes much narrow when field is along the $b$-axis.

\subsection{Muon-spin relaxation}

\begin{figure}[b]
	\includegraphics[width=\columnwidth]{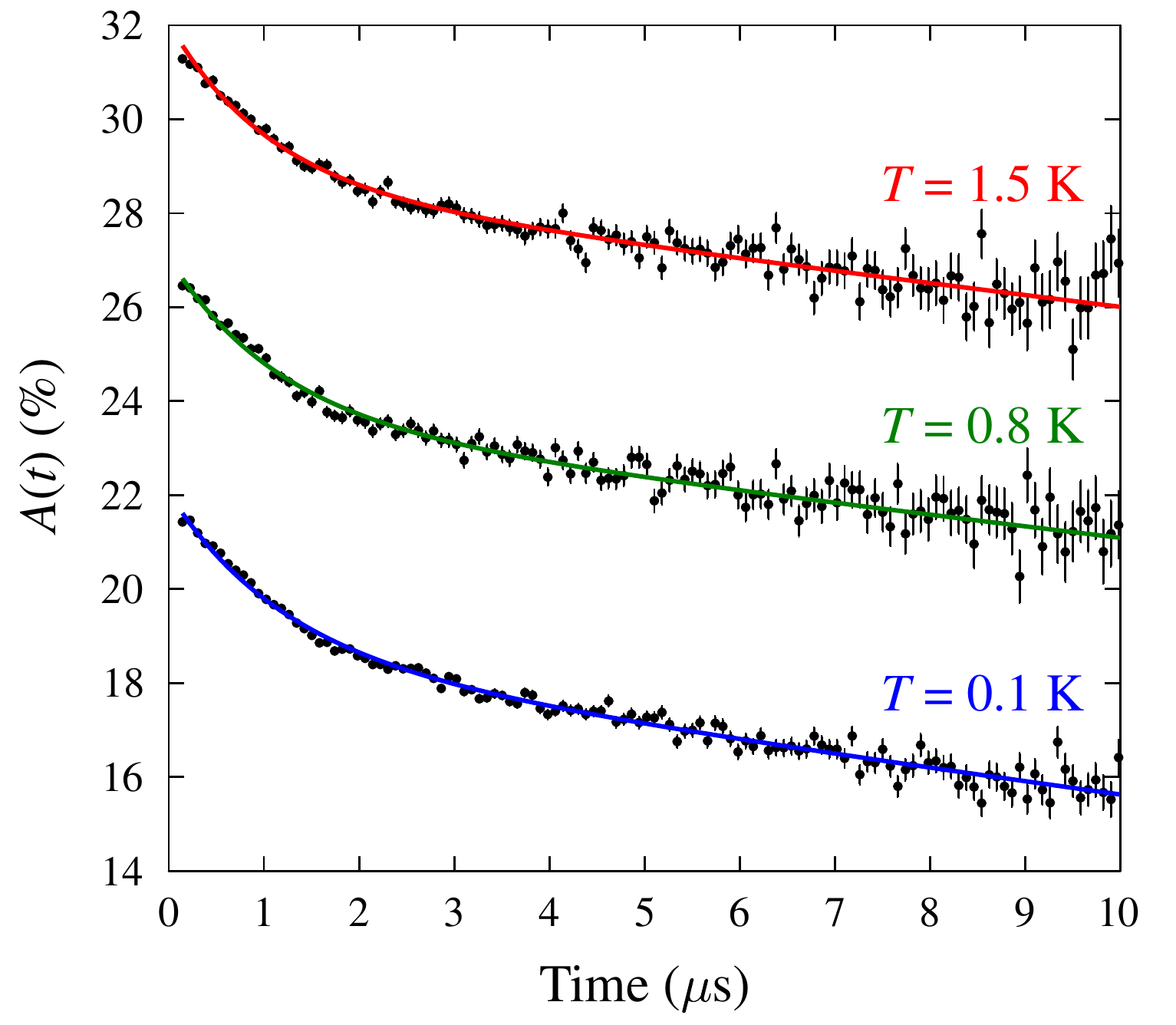}
	\caption{Zero field $\mu^+$SR spectra for CuVOF$_4$(H$_2$O)$_6\cdot$H$_2$O at three different temperatures.  For clarity, we show $A(t)$ + 5\% and $A(t)$ + 10\% for the data at $T=0.8$~K and $T=1.5$~K, respectively.  }
	\label{fig:zf_spectra}
\end{figure}

Zero field measurements were made on 
CuVOF$_4$(H$_2$O)$_6\cdot$H$_2$O for $0.1 \le T \le 1.5$~K. Example spectra at three different temperatures are shown in Fig.~\ref{fig:zf_spectra}. The spectra do not show any oscillations in the asymmetry, that would be characteristic of magnetic order, down to 0.1 K. These spectra were fit to a stretched exponential function
\begin{equation}
A(t)=A_\textrm{1}e^{-\lambda_1 t}+A_{2}e^{-\lambda_2 t},
\label{eq:lf_asymmetry}
\end{equation}
where the component with amplitude $A_1$ and relaxation rate $\lambda_2$ is due to muons that are sensitive to fluctuating electronic moments. A slowly-relaxing component with amplitude $A_2$ and $\lambda_2$ has contributions from several sources: muons that stop in the silver sample holder or cryostat tails, or from muons stopping in the sample, but at positions where they are not sensitive to the electronic moments. 
\begin{figure}[t]
	\includegraphics[width=\columnwidth]{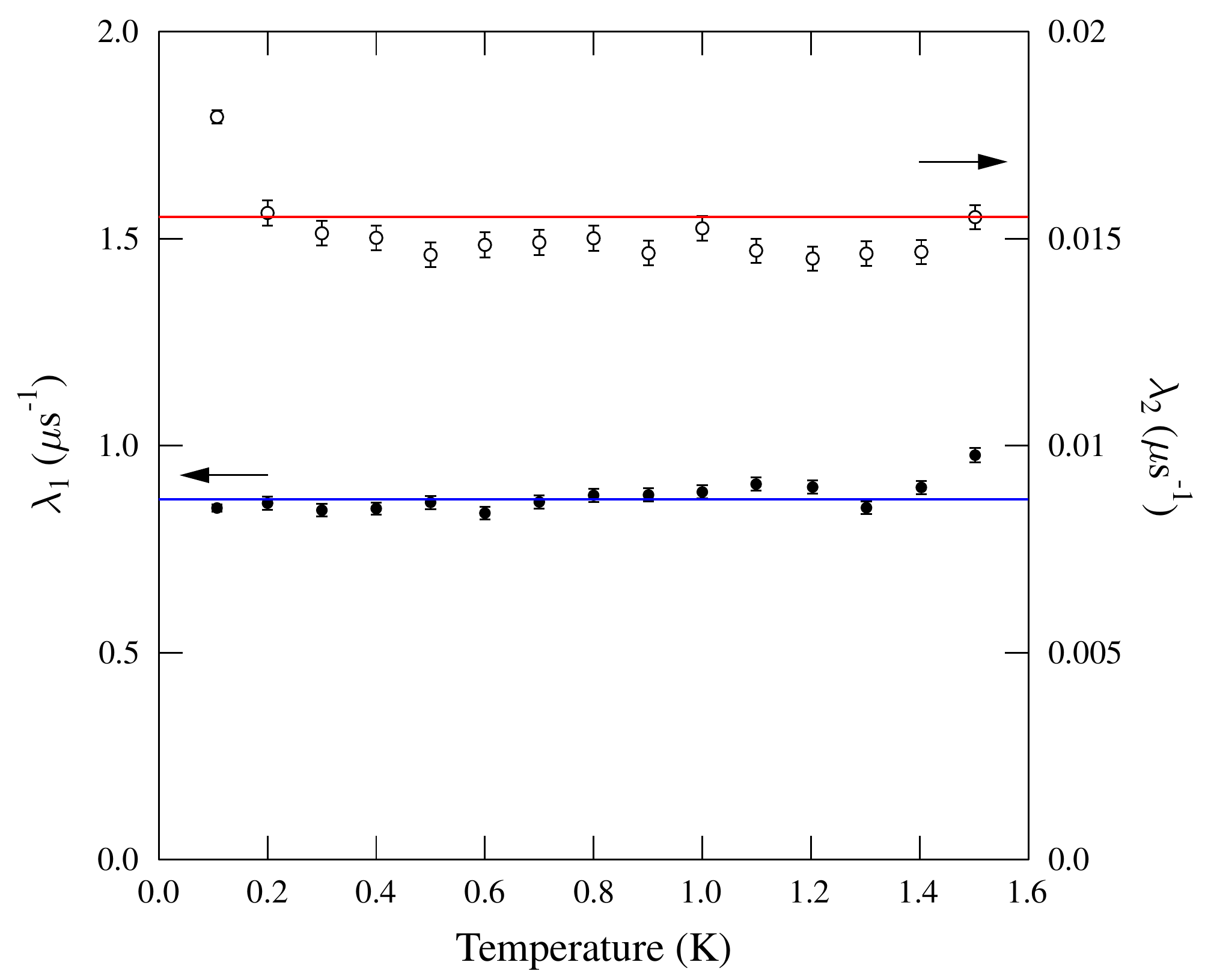}
	\caption{Temperature dependence of the relaxation rates in Eq. \ref{eq:zf_asymmetry}.  Both $\lambda_1$ (filled circles) and $\lambda_2$ (empty circles) are seen to be approximately constant. Note the different axis for each relaxation rate. }
	\label{fig:zf_lambda}
\end{figure}
\begin{figure}[b]
	\includegraphics[width=\columnwidth]{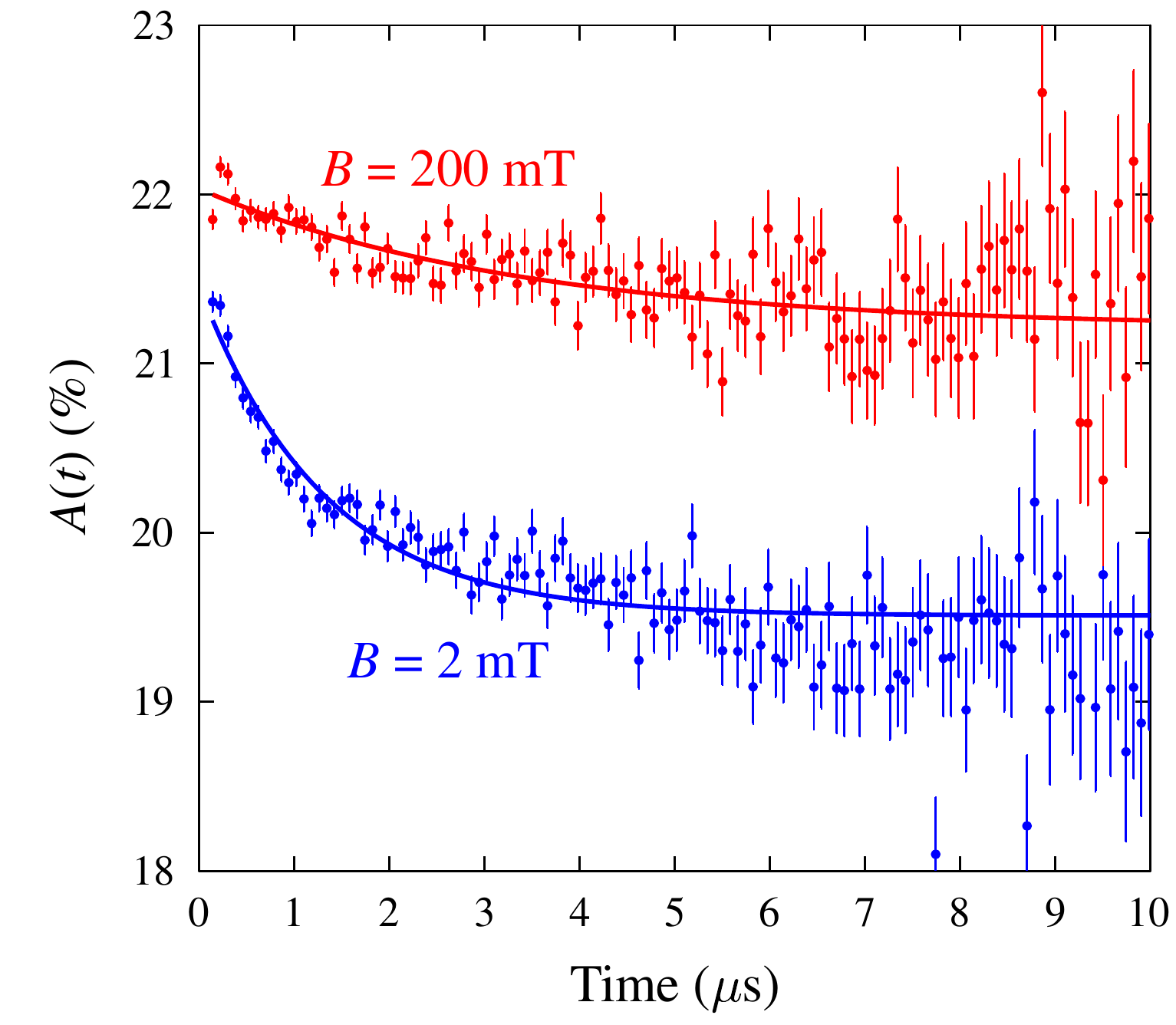}
	\caption{Longitudinal field $\mu^+$SR spectra for CuVOF$_4$(H$_2$O)$_6\cdot$H$_2$O at two different fields.  Spectra are displaced vertically for clarity.}
	\label{fig:lf_spectra}
\end{figure}
The amplitudes $A_1$ and $A_2$ reflect the proportion of muons stopping in each of the distinct magnetic environments and were found not to vary significantly with temperature, so were fixed to their average values 3.35\% and 18.7\% respectively in the fitting procedure. As shown in Fig.~\ref{fig:zf_spectra}, there is no significant change in the shape of the spectra with changing temperature.  The relaxation parameters obtained from a fitting of the above form at each temperature are shown in Fig.~\ref{fig:zf_lambda} and also show little temperature dependence.

\begin{figure}[t]
	\includegraphics[width=\columnwidth]{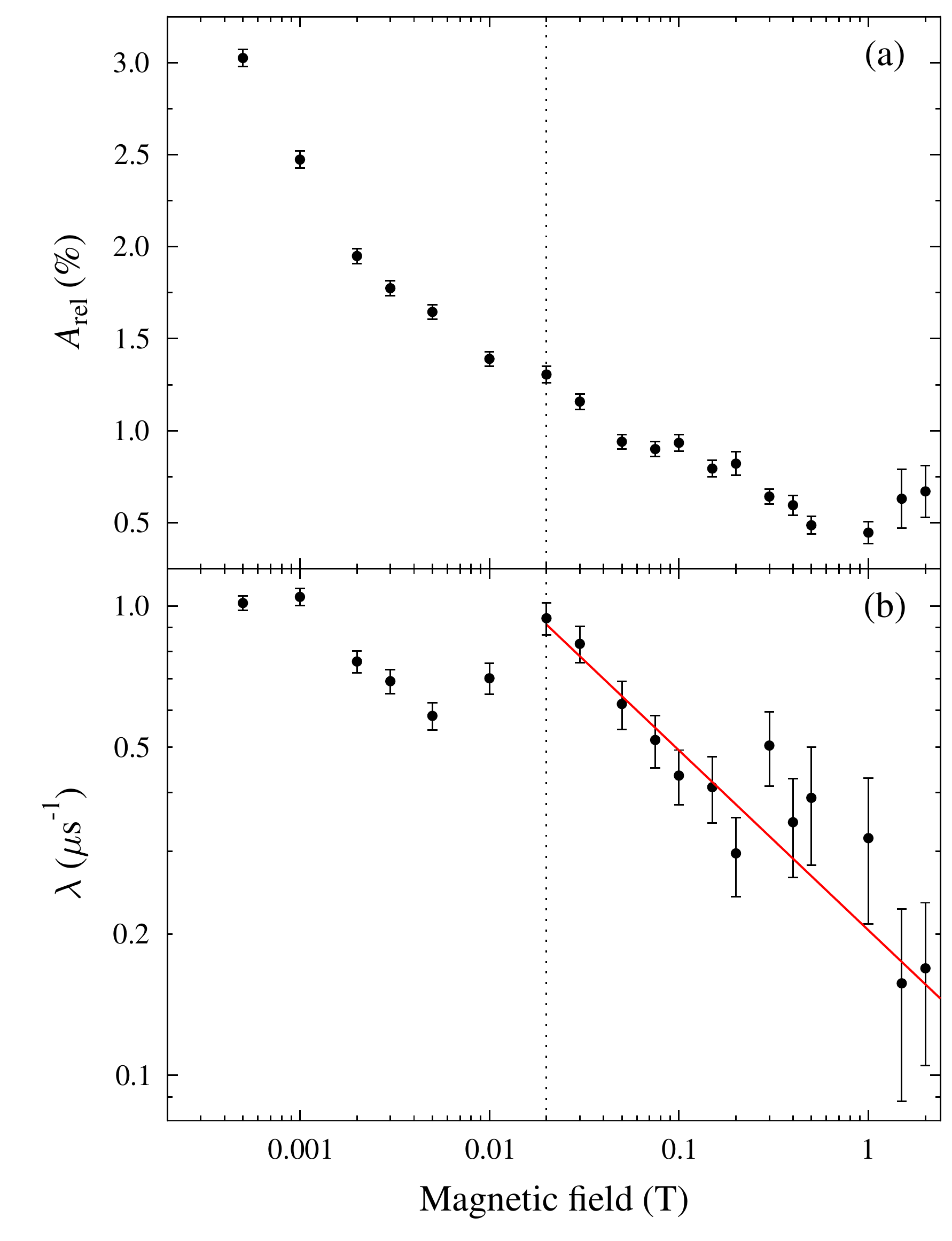}
	\caption{(a) Relaxing asymmetry and (b) relaxation rate for CuVOF$_4$(H$_2$O)$_6\cdot$H$_2$O as a function of applied longitudinal magnetic field.  In (b) we also show a fit to power law, appropriate for diffusive spin transport.}
	\label{fig:lf_lambda}
\end{figure}

We also carried out longitudinal field (LF) measurements at $T=1.2$~K for $0.5 \le B \le 2000$~mT. Example spectra at two different fields are shown in Fig.~\ref{fig:lf_spectra}. We see that the effect of an applied LF is to quench the relaxation of the slowly-relaxing component of the asymmetry, which is likely to have a significant contribution from quasistatic nuclear moments. We also see that the amplitude of the faster-relaxing component is reduced. In order to examine the field-dependence of the faster relaxation rate, the spectra were fitted to
\begin{equation}
A(t)=A_{\textrm{rel}}e^{-\lambda t}+A_{\textrm{bg}},
\label{eq_lf}
\end{equation}
where the relaxing asymmetry $A_{\textrm{rel}}$ reflects the dynamics of the electron spins and $A_{\textrm{bg}}$ represents a constant background asymmetry. The background asymmetry increases as a function of field, so was allowed to vary in the fits.  The relaxing asymmetry $A_{\textrm{rel}}$ [see Fig.~\ref{fig:lf_lambda}(a)] and the relaxation rate $\lambda$ [Fig.~\ref{fig:lf_lambda}(b)] were both found to decrease with increasing field.

For fields below around 20 mT, the relaxation due to nuclear moments is not fully quenched and Eq. \eqref{eq:lf_asymmetry} provides a relatively poor fit to the data, with the residual nuclear relaxation influencing the values of $\lambda$ obtained in these fits. The field-dependence of relaxing asymmetry is also particularly strong for these lower fields. On the other hand, for fields above 20 mT, the background relaxation is sufficiently quenched and Eq. \eqref{eq_lf} describes the spectra well. As shown in Fig.~\ref{fig:lf_lambda}, the field-dependence of the relaxation rate is well-described by a power-law fit of the the form $\lambda=aB^{-n}$ with  $n=0.38(4)$ for $20 \le B \le 2000$~mT.

%

\subsection{Density functional theory}

\begin{figure}[b]
    \centering
    \includegraphics[width = \columnwidth]{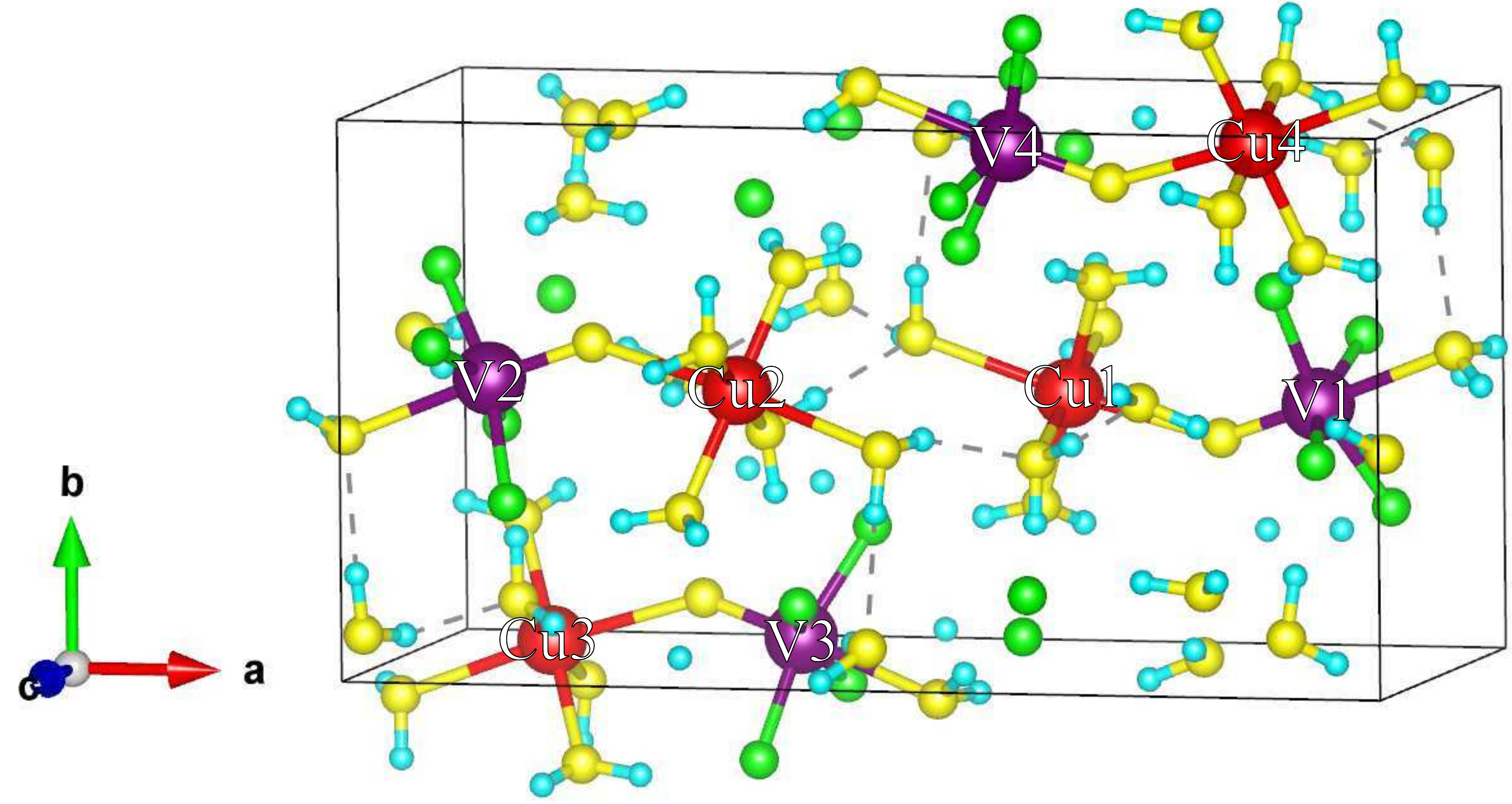}
    \caption{Unit cell of CuVOF$_4$(H$_2$O)$_6 \cdot$H$_2$O with labels corresponding to pairs of spins considered by Eq.~\ref{eq:ising} and as highlighted in Table~\ref{table:coupling_constants}.
    }
    \label{fig:CuV_labels}
\end{figure}

A commonly-used approach to calculating exchange constants using density functional theory involves computing the total energies of a series of spin configurations. The singlet-triplet gap $J_0$ for two electrons can be obtained from the energy difference between the single-determinant broken symmetry (BS) and triplet (T) states via $J_0=E_\mathrm{T}-E_\mathrm{BS}$ \cite{ruiz-1999}. This approach can be extended to polynuclear systems by expressing the energy differences between different spin configurations as the sum of pairwise interactions \cite{ruiz-2003}. The total energy for each configuration is then expressed in terms of an Ising model

\begin{equation}\label{eq:ising}
	E=\frac{1}{2}\sum_{i<j} J_{ij} \sigma_i \sigma_j+E_0,
\end{equation}
where the Ising spin operators $\sigma_{i,j}=\pm1$ and $E_0$ represents the nonmagnetic contribution to the total energy. This leads to a set of linear equations that can be solved to obtain the $J_{ij}$ for each pair of spin centres. 

\begin{table} 
	\begin{tabular}{lr}
		\hline
		\hline
		& \vspace{-0.3cm} \\
		Spin pair & $J_{ij}$ \\ 
		\hline \\
		& \vspace{-0.7cm} \\
		Cu1--V1 & $J_0$ \\
		Cu1--V3 & $J''$ \\
		Cu1--V4 & $4J'$ \\
		Cu2--V2 & $J_0$ \\
		Cu2--V3 & $4J'$ \\
		Cu2--V4 & $J''$\\
		Cu3--V1 & $J''$ \\
		Cu3--V2 & $4J'$\\
		Cu3--V3 & $J_0$ \\
		Cu4--V1 & $4J'$\\
		Cu4--V2 & $J''$ \\
		Cu4--V4 & $J_0$ \\	
		\hline
		\hline
	\end{tabular}
	\caption{Ising exchange constants assigned to each pair of spins in Eq. \eqref{eq:ising}. We set $J_{ij}=0$ for all other pairs.}\label{table:coupling_constants}
\end{table}

\begin{table} 
	\begin{tabular}{lr}
		\hline
		\hline
		& \vspace{-0.3cm} \\
		Spin configuration & Energy \\ 
		\hline \\
		& \vspace{-0.7cm} \\
		11111111 & $E_\mathrm{FM}=2J_0+8J'+2J''+E_0$ \\ 
		11110000 & $E_\mathrm{AFM1}=-2J_0-8J'-2J''+E_0$ \\ 
		11000011 & $E_\mathrm{AFM2}=-2J_0+8J'+2J''+E_0$ \\ 
		11001100 & $E_\mathrm{AFM3}=2J_0-8J'-2J''+E_0$ \\ 
		10011001 & $E_\mathrm{AFM4}=2J_0+8J'-2J''+E_0$ \\ 
		10010110 & $E_\mathrm{AFM5}=-2J_0-8J'+2J''+E_0$ \\ 
		\hline
		\hline
	\end{tabular}
	\caption{Spin configurations and their associated energies given by Eq. \eqref{eq:ising}.}\label{table:ising_energies}
\end{table}

We have used the DFT total energy approach to calculate the exchange constants corresponding to each of the exchange pathways in this system. The coupling constants $J_{ij}$ assigned to each pair of spins for the structure in Fig.~\ref{fig:CuV_labels} are shown in Table.~\ref{table:coupling_constants}. These comprise an intradimer exchange $J_0$ between Cu and V ion belonging to the same dimer, an interdimer exchange $J''$ between Cu and V ions belonging to nearest-neighbour dimers within the $bc$ plane and a further interdimer exchange $J'$ between dimers along the $a$ axis. By considering a ferromagnetic configuration and a series of antiferromagnetic configurations and mapping the DFT-calculated total energies to Eq. \eqref{eq:ising} we obtain the set of linear equations shown in Table \ref{table:ising_energies}. We denote each spin configuration by a list of 0s (spin down) and 1s (spin up) in the order Cu$_1$Cu$_2$Cu$_3$Cu$_4$V$_1$V$_2$V$_3$V$_4$. Solving these equations yields the following expressions for the exchange constants,
\begin{equation}\label{j_solutions}
	\begin{aligned}
		J_0=\frac{E_\mathrm{AFM3}-E_\mathrm{AFM1}}{4}, \\
		J'=\frac{E_\mathrm{AFM2}-E_\mathrm{AFM5}}{16}, \\
		J''=\frac{E_\mathrm{AFM4}-E_\mathrm{FM}}{4}.
	\end{aligned}
\end{equation} 

\begin{figure}[htb]
	\centering
	\includegraphics[width=0.8\columnwidth]{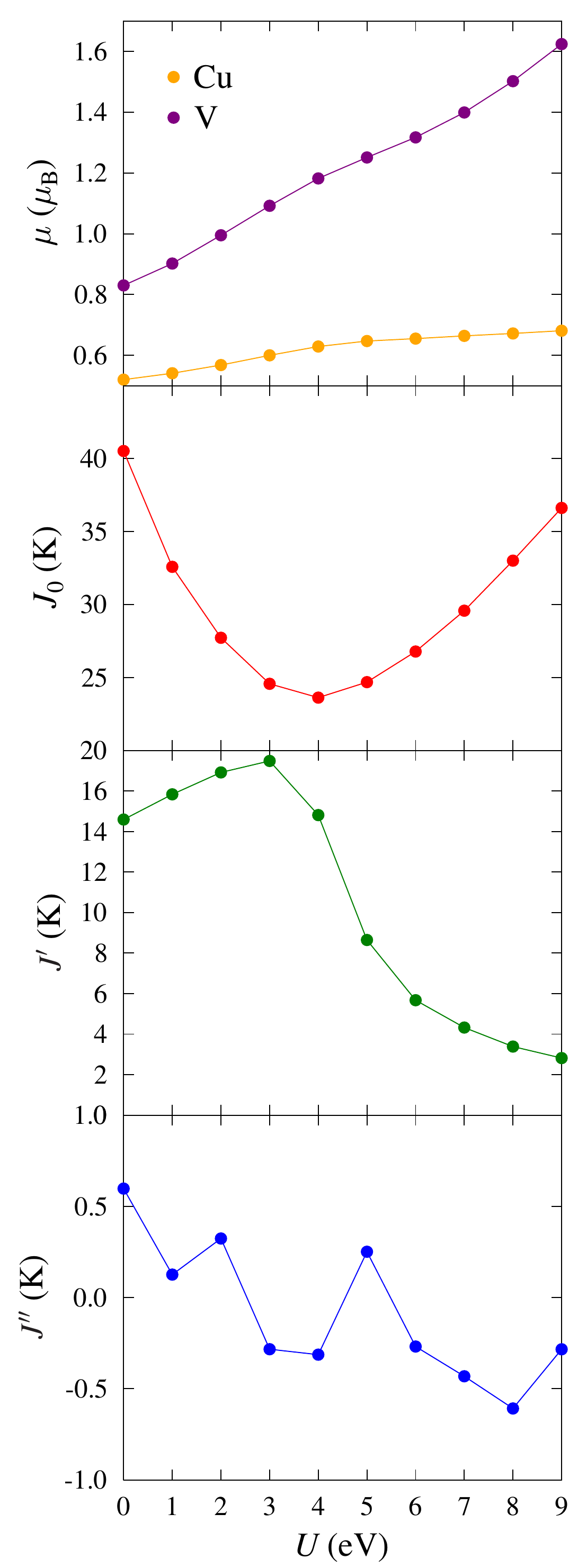}
	\caption{Ordered moments and exchange constants as a function of the Hubbard $U$ applied to both Cu and V $d$ orbitals.}
	\label{fig:JvsU}
\end{figure}

Magnetic exchange constants calculated using DFT can depend very strongly on the exchange-correlation functional employed \cite{martin}. The exchange constant $J \sim t^2/U$ where $t$ is the effective hopping integral and $U$ is the onsite Coulomb interaction. Functionals such as the local spin density approximation and the generalized-gradient approximation (GGA) underestimate the localization (and hence $U$) giving rise to an overestimate of $J_0$ \cite{martin}. Hybrid functionals such as B3LYP can be used to remedy this in some cases and this approach has been applied successfully to number of Cu-based molecular magnets \cite{DosSantos2016,Huddart2019}. Here, we instead using DFT+$U$, an extension to DFT where the strong on-site Coulomb interaction of localized electrons is treated with an additional Hubbard-like term.
To determine the total energy associated with each of these magnetic structures, we carried out spin-polarised DFT+$U$ calculations using the plane-wave basis-set electronic structure code \textsc{castep} \cite{CASTEP}. Calculations were carried out within the GGA using the PBE functional \cite{PBE}. We used a plane-wave cutoff energy of 1400 eV and a $1 \times 2 \times 2$ Monkhorst-Pack grid \cite{MPgrid} for Brillouin zone integration, results in an energy difference between the FM and AFM1 configurations that converges to around 0.2~meV per unit cell. To account for the strong on-site Coulomb repulsion of localized electrons, a Hubbard $U$ of equal magnitude was applied to both the Cu and V $d$ orbitals. The calculated exchange constants and the magnitude of the ordered moments (obtained using population analysis) for each of the ions in the AFM1 configuration are shown in Fig.~\ref{fig:JvsU}. 

As seen in Fig.~\ref{fig:JvsU}, the exchange constants depend very strongly on the value of $U$ chosen. We also see that the moments on the magnetic ions increase with increasing $U$. This is especially true for V, whose moment increases by a factor of 2 between $U=0$~eV and $U=9$~eV. The principal exchange $J_0$ initially decreases with increasing $U$, but reaches a minimum at $U=4$~eV and then recovers as $U$ increases further. On the other hand, the interdimer exchange $J'$ initially increases slightly with $U$, but for $U>3$~eV is rapidly suppressed with increasing $U$. The exchange constant $J''$ is very small for all $U$ and cannot be distinguished from zero within the uncertainties associated with these calculations. The principal exchange constant $J_0=24.7(6)$~K obtained using $U=5$~eV shows reasonable agreement with the experiment result while also having a physically plausible value of $U$. The corresponding interdimer exchange constant $J'=8.6(15)$~K is somewhat larger than the experiment result. A value of $J'$ can be obtained (at the expense of a larger $J_0$) by going to larger values of $U$, but these are somewhat unphysical. The uncertainties in the calculated $J_{ij}$ derives from limits on the convergence of the DFT energies with respect to the basis set \cite{IOThomas}. We note that while the hybrid functional B3LYP has been shown to accurately predict the exchange constants in some other Cu-based systems, calculations using this functional on C\MakeLowercase{u}(H$_2$O)$_5$VOF$_4\cdot$H$_2$O predict a ferromagnetic intradimer coupling, in clear contradiction with the experimental result.

The fact that both the intradimer and interdimer exchange constants are antiferromagnetic, i.e. $J_0,J'>0$ signifies that the configuration AFM1 is the ground state of the system.
In Fig.~8
~in the main text we show the spin density distribution for the ground state AFM1, obtained for $U=5$~eV (though we note that qualitatively similar results from $U=0$~eV).
As seen in Fig.~8
~in the main text, the spins of the Cu and V ions within a dimer are arranged antiferromagnetically, Cu and V spins belonging to neighboring dimers are also arranged antiferromagnetically.
The spin density distribution across a single dimer when the system is in its magnetic ground state is shown in Fig.~8(b)
~in the main text. There is significant spin density on the Cu and V ions; M\"{u}lliken population analysis shows that the total spin on the Cu and V ions is 0.647 $\hbar/2$ and 1.251 $\hbar/2$, respectively. The O atom joining the Cu and V within a dimer has a total spin of 0.234 $\hbar/2$, with this spin density having the opposite sign to the V ion within the dimer (this is also true for ferromagnetic dimer configurations). The Cu coordination octahedron experiences a Jahn-Teller (JT) distortion along the axis parallel to the Cu--O--V direction. This places the magnetic orbitals along the plane perpendicular to this axis, and we see this reflected in the shape of the isosurfaces of the spin density for the Cu ion. These orbitals lie along the Cu--O bonds and induces a total spin $\approx 0.09~\hbar/2$ on each the O atoms within this plane. (The combination of this delocalization of the Cu spin density across its coordination octahedron and the fact that the spin on central O atoms point in the opposite direction to the V spin means that the net spin for the whole system is close to zero, despite the greater spin density on the V ion compared to the Cu ion.) On the other hand, any spin density on the apical O atom on the outside of the dimer is very small. In light of this JT distortion and its effect on the spin density for O atoms along the JT axis, we should therefore consider the spin density on the O atom at the center of the dimer as instead resulting from an antiferromagnetic interaction with the V atom, with which it shares a short bond ($\approx 1.6$ \AA). The spin density on the F atoms is fairly small ($< 0.01~\hbar/2$), which can be explained by the fact that the V---F bonds lie along nodes of the spin density on the V.
 
The spin density on the central O atom and its antiferromagnetic alignment with the spin of the V ion is likely to play a key role in promoting the intradimer exchange, though it is not clear how this exchange continues along the dimer to the Cu ion, given the fact that this O atom lies along the JT axis of Cu. The three-dimensional (3D) structure is built up by connecting these dimer units through F$\cdots$H---O hydrogen bonds. This pathway links O atoms in the JT plane, that are well-coupled to the Cu spin, to the F atoms bonded to the V spin and this gives rise to interdimer exchange $J'$ and $J''$, within the $bc$ plane and along the $a$ axis, respectively. There are also hydrogen bonds between O atoms in the JT plane with apical O atoms on Cu ions in an adjacent dimer, but this exchange pathway might be expected to be weaker owing to the very small spin density on the apical O atom.  

\begin{figure*}[htb]
	\includegraphics[width=\textwidth]{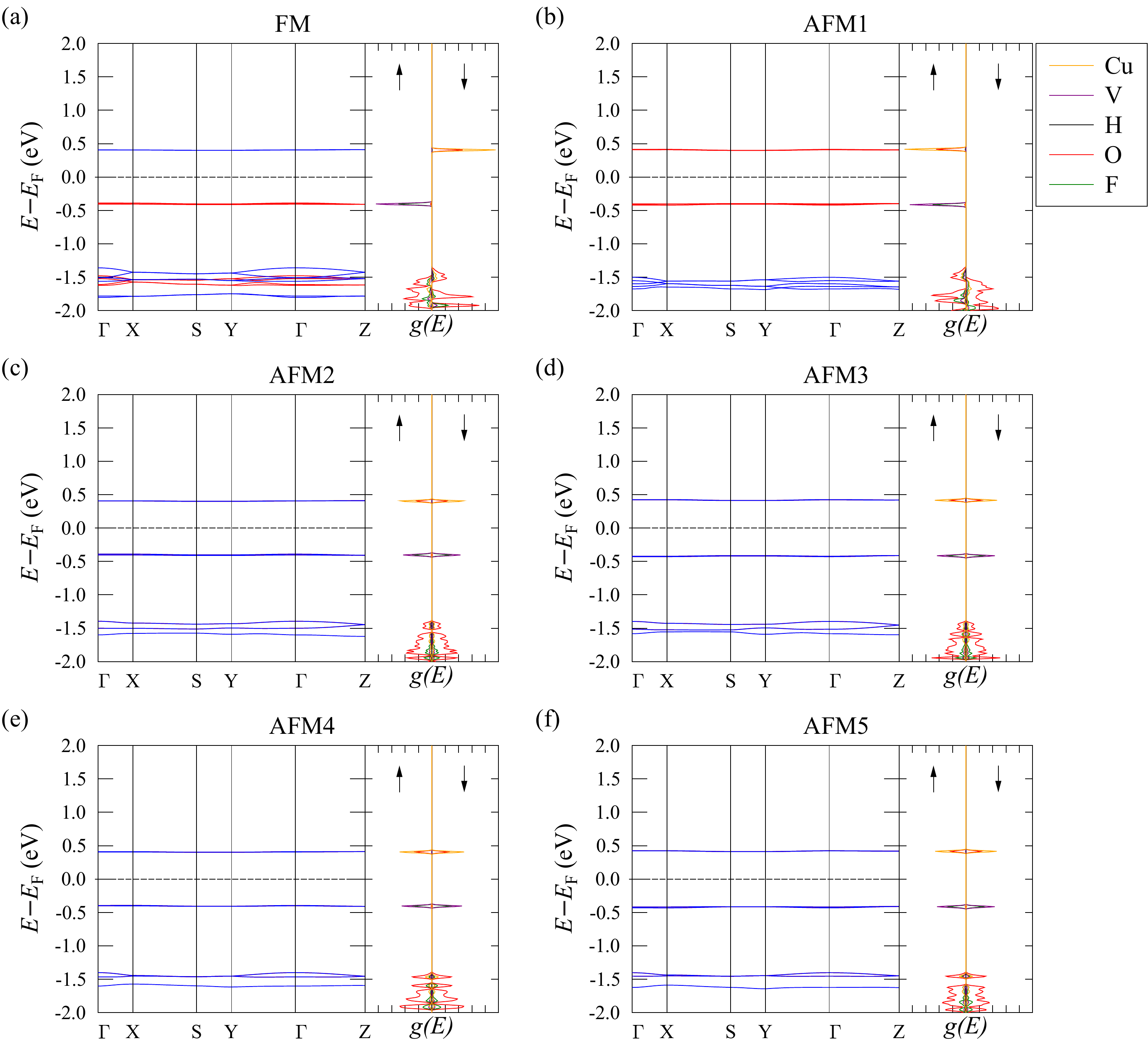}
	\caption{Spin-polarized band structures for each of the magnetic configuration and the density of states for each spin channel. Bands corresponding to spin-up and spin-down are indicates by red and blue lines respectively. The density of states are shown projected onto atomic species.}
	\label{fig:spectral}
\end{figure*}

Spin-polarized band structures and density of states (DOS) for each of these magnetic configurations are shown in Fig.~\ref{fig:spectral}. For calculations of the density of states a finer $5\times 10 \times 10$ for Brillouin zone sampling was used. For all configurations the band structure is characterized by two sets of flat bands, one just above and just below the Fermi energy. The lack of significant band dispersion for these states implies that they are localized to the Cu and V. As seen from the projected density of states (PDOS) these correspond to the Cu and V ions, respectively. The PDOS also show the hybridization between the Cu and V orbitals with those from O and F respectively. Note that these O atoms are those lying in the JT plane; the orbitals on both the apical and central O atoms along the JT axis instead lie further below the Fermi energy. For small values of $U$, these two sets of bands are sufficiently close to the Fermi energy ($<0.1$~eV away for $U=0$) that they are all fractionally occupied once smearing is taken into account. However, the band gap increases with increasing $U$, such that for $U=5$~eV the occupation of the bands above the Fermi energy is almost zero, with the bands below being close to fully occupied.

The band structures and PDOS for configurations AFM2, AFM3, AFM4 and AFM5, shown in Figs.~\ref{fig:spectral}, are fairly typical for an antiferromagnetic system, with degenerate up and down bands and equal DOS in the up and down spin channels. For the FM state the Cu and V PDOS occupy opposite spin channels. The spin-up V orbitals are occupied and the spin-down Cu orbitals are unoccupied, resulting in an overall spin polarisation. For the AFM1 configuration the Cu and V PDOS instead occupy the same spin channel. This has an effect on which spin channel the Cu PDOS that lie well below the Fermi energy (shown in Fig.~9
~in the main text) occupy, such that the majority of occupied Cu orbitals are spin-down, giving an overall antiferromagnetic state.

\subsection{Tight-binding model}

To more directly assess the relative strengths of the various exchange pathways we developed a description of the system in terms of a tight-binding model. To obtain an appropriate set of basis function we constructed maximally-localized Wannier functions \cite{wannier1,wannier2}. First-principles calculations were carried out using \textsc{Quantum ESPRESSO} \cite{QE-2009,QE-2017}, with Wannier functions being computed using the interface with \textsc{Wannier90} \cite{w90}. As with our earlier calculations using \textsc{castep}, these calculations were carried out within the GGA using the PBE functional \cite{PBE}. To simplify our analysis, the system was treated as non-spin-polarized. Furthermore, values of the Hubbard $U$ are generally not transferable, with their values depending on many physical or technical details (such as the pseudopotentials used), and hence the Hubbard $U$ was omitted from these calculations. Pseudopotentials were obtained from the SSSP efficiency library \cite{SSSP}, which specifies a projector-augment wave (PAW) pseudopotential for O \cite{Pslibrary}, ultrasoft (US) pseudopotentials for F, V, and Cu \cite{GBRV}, and an US pseudopotential for H \cite{DalCorso}. Energy cutoffs of 60 Ry and 480 Ry where used for the wavefunctions and density, respectively, and $3 \times 4 \times 5$ Monkhorst-Pack grid \cite{MPgrid} was used for Brillouin zone sampling. 

\begin{figure}[b]
	\includegraphics[width=\columnwidth]{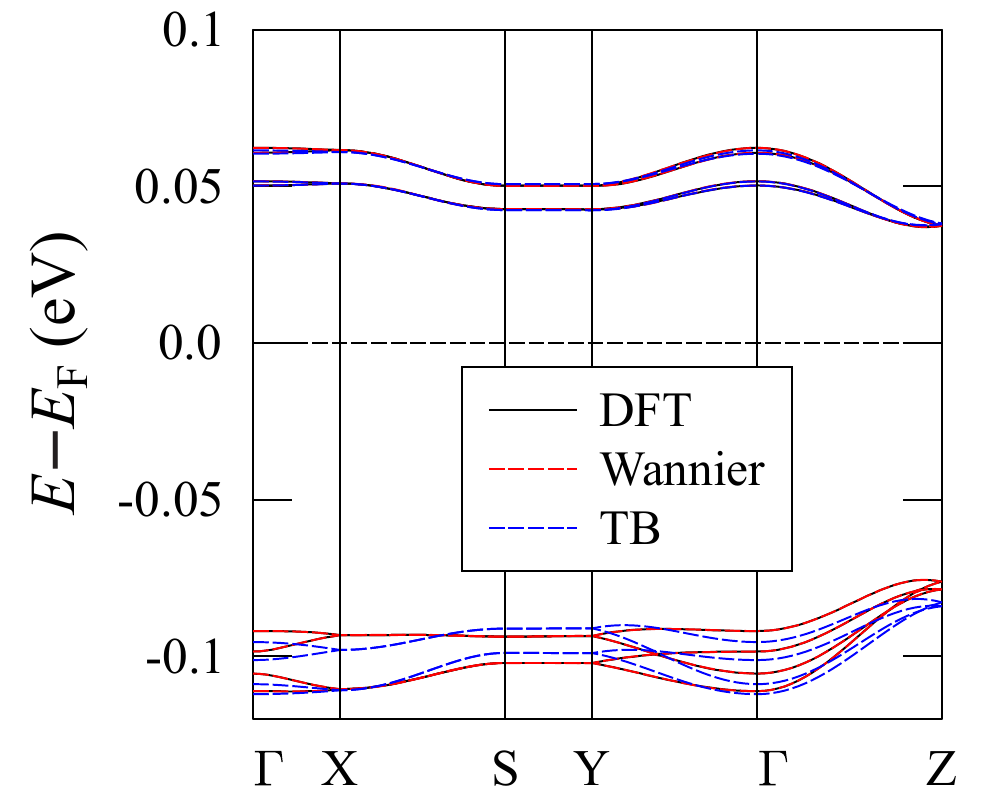}
	\caption{Electronic band structures calculated using DFT (black solid line), a Wannier reconstruction (red dashed line) and a tight-binding model limited to nearest-neighbor hopping  (blue dashed line).}
	\label{fig:bands}
\end{figure}

\begin{figure}[b]
	\includegraphics[width=\columnwidth]{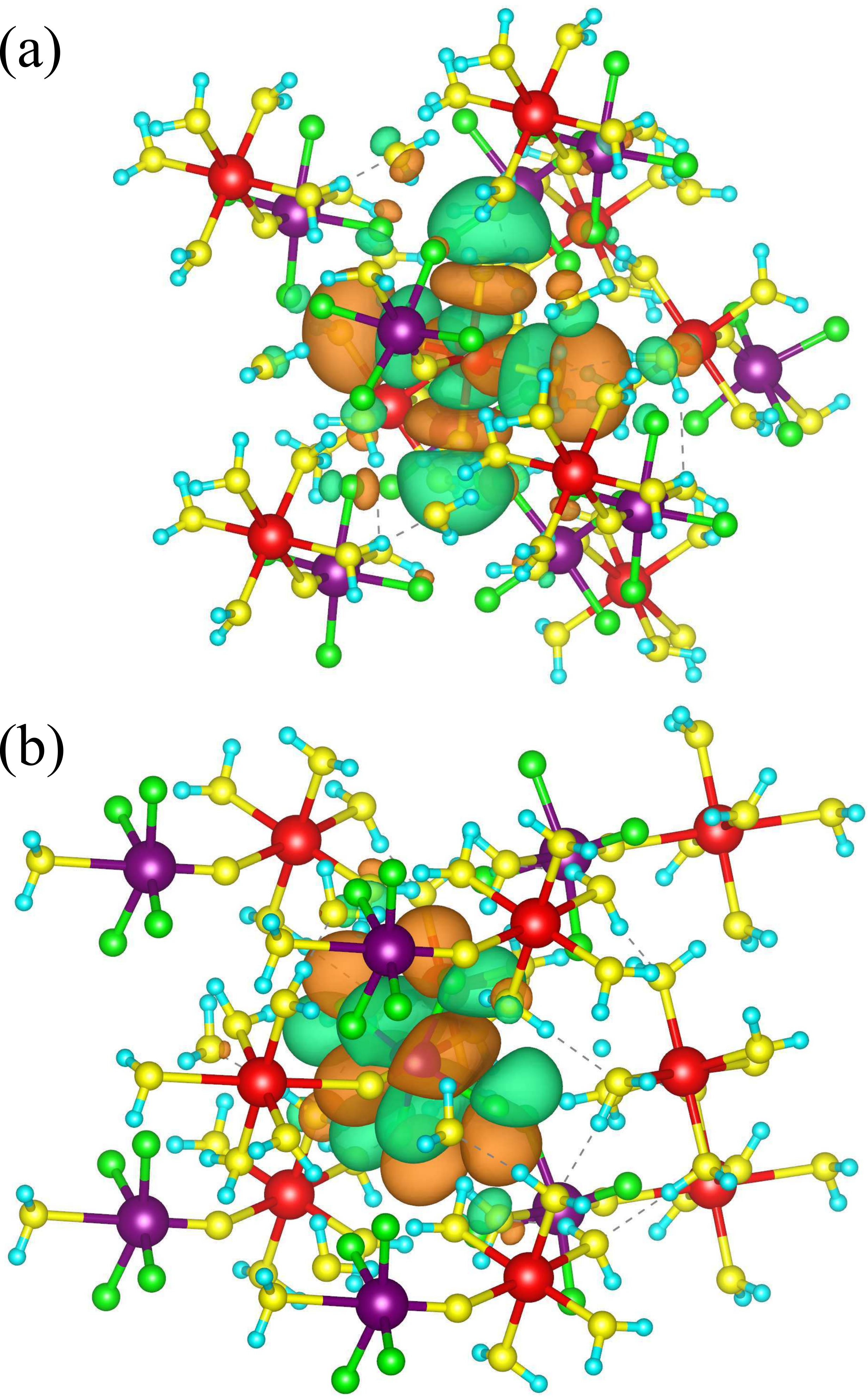}
	\caption{The two distinct classes of maximally-localized Wannier functions, which are localized either on (a) Cu or (b) V.}
	\label{fig:wannier}
\end{figure}

The DFT band structure in the vicinity of the Fermi energy is shown in \ref{fig:bands}. It comprises two sets of four bands, either $\approx 0.05$ eV above or $\approx 0.1$ eV below the Fermi energy and well-separated from the other bands. We note that due to the smearing employed in these calculations, the set of bands just above the Fermi energy are partially occupied. (The occupancies of the bands just below and above the Fermi energy are around 0.74 and 0.26, respectively). We therefore constructed a set of eight Wannier functions from the orbitals corresponding to these bands. The resulting Wannier functions are shown in Fig.~\ref{fig:wannier}, and fall into two distinct categories. These functions are either localized on Cu [Fig.~\ref{fig:wannier}(a)] or V [Fig.~\ref{fig:wannier}(a)]. There are four functions of each type, each localized on a different Cu or V ion within the unit cell. 

We proceeded to construct a single-particle Hamiltonian using the set of Wannier orbitals $\ket{i\textbf{R}}$, centered on the Cu or V ion $i$ in the unit cell whose origin is at $\textbf{R}$. The overlaps between the Wannier orbitals
\begin{equation}
	h_{ij}(\textbf{R})=\bra{i\textbf{0}}\hat{\mathcal{H}}\ket{j\textbf{R}},
\end{equation}
where $\hat{\mathcal{H}}$ is the full microscopic Hamiltonian, lead to a model Hamiltonian
	\begin{equation}\label{wannier_ham}
		\hat{H}=\sum_{i,j,\textbf{R}}\hat{c}^{\dagger}_{i\textbf{0}}h_{ij}(\textbf{R})\hat{c}_{j\textbf{R}},
	\end{equation}
where $\hat{c}^{(\dagger)}_{j\textbf{R}}$ annihilates (creates) an electron at site $j$ in the unit cell at $\textbf{R}$. Diagonalizing $\hat{H}$ yields the band structure shown as a red dashed line in Fig.~\ref{fig:bands}. We see that the DFT band structure is reproduced accurately, demonstrating that the Wannier orbital construction was successful.

To assess the relative importance of the various exchange pathways, we constructed a tight-binding model based on Eq.~\ref{wannier_ham}, but limited to nearest-neighbor interactions,
	\begin{equation}
	\hat{H}_\mathrm{TB}=\sum_{i,j}\sum_{\{\textbf{R}\}_{ij}}\hat{c}^{\dagger}_{i\textbf{0}}h_{ij}(\{\textbf{R}\}_{ij})\hat{c}_{j\textbf{R}},
\end{equation}
where the sum over lattice vectors is restricted to the set $\{\textbf{R}\}_{ij}$ for which $\ket{i\textbf{0}}$ and $\ket{j\textbf{R}}$ are centered on nearest-neighbor ions. Solving this model results in the band structure shown as a blue dashed line in Fig.~\ref{fig:bands}. We see that while this model captures the main features of the band structure, its agreement with the DFT result is not as good as for the full Wannier reconstruction, which includes a significantly larger number of matrix elements $h_{ij}(\textbf{R})$.

The evaluation of the hopping strength for a given exchange pathway is complicated by the fact that, for each pair of ions $i$ and $j$, $h_{ij}(\{\textbf{R}\}_{ij})$ can take more than one distinct value. This is due to the nearest-neighbor distances along different directions varying slightly, resulting in different overlaps. To simplify our analysis we define a hopping strength $t_{ij}$ for each pair of ions as $t_{ij}=\max\{h_{ij}(\textbf{R}):\textbf{R}\}$. Denoting each Wannier function by the ion at its center, we have in the basis (Cu1,Cu2,Cu3,Cu4,V1,V2,V3,V4),
 
 \begin{widetext}
 \begin{equation}
 	\textbf{t}=\begin{pmatrix}
0.01 & 0.63 & -0.41 & 1.53 & 2.87 & -0.21 & -5.20 & -25.28 \\
0.63 & 0.01 & 1.53 & -0.41 & -0.21 & 2.87 & -25.28 & -5.20 \\
-0.41 & 1.53 & 0.01 & 0.63 & -5.20 & -25.28 & 2.87 & -0.21 \\
1.53 & -0.41 & 0.63 & 0.01 & -25.28 & -5.20 & -0.21 & 2.87 \\
2.87 & -0.21 & -5.20 & -25.28 & -0.79 & 0.76 & 0.06 & 0.61 \\
-0.21 & 2.87 & -25.28 & -5.20 & 0.76 & -0.79 & 0.61 & 0.06 \\
-5.20 & -25.28 & 2.87 & -0.21 & 0.06 & 0.61 & -0.79 & 0.76 \\
-25.28 & -5.20 & -0.21 & 2.87 & 0.61 & 0.06 & 0.76 & -0.79 \\
 	\end{pmatrix}\mathrm{meV}.
 \end{equation}
\end{widetext}
We first note that the hoppings between Cu ions (top-left quadrant) and between V ions (bottom-right quadrant) are generally smaller than those between Cu and V ions, justifying the omission of these exchange pathways from our earlier model for computing the exchange constants. For the hoppings resulting from the overlap between Cu and V ions, we can map these to the exchange constants calculated earlier, as these should be related via $J\sim t^2/U$. We find $t=2.87$~meV, $t'=-25.28$~meV and $t''=-5.20$~meV, corresponding to $J_0$, $J'$ and $J''$, respectively. Immediately of note is the fact that, of these three transfer integrals, the one corresponding to the principal exchange $J_0$ is actually the weakest. This can be rationalized by considering the shapes of the Wannier functions in Fig.~\ref{fig:wannier}. For both Cu- and V-centered Wannier functions, the wavefunctions are largest in the plane orthogonal to the intradimer direction, and hence their overlap within a dimer is small. On the other hand, the transfer integral $t'$ describing the overlap between these orbitals on adjacent dimers within the $bc$ plane is much larger to the strong overlap for these directions. To obtain a transfer integral that more accurately reflects the interdimer exchange, we must consider that the interdimer coupling 4$J'$ is due to exchange between a Cu ion on a dimer and a V ion on each of four nearest-neighbour dimers. As noted above, these interdimer distances are not all equal and hence $J'$ better represents the average coupling for exchange between dimers in the $bc$ plane. We therefore consider all of the distinct transfer integrals corresponding to these pairs of Cu and V ions. These are $t'_1=-25.28$ meV, $t'_2=-21.64$ meV, $t'_3=-7.72$ meV, $t'_4=1.45$ meV. Bearing in mind that $J \propto t^2$ we can formulate an average hopping strength for interdimer coupling in the $bc$ plane as
\begin{equation}
	|t'_\mathrm{avg}|=\sqrt{[(t'_1)^2+(t'_2)^2+(t'_3)^2+(t'_4)^2]/4}.
\end{equation}
The resulting effective hopping $|t'_\mathrm{avg}|=17.1$~meV is still significantly larger than the intradimer hopping. We note that treating the system as spin-polarized instead produces nearly identical results to these, due to the fact that, in the magnetic ground state, the bands close to the Fermi energy all occupy the same spin channel.



\providecommand{\noopsort}[1]{}\providecommand{\singleletter}[1]{#1}%